\documentclass[12pt]{article}
\usepackage{graphicx}
\usepackage{epsfig}
\usepackage{epstopdf}
\DeclareGraphicsExtensions{.pdf,.eps,.png,.jpg,.mps}

\def\lsim{\mathrel{\rlap{\lower3pt\hbox{\hskip0pt$\sim$}}
     \raise1pt\hbox{$<$}}}         
\def\gsim{\mathrel{\rlap{\lower4pt\hbox{\hskip1pt$\sim$}}
     \raise1pt\hbox{$>$}}}         

\usepackage{amsmath}
\usepackage{amsfonts}
\begin{document}
\begin{titlepage}

\centerline{\Large \bf Phynance}
\medskip

\centerline{Zura Kakushadze$^\S$$^\dag$\footnote{\, {\tt Email: zura@quantigic.com}. Emails pointing out any typos or other inadvertent errors that slipped through the cracks are more than welcome and will be greatly appreciated.}}

\bigskip

\centerline{\em $^\S$ Quantigic$^\circledR$ Solutions LLC}
\centerline{\em 1127 High Ridge Road \#135, Stamford, CT 06905\,\,\footnote{\, DISCLAIMER: This address is used by the corresponding author for no
purpose other than to indicate his professional affiliation as is customary in
publications. In particular, the contents of this paper
are not intended as an investment, legal, tax or any other such advice,
and in no way represent views of Quantigic® Solutions LLC,
the website \underline{www.quantigic.com} or any of their other affiliates.
}}
\centerline{\em $^\dag$ Department of Physics, University of Connecticut}
\centerline{\em 1 University Place, Stamford, CT 06901}
\medskip
\centerline{(May 6, 2014)}

\bigskip
\medskip
\centerline{\it Dedicated to the memory of my father}
\centerline{\it Jemal Kakushadze, Ph.D. (1940-2005)}

\bigskip
\medskip

\begin{abstract}
{}These are the lecture notes for an advanced Ph.D. level course I taught in Spring'02 at the C.N. Yang Institute for Theoretical Physics at Stony Brook. The course primarily focused on an introduction to stochastic calculus and derivative pricing with various stochastic computations recast in the language of path integral, which is used in theoretical physics, hence ``Phynance". I also included several ``quiz" problems (with solutions) comprised of (pre-)interview questions quantitative finance job candidates were sometimes asked back in those days. The course to a certain extent follows an excellent book ``Financial Calculus: An Introduction to Derivative Pricing" by M. Baxter and A. Rennie.
\end{abstract}
\end{titlepage}

\newpage

\tableofcontents
\setcounter{tocdepth}{3}
\listoffigures
\newpage

\section{Introduction: How Does ``Bookie the Crookie" Make Money?}

{}When odds are quoted in the form ``$n-m$ against'', it means that
the event has probability $m/(n+m)$, and a successful bet of $\$ m$
is rewarded with $\$ n$ (plus the stake returned).

{}Similarly, when the odds are quoted in the form ``$n-m$ on'', it is the
same as ``$m-n$ against''.

{}Suppose we have two horses, with the true odds $n-m$ against the first horse.
Suppose the gamblers bet total of $B_1$ on the first horse, and $B_2$ on the
other horse. Then if the first horse wins, the bookmaker makes a net profit
(this could be a gain or a loss) of
\begin{equation}
 P_1=B_2-{n\over m} B_1~,
\end{equation}
while if the second horse wins, the bookmaker makes a net profit of
\begin{equation}
 P_2=B_1-{m\over n} B_2~.
\end{equation}
The average long-term profit is
\begin{equation}
 \langle P\rangle = {m\over {n+m}}P_1+{n\over {n+m}} P_2=0~,
\end{equation}
so the bookmaker breaks even by quoting the true odds.

{}To make a long-term profit, the bookmaker sells more than 100\% of the race
by quoting somewhat different odds than the true odds. Thus, let the odds
quoted for the first and the second horses be $n_1-m_1$ against and
$n_2-m_2$ on, respectively. Now the average long-term profit is
\begin{eqnarray}
 \langle P\rangle =&& {m\over {n+m}}\left[B_2-{n_1\over m_1} B_1\right]+
 {n\over {n+m}} \left[B_1-{m_2\over n_2} B_2\right]=\nonumber\\
  &&{nB_1\over{n+m}}\left[1-{mn_1\over nm_1}\right]+
 {mB_2\over{n+m}}\left[1-{nm_2\over mn_2}\right]~.
\end{eqnarray}
Thus, the bookmaker can guarantee positive $\langle P\rangle$ by setting
$n_1,m_1$ and $n_2,m_2$ such that
\begin{eqnarray}
 &&{n_1\over m_1}<{n\over m}~,\\
 &&{m_2\over n_2}<{m\over n}~.
\end{eqnarray}
Note that the implied probabilities then are larger than the true
probabilities:
\begin{eqnarray}
 &&{m_1\over{n_1+m_1}}>{m\over{n+m}}~,\\
 &&{n_2\over{n_2+m_2}}>{n\over{n+m}}~,
\end{eqnarray}
so that the bookmaker is, in fact, selling more than 100\% of the race.
As the saying goes, lottery is a tax on people who don't know math.

\section{Bid, Ask and Spread}

{}Something similar to the bookmaker example discussed above occurs in financial markets. Let's consider a stock XYZ. There are the buyers, and there are the sellers. The buyers quote their bids, the sellers quote their asks (or offers), together with how many shares of the stock they want to buy/sell. Let $B$ be the highest bid price, and let $A$ be the lowest ask price. The difference $S \equiv A - B$ is called the bid-ask spread. Typically, $S > 0$.

{}If $S = 0$ (this is called {\em locked market}), then the lowest ask $A$ is the same as the highest bid $B$, and a transaction will occur at that price $P = A = B$, where a seller (or sellers) will transfer to a buyer (or buyers) their shares. The number of shares $V$ sold at that price equals $V = \mbox{min}(V_{Bid}, V_{Ask})$, where $V_{Bid}$ is the total number of shares quoted by the buyers at the price $P$ and $V_{Ask}$ is the total number of shares quoted by the sellers at the price $P$.

{}If $S < 0$ (this is called {\em crossed market}), then the lowest ask is below the highest bid, and a transaction will also occur, but the price $P$ at which it occurs will be in the range $A\leq P \leq B$ and it can depend on a variety of factors, {\em e.g.}, the precise algorithm employed by a given exchange for determining $P$ can depend on the timing of when various bids and asks where placed into the queue by the buyers and sellers. In fact, there might be more than one prices $P_i$ at which the transactions can occur with varying numbers of shares $V_i$ sold at those prices. Some buyers/sellers may receive what is known as {\em price improvement}, {\em e.g.}, a buyer bids 100 shares of XYZ at the price $B$ and his order is {\em filled} (this is market lingo) at a better price $P < B$.

{}So, one way to make money in the stock market is to be a market-maker, constantly selling at the ask and buying at the bid. Assuming the spread $S > 0$, if you buy $V$ shares of XYZ at the bid $B$ and then turn around and sell them at the ask $A$, your profit will be $V\cdot(A - B) = V\cdot S$. You have traded $2V$ shares (bought $V$ shares and sold $V$ shares), so your profit-per-share is $S/2$. (Typically, the spread is quoted in cents, and the profit-per-share is quoted in {\em cents-per-share}.) This is known as making {\em half-spread}. Similarly, if you go into the market and buy at the ask and sell at the bid -- this is called buying and selling {\em at market} (because you're paying the market prices) -- then you're incurring half-spread {\em transaction cost} on your trades, and the market-makers are making their half-spread on your transactions.

{}Nonetheless, plenty of people incur half-spread transaction cost on their trades because the way they make money is not by market-making but by capitalizing on stock price movements that are larger than the bid-ask spread. There is {\em technical analysis}, which is based on statistical analysis of market activity based on patterns and does not concern itself with the fundamentals of each company, which in contrast is what {\em fundamental analysis} does -- it makes investment decisions based on the fundamentals of the company, such as growth potential, earnings, {\em etc.} By its very nature, typically fundamental analysis operates on the time scales which are longer than those of technical analysis. Whatever the method, the money making motto is ``Buy low, sell high!" In practice, it's much harder to do than it sounds.

\section{Stocks, Bonds and Free Markets}

{}Stocks and bonds as well as other financial instruments are important ingredients of free market economy. Financial markets and the economy itself are products of human civilization, and, therefore, are not directly governed by the fundamental laws of nature ({\em i.e.}, laws of physics). Nonetheless, it is fascinating that they are based on certain universal principles, and there are reasons why the financial markets have been efficiently integrated into the free market economy notwithstanding the fact that the system is by no means perfect, which sometimes results in failures such as stock market bubbles and crashes.

{}One of the most fundamental principles of the free market economy is the interplay between supply and demand. Thus, regardless of what specifically is being traded, whether it is goods, commodities, stocks or other valuable instruments, buyers, who create demand, drive its price up, while sellers, who are suppliers, drive the price down. The supply and demand then determine the price. For instance, if sellers are asking an unreasonably high price not reflecting current demand levels, trades at this price are unlikely to occur in large quantities as the buyers will not be willing to pay more than they have to. Similarly, if the current supply level is low, then a buyer bidding at an unreasonably low price cannot expect to successfully complete a trade at that price -- most likely there will be other buyers bidding at higher price levels more acceptable to the suppliers.

{}Stock and bond markets as any other free market generally are expected to operate in this way -- buyers drive stock prices up, while sellers drive them down. This simple principle does indeed work in the financial markets, but what determines the supply and demand for a given financial instrument is quite nontrivial and is often times dictated by certain important details of how these markets are structured, which set the rules of the game. The purpose of this section\footnote{\, This section (with minor modifications) appeared some number of years ago as a standalone article in the online magazine Kvali.com.} is to elucidate some aspects of financial markets, in particular, why there exists demand for stocks and bonds, that is, why investors are willing to allocate their funds in these financial instruments. Nontrivial, and perhaps even controversial, issues arise in this regard as there is no fundamental law of nature that would dictate that any of these instruments should exist in the first place.

{}Let us begin with bonds. There are various types of bonds with different features, and we will not attempt to describe them all in detail; rather, we will focus on those that most bonds have in common. A bond is an obligation where the issuer of the bond promises to the purchaser to pay back the so-called face or par value of the bond or some other amount at some later time called maturity of the bond. Typically bonds also make periodic (mostly annual or semi-annual) coupon payments to the purchaser. Basically, the issuer of the bond borrows money from the purchaser and makes a promise that at maturity this money will be returned to the purchaser along with some additional amount, some of which might be paid before maturity, which is essentially the interest the purchaser earns. Thus, consider a simple example where a bond, which matures in exactly one year, has a face value of \$1,000. The purchaser pays this amount now to acquire the bond, and the issuer promises to pay back \$1,000 at maturity (that is, in one year from the purchase date) plus \$50 as a one-time coupon payment, which is also paid at maturity. The purchaser's investment of \$1,000, therefore, has 5\% annual return or yield. Note that if for some reason the price to purchase such a bond went up to, say, \$1,250, then the corresponding yield would go down to 4\% (assuming that the coupon payment is fixed), while if it dropped down to, say, \$500, then the corresponding yield would go up to 10\%. Thus, the higher the price the lower the yield, and {\em vice-versa}.

{}Bonds, being obligations, are typically relatively low risk investments. However, they do bear some risk, in particular, credit risk -- after all, the bond issuer can sometimes default, that is, declare bankruptcy, in which case it might not always be possible to receive the originally invested amount as well as some or all of the promised coupon payments. Bonds issued by governments of stable countries such as U.S. Treasury bonds are virtually risk free -- government debt is a very low risk investment because it is backed by the taxation power of the government.\footnote{\, Nonetheless, S\&P's downgrade of the U.S. credit rating from AAA (outstanding) to AA+ (excellent) on August 5, 2011 is a fact!} Indeed, if the government debt is not unreasonably high, the government can exercise its ability to increase taxes to pay down its debt. Municipal bonds are issued by State and local governments, typically to raise money for developing local infrastructure (building roads, hospitals, {\em etc.}). State issued bonds can also be backed by the taxation power of a State. In the United States interest earned from such bonds is exempt from State taxes, albeit Federal taxes must still be paid on such interest income. State issued bonds, therefore, typically have lower yields than other comparable bonds (with the same credit risk) -- this is because otherwise it would be more advantageous to invest into State issued bonds than in the comparable bonds as the former earn interest taxed at a lower rate, so increased demand on such bonds would drive their prices up, and, consequently, yields down, until it is no longer more advantageous to invest in the State issued bonds over the comparable bonds.
Other Municipal bonds, such as those issued by local governments, usually bear higher risk as (at least partially) they are typically backed by future returns of the investment for which the money is raised by issuing the bonds. For instance, if a town needs to build a new hospital, to raise required funds it could issue bonds backed by future returns from the hospital. However, not all such undertakings are always successful, hence higher risk associated with such bonds. Higher risk bonds typically have higher yields. This is an example of a more general principle -- higher risk investments should have higher expected returns. Indeed, if one could enjoy the same return from a lower risk investment as from a riskier one, one would clearly tend to choose the former. Since the demand for lower risk investments would then be higher than for their higher risk counterparts, the price one would end up paying for a lower risk investment would also be higher, while the corresponding yield would be lower. Put another way, the ratio of the return over the associated risk should generally be approximately the same for all investments or else the supply and demand paradigm will eventually make sure that it is.

{}Not only various governments but also private sector corporations can issue bonds to raise money to develop a new product, open a new factory plant, {\em etc.} Corporate bonds have higher yields as they are riskier than government issued bonds -- their credit risk is higher. Corporate bonds are backed by the ability of a corporation to generate earnings from sales of products and/or services, so if the business is not doing too well, the credit rating or the corporation goes down, the prices of its bonds also go down, and the yields go up. Since corporations can default, the ability of their bond holders to collect at least portions of their original investments in the case of bankruptcy is important. In fact, corporate bond holders are the first ones in line to partially if not completely get their money back from the proceeds of liquidation of the corporation after its default. The stock holders, on the other hand, have lower priority in the liquidation process and may receive nothing even if the bondholders are completely or partially compensated.

{}Stocks and corporate bonds are different in many more ways than the one just mentioned. When a corporation issues bonds, it borrows money from bond holders, that is, its outstanding bonds count toward its debt. There is an alternative and somewhat easier way for a corporation to raise money -- it can issue stock. There are two main types of stock, preferred stock and common stock. The preferred stock can roughly be thought of as a hybrid between a corporate bond and the common stock. In the following we will mostly focus on the common stock, and for the sake of brevity we will omit the adjective ``common". Let us, however, mention that, once the corporation defaults, in the liquidation process bond holders, preferred stock holders and common stock holders are compensated with the decreasing priority.

{}Outstanding stock is not a form of corporate debt, in particular, stocks are not obligations, they have no maturity, and the corporation does not promise to pay back the stockholders their originally invested amount any time later. Instead, stockholders or shareholders are owners of the corporation in the proportion to the total stock issued by the corporation. Some of this stock, which is called treasury stock, can be owned by the corporation itself. In fact, the total value of the corporation, which is referred to as its market capitalization, is determined by the number of issued shares multiplied by the current market price of one share. The latter, in turn, depends on the free market supply and demand levels for the shares of the corporation. It is important to note that stocks can and do become undervalued or overpriced in the free market, and the reasons for this are manifold. We will return to this point once we discuss some of the factors that are expected to determine what the ``fair" price of a given stock should be.

{}So, what compels investors to allocate their funds in stocks? Thus, unlike bonds, stocks do not pay coupons, that is, shareholders do not earn interest. Some stocks do pay dividends, however. Typically the annualized stock dividend is a low single digit percentage of the current stock price. Whether the stock pays a dividend is decided by the corporate governing body (the board of directors), and the amount of dividend can be changed (including to paying no dividend at all) without the shareholders' approval. For instance, if the business is not doing too well, the corporation may decide to no longer pay out dividends. This usually will result in a decline in the stock price as the demand for the stock most likely will decrease.

{}Even though shareholders do not earn interest, do not essentially have any guarantees as to recovering their investment in the future, and may not even be paid any dividends, they are (partial) owners of the corporation. This ownership entitles them to certain rights such as a right to vote for various corporate decisions including electing the board of directors. Also, if another entity (such as another corporation) intends to acquire the corporation, the current stock holders can vote for or against such a takeover depending on whether it is in their interests or not. This is one of the key reasons why some investors are willing to become shareholders. Thus, imagine that a corporation is doing well, and has good revenues as well as earnings. If, for some reason, the market price for its shares is unreasonably low, another entity could buy enough shares in the open market and attempt a hostile takeover of the corporation -- each share gives this hostile entity one vote, and all it needs is 51\% of the votes for a successful takeover. This might not be in the interests of the corporation, which includes its board of directors, who are typically shareholders themselves, its officers as well as all other shareholders. The board of directors, which is expected to act in the interests of at least most shareholders (after all, it was elected by the majority of shareholders' votes), in this case is likely to decide that the corporation should buy back some of the outstanding shares in the open market, which will ultimately result in an increase in the stock price. This buy-back mechanism then is expected to ensure that the stock price grows as the revenues and more importantly earnings of the corporation grow -- the corporation must pay cash to buy back some of its outstanding shares, and the ability to do so is directly linked to its earnings.

{}Thus, it is the earnings of the corporation that are expected to determine the price for its shares. Therefore, if an investor believes that the corporation has strong fundamentals, {\em i.e.}, the ability to generate earnings in the future, he or she might decide to become a shareholder. Generally, such an investment bears higher risk than a comparable bond investment. Thus, a typical stock price has annual volatility, which is a measure of how much it fluctuates, of 30-35\%, while bonds usually have volatility in the 5-7\% range. (These figures can vary depending on the economic cycle.) Since stocks are higher risk investments, they should have adequately higher returns, and historically on average this indeed appears to be the case.

{}As we already mentioned, even though the stock market system has worked over many decades, it is by no means perfect. Thus, corporations are expected to buy back their stock if its price falls too low, but there is no actual law or rule that they must do so. If such a rule were in place, corporations would be much less inclined to exaggerate their earnings. Thus, imagine that a corporation had to buy back some of its outstanding stock according to its reported earnings levels (say, in some proportion to earnings per share). If it exaggerated its earnings, the corporation would then have to buy back more outstanding stock (and at a higher price as the market demand on its stock would be artificially inflated), that is, the corporation would have to pay more cash than if it reported its earnings correctly. This would clearly be difficult to do if the corporation did not actually have the cash. The lack of such a rule (or an analogous regulation) might be (at least partially or indirectly) contributing into stock market bubbles.

{}Thus, many of the new internet companies during the .com boom never intended to buy back their stock, and the stock prices soared to more than unreasonably high levels as many investors were betting their money on the future potential of these companies to generate earnings, which was often times exaggerated by the companies themselves without any evident strong fundamentals present at the time. In fact, in many cases stocks trade at prices that factor in a potential for growth, and not just the current earnings levels. Sometimes such optimistic bets do not pay off, and the investors bear losses. On the other hand, many companies do meet or even outperform investors' expectations (typically these are companies with strong fundamentals), in which case such investments pay off well.
Another important point is that the stock market does not like uncertainty. If, for instance, there is a possibility that the economy might not do well in the nearest future, or, say, there could be a war and its outcome is somewhat uncertain, many investors tend to get out of their stock positions, which can sometimes lead to panic selling, and stock market crashes. Thus, the stock market sentiment goes a long way, and stock prices are substantially affected by what various investors think at any given time. This is partly responsible for the fact that stocks are more volatile than some other financial instruments such as bonds. This volatility makes the stock market game rather exciting, at least for some investors.

{}In some sense stock market is analogous to foreign currency exchange -- corporations are like countries, and stocks are like their currencies. Trading stocks is then like reallocating funds between different currencies. However, this resemblance does not go all the way -- there are important differences as well. Thus, convertible currencies are backed by reserves of the countries as well as by laws ensuring that they can be used to purchase goods, services, {\em etc.} For instance, all U.S. Federal Reserve Notes (that is, cash) regardless of denomination have the following crucial statement on their faces: ``This is legal tender for all debts, public and private". This statement is backed by the U.S. Federal law. Stockholders do not enjoy such a privilege -- you cannot exchange stocks for a bowl of soup at a local deli, you must first sell them on a national stock exchange for {\bf{\em cash}}!

{}There are many rules and regulations that stock markets must follow. These rules have been evolving by learning from the past experiences as well as to ensure that investors' interests are most adequately protected from potential fraud, market manipulation, misinformation (such as exaggerated corporate earnings), {\em etc}. The stock market is an important ingredient of the free market economy. And there is a fine line between regulation and overregulation; it's a balancing act.

\section{Arbitrage Pricing}

{}Suppose we have a stock $S$ and a cash bond with continuously compounded\footnote{\, As mentioned in the previous section, usually bonds pay coupons annually or semi-annually. Continuous compounding with constant interest rate $r$ means that, if we have \$1 at time $t$, at time $t + \Delta t$, where $\Delta t$ is small, it earns additional $r\Delta t$ dollars in interest, and this occurs continuously. The net result is that \$1 at $t=0$ turns into $\exp(rt)$ dollars at time $t$. The reason why interest exists in the first instance is because of the time value of money: typically, barring deflation, \$1 today is worth more than \$1 a year from now. The ``fundamental" reason for this is related to economic growth and the fact that investing, {\em e.g.}, in businesses is expected to generate returns -- which is one reason why interest rates are low when the economy is bad. More prosaically, the time value of money can be traced to human mortality and the fact that time is the most valuable commodity as it is in finite and rather short supply for each individual human being -- all the eternity notwithstanding.} constant interest rate $r$. Let the stock price at time $t=0$ be $S_0$.

{}Consider a {\em forward} contract, where one of the two parties agrees
to sell the other the stock at some future time $T$ (which is known as {\em expiry/delivery date/maturity} of the contract) for the {\em strike price}
$k$ on which they agree now, that is, at $t=0$. The forward price is actually
independent of the stock movements between $t=0$ and $t=T$, and is given by:
\begin{equation}
 k=S_0\exp(rT)~.
\end{equation}
The reason for this is {\em arbitrage}. Generally, arbitrage is a mechanism for making ``correct" market prices, known as {\em arbitrage pricing}. In its {\em idealized} form arbitrage means that, if the price of something is not ``correct", {\em i.e.}, it is not priced according to arbitrage pricing, there is a risk-free way of making profit.\footnote{\, The real life usually is much trickier than the idealized form of arbitrage. There are many things that can go wrong in reaping this ``risk-free" profit, making it not so risk-free. Furthermore, in real life there are transaction costs, which are {\em ignored} in the argument below. Even if there was risk-free profit to be made on paper, in real life such profit could be reduced to breaking even or even loss by transaction costs. In fact, some people make money by essentially exclusively becoming a transaction cost to others' trading, an example being notorious high frequency traders.}

{}Thus, suppose a bank was offering a forward with
a strike price $k>S_0\exp(rT)$. Then at $t=0$ we could borrow $S_0$ dollars
by selling cash bonds, and purchase one unit of stock. At time $T$ we could
sell our stock to that bank for $k$ dollars, repay our debt, which is now
$S_0 \exp(rT)$, and make a risk-free profit of $k-S_0\exp(rT)$ dollars.

{}Next, suppose a bank was offering a forward with
a strike price $k<S_0\exp(rT)$. Then at $t=0$ we could sell
one unit of stock, and buy $S_0$ worth of cash bonds.
At time $T$ our
bonds are worth $S_0 \exp(rT)$, and we could buy one unit of stock
from that bank for $k$ dollars, hence making a risk-free profit of
$S_0\exp(rT)-k$ dollars.

{}So, now that we have figured out the arbitrage pricing for our forward, we come to the simplest example of what is known as {\em hedging}, which is investing to reduce the risk of adverse price movements in a given asset. Typically, a hedge consists of taking an offsetting position in another asset. So, suppose a bank enters into the above forward contract to deliver the stock at maturity $T$ at the strike price $k = S_0 \exp(rT)$. To hedge its exposure to adverse price movements of the stock, which could increase in price by the time $T$, the bank would borrow $S_0$ dollars worth of cash bonds at $t = 0$ and buy the stock at price $S_0$ with that cash. At time $T$ the bank delivers the stock to the other party of the forward contract, collects $k$ dollars from said party, and pays off its debt, which is worth exactly $k$ dollars at time $T$ because of the accrued interest. The bank breaks even.

{}But banks are for-profit organizations, they are not in the business of breaking even. So, how does a bank make money in this particular example? Just as the bookmaker, the bank must charge a premium to make money. So, the effective strike price in the forward contract must be $k^\prime > k$, and the bank makes profit equal the difference $k^\prime - k$ (in reality, less any other transaction costs, such as those associated with purchasing the stock, and any costs of carry and/or other expenses -- the bank has to pay its employees salary, rent, {\em etc.} -- which we will not delve into here). The difference between $k^\prime$ and $k$ may be structured as a commission or some other way in the actual forward contract. To the other party to the contract, the difference between $k^\prime$ and $k$ is then basically a transaction cost. As mentioned above, in many cases profit is made in the form of transaction cost, one way or another.

{}Forwards are the simplest forward-looking contracts. Complexity is added once derivatives such as call and put options are considered. We will discuss these in more detail in subsequent sections. Here we simply define the simplest of such contracts to motivate further developing the mathematical machinery in the subsequent sections. A European call option is a right (but not obligation) to buy a stock at the maturity
time $T$ for the strike price $k$ agreed on at time $t=0$. The claim for the call option $f^c(S_T,k)=(S_T-k)^+$. Here $(x)^+=x$ if $x>0$, and $(x)^+=0$ if $x\leq 0$. By the ``claim" we mean how much the option is worth at maturity $T$. If the stock price at maturity $S_T > k$, then the option holder gains $S_T - k$ (excluding the cost paid for the option at $t=0$). If the price at maturity $S_T \leq k$, then there is no profit to be made from the option as it makes no sense to exercise it if $S_T < k$ (as it is cheaper to buy the stock on the market) and it makes no difference if $S_T = k$ -- all this is assuming no transaction costs. Similarly, a European put option is a right (but not obligation) to sell a stock at the maturity
time $T$ for the strike price $k$ agreed on at time $t=0$. The claim for the put option is given by $f^p(S_T,k)=(k-S_T)^+$. To understand how to price these and other derivatives, we need some more mathematical tools.

\section{Binomial Tree Model}

{}One such tool is the binomial tree model. At time $t=0$ the stock price is $S_0$. At time $t=\delta t$ the stock
price can take two values: $S_+$ and $S_-$. At $t=0$ the bond is worth
$B_0$, and at time $t=\delta t$ it is worth $B_0\exp(r\delta t)$.

{}Suppose we have a clam $f$, which at time $t=\delta t$ takes two values
$f_+$ and $f_-$ according to the stock price.\footnote{\, Here the claim $f$ is completely arbitrary and can correspond to the most exotic derivatives imaginable. The discussion below is completely general.} We can synthesize this
derivative as follows. Let $(\phi,\psi)$ be a general portfolio of $\phi$
units of stock $S$ and $\psi$ units of the cash bond $B$. Further, let
\begin{eqnarray}
 &&\phi S_+ +\psi B_0\exp(r\delta t)=f_+~,\\
 &&\phi S_- +\psi B_0\exp(r\delta t)=f_-~,
\end{eqnarray}
so that we have
\begin{eqnarray}
 &&\phi ={{f_+ -f_-}\over{S_+-S_-}}~,\\
 &&\psi=B_0^{-1}\exp(-r\delta t){{S_+f_--S_-f_+}\over{S_+-S_-}}~.
\end{eqnarray}
Thus, if we buy this portfolio at $t=0$, we will guarantee the correct outcome
for the derivative.

{}The price of this portfolio at time $t=0$ is given by
\begin{equation}
 V=S_0\phi+B_0\psi=S_0{{f_+ -f_-}\over{S_+-S_-}}+
 \exp(-r\delta t){{S_+f_--S_-f_+}\over{S_+-S_-}}~.
\end{equation}
In the case of a forward we have $f=S-k$, so that
\begin{equation}
 V=S_0-\exp(-r\delta t) k~,
\end{equation}
which vanishes for $k=S_0\exp(r\delta t)$ as it should according to the
arbitrage pricing.

{}In fact, the above price for a general derivative $f$ is
precisely the arbitrage price. This can be seen as follows. Suppose a bank
was offering to buy or sell the derivative for a price $P$ less than $V$.
We can buy the derivative from that bank, and sell the $(\phi,\psi)$ portfolio
to exactly match it with a net profit $V-P$. At the maturity time the
derivative would exactly cancel the value of the portfolio (which replicates
the claim $f$) regardless of the stock price. So we are making a risk-free
profit $V-P$.

{}Similarly, if a bank was offering the above derivative at a price $P>V$,
we could sell this derivative to that bank, and buy the $(\phi,\psi)$
portfolio. At the end of the day we have a risk-free profit $P-V$.

{}The hedge in replicating the claim $f$ at time $t=\delta t$ is in that one purchases the $(\phi, \psi)$ portfolio at $t = 0$, which reproduces the claim $f$ no matter whether the price goes from $S_0$ at time $t=0$ to $S_+$ or $S_-$ at time $t=\delta t$. Put differently, arbitrage and hedging are two sides of the same coin.

\subsection{Risk-neutral Measure}

{}We can rewrite the price $V$ as
\begin{equation}
 V=\exp(-r\delta t)\left[qf_+ +(1-q)f_-\right]~,
\end{equation}
where
\begin{equation}
 q\equiv {{S_0\exp(r\delta t)-S_-}\over {S_+-S_-}}~.
\end{equation}
The set ${\bf Q}\equiv
\{q,1-q\}$ is called the {\em risk-neutral measure}. The fact that $q>0$
follows from the fact that otherwise we have $S_0\exp(r\delta t)\leq S_-<S_+$,
which would guarantee unlimited risk-free profit by selling the cash bond and
buying stock at $t=0$. On the other hand, we also have $q<1$ as otherwise
we have $S_-<S_+\leq S_0\exp(r\delta t)$, which would guarantee unlimited
risk-free profit by selling the stock and buying the cash bond at $t=0$. That is, arbitrage pricing requires that $S_- < S_0\exp(r\delta t) < S_+$.

{}Thus, as we see, the price of the derivative is given by
the expectation of the {\em discounted} \footnote{\, Intuitively, we can understand why the expectation is of the discounted claim and not the claim itself from the time value of money argument: the claim $f$ is at a future time $t=T$, whereas $V_0$ is computed at the present time $t=0$, so we must discount the claim to arrive at its current worth.} claim $\exp(-r\delta t)f$ with respect
to the risk-neutral measure ${\bf Q}$:
\begin{equation}
 V=V_0=B_0\langle B_T^{-1} f\rangle_{\bf Q}~,
\end{equation}
where the maturity time $T=\delta t$.

{}The above results are straightforwardly generalized to the case of a binomial
tree with multiple time-ticks. Starting from the last time-tick we can
reconstruct the claim $f$ at earlier times via
\begin{equation}
 f_{\rm{\scriptstyle{now}}}=\exp(-r\delta t)\left[qf_{\rm{\scriptstyle{up}}}+(1-q)
 f_{\rm{\scriptstyle{down}}}\right]~,
\end{equation}
where
\begin{equation}
 q={{\exp(r\delta t)S_{\rm{\scriptstyle{now}}}-S_{\rm{\scriptstyle{down}}}}\over
 {S_{\rm{\scriptstyle{up}}}-S_{\rm{\scriptstyle{down}}}}}~.
\end{equation}
The trading strategy is given by:
\begin{eqnarray}
 &&\phi={{f_{\rm{\scriptstyle{up}}}-f_{\rm{\scriptstyle{down}}}}\over
 {S_{\rm{\scriptstyle{up}}}-S_{\rm{\scriptstyle{down}}}}}~,\\
 &&\psi=B_{\rm{\scriptstyle{now}}}^{-1}\left[f_{\rm{\scriptstyle{now}}}-\phi
 S_{\rm{\scriptstyle{now}}}\right]~.
\end{eqnarray}
The price of the derivative is given by
\begin{equation}
 V_t=B_t\langle B_T^{-1} f\rangle_{\bf Q}~,
\end{equation}
where ${\bf Q}$ is the corresponding risk-neutral measure.

\subsection{An Example: Baseball World Series}

{}Suppose 2 teams play a series of up to $(2n+1)$ games -- think Baseball World Series with 7 games -- in which the
first team to win $(n+1)$
games wins the series and then no other games are played.
Suppose that you want to bet on each individual game in such a way that when
the series ends you will be ahead \$100 if your team wins the series, or behind
by exactly \$100 if your team loses the series, no matter how many games it
takes. How much would you bet on the first game?

{}This can be thought of as a derivative pricing question. Indeed, we can view the
series as a binomial process with the known claim at the end of the series.
Thus, to solve this problem we can draw a binary tree and work
backwards. Let us put ``$+$'' if our team wins, and put ``$-$'' if our team
loses. We will put ``0'' at the root of the binary tree (the beginning of the
series). The longest branches of the tree have $2(n+1)$ nodes (corresponding to
all $(2n+1)$ games played), the node before the
last one having $n$ $+$'s and $n$ $-$'s, no matter in what order. If, however,
$(n+1)$ $+$'s or $-$'s occur before $(2n+1)$ games, the corresponding branch
is shorter as the series ends. For us it will be convenient to have all
branches of the same length (that is, containing $2(n+1)$ nodes). To achieve
this, we will continue a terminated branch so that it has $2(n+1)$ nodes, and
at the last nodes put the claim of $+\$100$ or $-\$100$ depending upon whether
$(n+1)$ $+$'s or $(n+1)$ $-$'s occurred first in this branch.

{}Next, note that we can confine our attention
to only a half of the binary tree, say, the half that corresponds to our team
winning the first game -- indeed, the other half is the same as this half up
to exchanging $+$'s and $-$'s. So, our truncated tree will now have branches
of uniform length containing $(2n+1)$ nodes, and the first node has $+$ in it,
which corresponds to our team winning. Let $X$ be the bet we made on the first
game. Then this is exactly how much money we have in the first node of the
truncated binary tree (where our team won the first game). Thus, we can
view $X$ as the value of the claim $f({\cal F}_1)$, where $i=1,2,\dots,(2n+1)$
numbers the game, while ${\cal F}_i$ is a particular {\em filtration} (or {\em history}) up
to the $i$th game. Thus, ${\cal F}_1=\{+\}$, ${\cal F}_2=\{++\},\{+-\}$,
and so on (we are focusing on the truncated tree). In particular,
$X=f({\cal F}_1)$. On the other hand, we also know that
$f({\cal F}_{2n+1})=+\$100$ if in
${\cal F}_{2n+1}$ $(n+1)$ $+$'s occur first,
and $f({\cal F}_{2n+1})=-\$100$ if ${\cal F}_{2n+1}$ $(n+1)$
$-$'s occur first. Thus, we would like to deduce the initial value of the claim
from the known final values of it -- so this is indeed a pricing question.

{}To determine $X$, we do not actually need the details of the underlying
market instruments we are trading to replicate the final claim. All we need is
the risk-neutral measure ${\bf Q}$. The elements of this measure are all $1/2$,
in particular, they are independent of the actual probabilities for our team to
win or lose at any given time (assuming that they are neither 0 nor 1). Indeed,
suppose at any given time we purchase a bet for $Y$ dollars (by holding
the zero interest rate cash bond short $Y$ dollars). If our team wins, we get
$2Y$ dollars back (the reward of $Y$ dollars plus the stake returned). If our
team loses, we get nothing back. This implies that the risk-neutral
probability for this bet is indeed $q=1/2$. Now we can immediately write down
the value of the claim at time $i=1$:
\begin{equation}
 X=f({\cal F}_1)=\langle f({\cal F}_{2n+1})\rangle_{\bf Q}=
 \left({1\over 2}\right)^{2n} \sum_{k=0}^{2n} {{(2n) !}\over{k!(2n-k)!}}
 (\$100 ~\epsilon_k)~.
\end{equation}
Here
\begin{equation}
 {{(2n)!}\over{k!(2n-k)!}}
\end{equation}
is the number of $k$ $+$'s we can place in $2n$ slots in an arbitrary order
(here we are taking into account that to specify ${\cal F}_{2n+1}$ we only
need to specify the last $2n$ entries as the first entry is always $+$ for the
truncated tree), and $\epsilon_k=+1$ if $k\geq n$, while $\epsilon_k=-1$ if
$k<n$. We then have:
\begin{equation}
 X={{(2n)!}\over{2^{2n}(n!)^2}}
 ~\$100~.
\end{equation}
Thus, for $n=0$ we have $X=\$100$, for $n=1$ $X=\$50$, for
$n=2$ $X=\$75/2$, for $n=3$ $X=\$125/4$, and so on.

\section{Martingales}

{}A {\em filtration} ${\cal F}_i$ is the history of a stock (or some other
process) up until the tick-time $i$ on the tree.

{}A {\em claim} $X$ on the tree is a function of the filtration ${\cal F}_T$
for some horizon time $T$.

{}The conditional {\em expectation} operator $\langle \cdot
\rangle_{{\bf Q}, {\cal F}_i}$ is defined along the latter portion of paths
that have initial segments ${\cal F}_i$.

{}A {\em previsible} process $\phi_i$ is a process on the tree whose values at
any tick-time $i$ depend only on the history up to one tick-time earlier,
${\cal F}_{i-1}$.

{}A process $M_i$ is a {\em martingale} with respect to a measure ${\bf P}$
and a filtration ${\cal F}_i$ if
\begin{equation}
 \langle M_j\rangle_{{\bf P},{\cal F}_i}=M_i~,~~~\forall i\leq j~.
\end{equation}
Note that for a martingale its expectation is independent of time:
\begin{equation}
 \langle M_j\rangle_{\bf P}=\langle M_j\rangle_{{\bf P},{\cal F}_0}=M_0~,
\end{equation}
that is, it has no {\em drift}.

\subsection{The Tower Law}

{}Let $X=X_T$ be a claim. Then the process
\begin{equation}
 N_j\equiv\langle X\rangle_{{\bf P},{\cal F}_j}
\end{equation}
is a ${\bf P}$-martingale.

{}This follows from the {\em tower law}:
\begin{equation}
 \left\langle\langle X\rangle_{{\bf P},{\cal F}_j}
 \right\rangle_{{\bf P},{\cal F}_i}=
 \langle X \rangle_{{\bf P},{\cal F}_i}~,~~~i\leq j~.
\end{equation}
To prove the tower law, let us represent a filtration ${\cal F}_i$ as follows:
\begin{equation}
 {\cal F}_i=\{\epsilon_1,\dots,\epsilon_i\}~,
\end{equation}
where $\epsilon_k=\pm$. Let the probability of the path starting from
the event corresponding to ${\cal F}_i$ and ending with the event corresponding
to ${\cal F}_j$, $i\leq j$, be $P_{\epsilon_1,\dots,\epsilon_i}(\epsilon_{i+1},
\dots,\epsilon_j)$. Then we have:
\begin{eqnarray}
 \langle N_j \rangle_{{\bf P},{\cal F}_i}=&&\sum_{\epsilon_{i+1},\dots,
 \epsilon_j} P_{\epsilon_1,\dots,\epsilon_i}(\epsilon_{i+1},
 \dots,\epsilon_j) N_j=\nonumber\\
 &&\sum_{\epsilon_{i+1},\dots,
 \epsilon_j} P_{\epsilon_1,\dots,\epsilon_i}(\epsilon_{i+1},
 \dots,\epsilon_j)\sum_{\epsilon_{j+1},\dots,\epsilon_T}
 P_{\epsilon_1,\dots,\epsilon_j}(\epsilon_{j+1},\dots,\epsilon_T) X_T=
 \nonumber\\
 &&\sum_{\epsilon_{i+1},\dots,
 \epsilon_T} P_{\epsilon_1,\dots,\epsilon_i}(\epsilon_{i+1},
 \dots,\epsilon_j)
 P_{\epsilon_1,\dots,\epsilon_j}(\epsilon_{j+1},\dots,\epsilon_T) X_T=
 \nonumber\\
 &&\sum_{\epsilon_{i+1},\dots,
 \epsilon_T} P_{\epsilon_1,\dots,\epsilon_i}(\epsilon_{i+1},
 \dots,\epsilon_T) X_T=\nonumber\\
 &&\langle X\rangle_{{\bf P},{\cal F}_i}=N_i~.
\end{eqnarray}
Here we have used
$P_{\epsilon_1,\dots,\epsilon_i}(\epsilon_{i+1},\dots,\epsilon_j)
P_{\epsilon_1,\dots,\epsilon_j}(\epsilon_{j+1},\dots,\epsilon_T)=
P_{\epsilon_1,\dots,\epsilon_i}(\epsilon_{i+1},\dots,\epsilon_T)$.

\subsection{Martingale Measure}

{}Let $S$ be the stock process, and $B$ be the cash bond process. Define
the {\em discounted} stock process $Z_i\equiv B_i^{-1} S_i$. Let us
determine the martingale measure ${\bf Q}$ for $Z$.

{}Under the martingale measure ${\bf Q}$ we have
\begin{equation}
 \langle Z_j\rangle_{{\bf Q},{\cal F}_i}=Z_i~,~~~i\leq j~.
\end{equation}
Note that by $Z_i$ on the r.h.s. we mean the value of $Z_i$
corresponding to the filtration ${\cal F}_i$. Let this value be denoted
by $Z_*({\cal F}_i)$.
In particular,
\begin{equation}\label{expZ}
 \langle Z_{i+1}\rangle_{{\bf Q},{\cal F}_i}=Z_*({\cal F}_i)~.
\end{equation}
Let ${\cal F}_i=\{\epsilon_1,\dots,\epsilon_i\}$, and ${\cal F}_{i+1}^{\pm}=
\{\epsilon_1,\dots,\epsilon_i,\pm\}$. Then
we have
\begin{eqnarray}
 \langle Z_{i+1}\rangle_{{\bf Q},{\cal F}_i}=&&Q_{{\cal F}_i}(+)
 Z_*({\cal F}_{i+1}^+)+Q_{{\cal F}_i}(-)Z_*({\cal F}_{i+1}^-)=\nonumber\\
 &&Q_{{\cal F}_i}(+)
 Z_*({\cal F}_{i+1}^+)+\left[1-Q_{{\cal F}_i}(+)\right]Z_*({\cal F}_{i+1}^-)~.
\end{eqnarray}
On the other hand, we have (\ref{expZ}). Thus, we have
\begin{equation}
 Q_{{\cal F}_i}(\pm)={{Z_*({\cal F}_i)-Z_*({\cal F}_{i+1}^\mp)}\over
 {Z_*({\cal F}_{i+1}^\pm)-Z_*({\cal F}_{i+1}^\mp})}~.
\end{equation}
This determines $Q_{{\cal F}_i}(\epsilon_{i+1})$. We can now determine all the
other ${\bf Q}$-probabilities. Thus,
\begin{equation}
 Q_{{\cal F}_i}(\epsilon_{i+1},\epsilon_{i+2})=
 Q_{{\cal F}_i}(\epsilon_{i+1})Q_{{\cal F}_{i+1}}(\epsilon_{i+2})~,
\end{equation}
and so on.

{}Finally, let us rewrite the ${\bf Q}$-probabilities in terms of values of
$S$. Let us assume that $B_i=B_0\exp(rt_i)$, $t_{i+1}-t_i\equiv\delta t$.
Then we have
\begin{equation}
 Q_{{\cal F}_i}(\pm)={{\exp(r\delta t)
 S_*({\cal F}_i)-S_*({\cal F}_{i+1}^\mp)}\over
 {S_*({\cal F}_{i+1}^\pm)-S_*({\cal F}_{i+1}^\mp})}~,
\end{equation}
which is a formula we have derived earlier for the risk-neutral measure. That is, the risk-neutral measure is the martingale measure.

\subsection{Binomial Representation Theorem}

{}Suppose we have a binomial tree with two processes $S$ and $N$. Then
we have
\begin{equation}
 \Delta N_i=\phi_i \Delta S_i + k_i~,
\end{equation}
where $\Delta N_i\equiv N_i-N_{i-1}$, $\Delta S_i\equiv S_i-S_{i-1}$, and
both $\phi$ and $k$ are {\em previsible} processes.

{}To show this, consider a particular filtration ${\cal F}_{i-1}$. Let
$S_*({\cal F}_{i-1})\equiv S_*$, $S_*({\cal F}_i^\pm)\equiv S_\pm$, and
similarly for $N$. Then $\Delta S_i$ takes two values $S_\pm-S_*$, and
$\Delta N_i$ takes two values $N_\pm-N_*$. Let
\begin{equation}
 \phi_i={{N_+-N_-}\over{S_+-S_-}}~.
\end{equation}
Note that $\phi_i$ is previsible by definition. We must now show that $k_i$
is also previsible. To do this, let us show that $k_*({\cal F}_i^+)=
k_*({\cal F}_i^-)$:
\begin{eqnarray}
 k_*({\cal F}_i^+)-k_*({\cal F}_i^-)=&&\left[(N_+-N_*)-\phi_i(S_+-S_*)\right]
 -\left[(N_--N_*)-\phi_i(S_--S_*)\right]=\nonumber\\
 &&(N_+-N_-)-\phi_i(S_+-S_-)=0~.
\end{eqnarray}
This implies that $k_i$ is indeed previsible. Indeed, at the node $i$, $k_i$ is independent of $\pm$, therefore it depends only on ${\cal F}_{i-1}$.

{}Now suppose that both $S$ and $N$ are ${\bf Q}$-martingales. Then $k_i$ is
identically zero. Indeed, since both $\phi_i$ and $k_i$ are previsible, we
have
\begin{equation}
 \langle \Delta N_i\rangle_{{\bf Q},{\cal F}_{i-1}}=
 \phi_i \langle \Delta S_i\rangle_{{\bf Q},{\cal F}_{i-1}}+k_i~.
\end{equation}
However,
\begin{equation}
 \langle \Delta N_i\rangle_{{\bf Q},{\cal F}_{i-1}}=
 \langle N_i\rangle_{{\bf Q},{\cal F}_{i-1}}-
 \langle N_{i-1}\rangle_{{\bf Q},{\cal F}_{i-1}}=
 \langle N_i\rangle_{{\bf Q},{\cal F}_{i-1}}-N_*=0~,
\end{equation}
and similarly for $\Delta S_i$. This then implies that $k_i\equiv 0$.

{}Thus, for any two ${\bf Q}$-martingales $S$ and $N$ we have
\begin{equation}
 \Delta N_i=\phi_i\Delta S_i~,
\end{equation}
where $\phi_i$ is previsible and plays the role of a ``discrete derivative". This leads to the {\em binomial representation
theorem} for ${\bf Q}$-martingales:
\begin{equation}
 N_i=N_0+\sum_{k=1}^i \phi_k \Delta S_k~.
\end{equation}

\subsection{Self-financing Hedging Strategies}

{}Before going into the details of self-financing hedging strategies, let us mention that in finance one can take a long position, {\em e.g.}, by purchasing a stock, and a short position, which means that one ``owns" a negative number of shares of the stock to be covered at some later time. With a long position, if the stock price goes up, the position has a gain, and if the stock price goes down, the position bears a loss. With the short position it is the opposite, if the stock price goes up, the position bears a loss, and if the stock price goes down, the position has a gain. To take a long position, one needs to borrow money to buy stock. When taking a short position, one receives the cash value equivalent to the price of the shorted stock at the time or shorting. In real life there are transaction costs associated with this, {\em e.g.}, the interest rate at which the received cash accrues interest when the stock is shorted is typically lower than the interest accrued on the borrowed cash when a long position is taken. Below we ignore any such discrepancies and transaction costs and consider the idealized situation where long and short positions are treated on an equal footing.

{}Let us construct a hedge for the claim $X=X_T$ on the stock $S$ in the
presence of the cash bond $B_i$. First, let us define the discounted
stock process $Z_i\equiv B_i^{-1}S_i$, and the discounted claim
$Y_T\equiv B_T^{-1} X_T$. Let
\begin{equation}
 E_i\equiv\langle Y_T\rangle_{{\bf Q},{\cal F}_i}~.
\end{equation}
Note that $E_i$ is a ${\bf Q}$-martingale. Moreover, $E_T=Y_T$, so
at the end of the day $E_i$ replicates the discounted claim $Y$. Let ${\bf Q}$
be the martingale measure for $Z$. Then there exists a previsible process
$\phi$ such that
\begin{equation}
 E_i=E_0+\sum_{k=1}^i \phi_k \Delta Z_k~.
\end{equation}
The previsible process $\phi$ is determined from
\begin{equation}
 \phi_i={{E_*({\cal F}_i^+)-E_*({\cal F}_i^-)}\over
 {S_*({\cal F}_i^+)-S_*({\cal F}_i^-)}}~.
\end{equation}
Next, define the following previsible process:
\begin{equation}
 \psi_i=E_{i-1}-\phi_i Z_{i-1}~.
\end{equation}
Finally, define a portfolio $\Pi_i=(\phi_{i+1},\psi_{i+1})$ consisting of
holding $\phi_{i+1}$ units of stock and $\psi_{i+1}$ units of the cash bond
at time $i$. This portfolio is worth
\begin{equation}
 V_i=\phi_{i+1}S_i+\psi_{i+1}B_i=B_iE_i~.
\end{equation}
If we hold this portfolio across the next time-tick, it is worth
\begin{equation}
 {\widehat V}_i=\phi_{i+1}S_{i+1}+\psi_{i+1}B_{i+1}~.
\end{equation}
Now (note that $V_{i+1}$ is the worth of the portfolio $\Pi_{i+1}$, whereas ${\widehat V}_i$ is the worth of the portfolio $\Pi_i$ by the tick $i+1$),
\begin{eqnarray}
 V_{i+1}-{\widehat V}_i=&&\left[\phi_{i+2}-\phi_{i+1}\right]S_{i+1}+
 \left[\psi_{i+2}-\psi_{i+1}\right]B_{i+1}=\nonumber\\
 &&\left[\phi_{i+2}-\phi_{i+1}\right]S_{i+1}+
 \left[E_{i+1}-\phi_{i+2}Z_{i+1}-E_i +\phi_{i+1}Z_i\right]B_{i+1}=\nonumber\\
 &&\left[\phi_{i+2}-\phi_{i+1}\right]S_{i+1}+
 \left[\Delta E_{i+1}-\phi_{i+2}Z_{i+1}+\phi_{i+1}Z_i\right]B_{i+1}=\nonumber\\
 &&\left[\phi_{i+2}-\phi_{i+1}\right]S_{i+1}+
 \left[\phi_{i+1}\left(Z_{i+1}-Z_i\right)-
 \phi_{i+2}Z_{i+1}+\phi_{i+1}Z_i\right]B_{i+1}=\nonumber\\
 &&\left[\phi_{i+2}-\phi_{i+1}\right]S_{i+1}+
 \left[\phi_{i+1}-\phi_{i+2}\right]Z_{i+1}B_{i+1}=0~.
\end{eqnarray}
That is, the value of the portfolio $\Pi_i$ by the end of the next time-tick,
that is, by time $i+1$ is precisely the same as that of the portfolio
$\Pi_{i+1}$. So we can sell the portfolio $\Pi_i$ at the end of this time-tick,
and buy the portfolio $\Pi_{i+1}$ without any loss or gain. The worth of the
final portfolio $\Pi_T$ is
\begin{equation}
 V_T=B_T E_T=B_T Y_T =X_T~.
\end{equation}
So this hedging strategy replicates the claim $X$ at the maturity time $T$.
On the other hand, note that the price of the portfolio $\Pi_0$ is
given by
\begin{equation}
 V_0=B_0 E_0=B_0\langle B_T^{-1} X\rangle_{\bf Q}~.
\end{equation}
This is the arbitrage price for the claim $X$ at time $t=0$.

\subsection{The Self-financing Property}

{}Let us take two arbitrary previsible processes $\phi_i$ and $\psi_i$,
and compute the value of the corresponding $\Pi_i$ portfolio:
\begin{equation}
 V_i=\phi_{i+1} S_i+\psi_{i+1} B_{i+1}~.
\end{equation}
In general the change in this value over one time-tick is given by:
\begin{equation}
 \Delta V_i\equiv V_{i+1}-V_i=\Delta \phi_{i+1} S_i+\Delta\psi_{i+1} B_i +
 \phi_{i+1}\Delta S_i +\psi_{i+1} \Delta B_i~,
\end{equation}
where $\Delta \phi_{i+1}\equiv \phi_{i+2}-\phi_{i+1}$, and
$\Delta \psi_{i+1}\equiv \psi_{i+2}-\psi_{i+1}$.

{}The self-financing property means that
\begin{equation}
 \Delta V_i=
 \phi_{i+1}\Delta S_i +\psi_{i+1} \Delta B_i~,
\end{equation}
that is, the change in the value of the strategy is solely due the changes
in the stock and bond values, {\em i.e.}, there is no cash flowing in or out of the strategy at any time. The condition for the strategy to be
self-financing is then
\begin{equation}
 \Delta \phi_{i+1} S_i+\Delta\psi_{i+1} B_i=0~.
\end{equation}
This condition is satisfied by the strategy discussed in the previous
subsection.

\section{Discrete {\em vs.} Continuous Models}

{}Thus far we considered a binomial tree model, which is discrete. While numerically one often deals with discrete models, such as binomial/trinomial trees, {\em etc.}, there is certain advantage to considering continuous models. One advantage of continuous models is that certain calculus methods can be applied, analytic computations are more streamlined, and the intuitive understanding is more easily developed. This is analogous to the difference between the pre-Newtonian physics and a much more streamlined Newtonian description based on continuous methods and calculus.

{}Consider the following discrete model:
\begin{eqnarray}
 &&B_t=\exp(rt)~,\\
 &&S_{t+\delta t}= S_t \exp(\mu\delta t +\sigma\epsilon_t\sqrt{\delta t})~,
\end{eqnarray}
where $\epsilon_t=\pm 1$ with the equal probabilities: $P(\epsilon_t)=1/2$
($r,\mu,\sigma$ are assumed to be constant). {\em I.e.}, time $t$ takes values in a semi-infinite discrete set $t = k \delta t$, $k \in \left[0,\infty \right)$.

{}Let
\begin{equation}
 X_t\equiv\sqrt{\delta t\over t}
 \sum_{k=0}^{t/\delta t-1}\epsilon_{k\delta t}~.
\end{equation}
Note that $\sqrt{t}X_t$ is nothing but a random walk on a discrete binomial tree. The quantity $X_t$ takes values with binomial distribution. As $\delta t\rightarrow 0$, $X_t$ becomes a normal random variable\footnote{Meaning, its distribution is Gaussian.} with
mean zero ($\langle X_t\rangle = 0$) and variance 1 ($\langle X_t^2\rangle - \langle X_t\rangle^2 = 1$) -- this is the {\em Central Limit Theorem}. The stock can then be written as
\begin{equation}
 S_t=S_0\exp(\mu t+\sigma \sqrt{t} X_t)~.
\end{equation}
So $\ln (S_t)$ is normally distributed with mean $\ln(S_0)+\mu t$ and
variance $\sigma^2 t$.

{}Let us compute the martingale measure ${\bf Q}$. We have:
\begin{eqnarray}
 q=&&{{S_t\exp(r\delta t)-S_{t+\delta t}^-}
 \over
 {S_{t+\delta t}^+-S_{t+\delta t}^-}}={{\exp([r-\mu]\delta t)-\exp(-\sigma
 \sqrt{\delta t})}\over{\exp(\sigma \sqrt{\delta t})-
 \exp(-\sigma\sqrt{\delta t})}}=\nonumber\\
 &&{1\over 2}\left[1-\sqrt{\delta t}~
 {{\mu+{1\over 2}\sigma^2-r}\over \sigma}+{\cal O}(\delta t)\right]~.
\end{eqnarray}
Note that this measure is independent of $t$.

{}Under this measure we have
\begin{eqnarray}
 \langle X_t\rangle_{\bf Q}=&&\sqrt{\delta t\over t}
 \sum_{k=0}^{t/\delta t-1}\langle \epsilon_{k\delta t}\rangle_{\bf Q}=\nonumber\\
 &&\sqrt{t\over \delta t}(2q-1)=-\sqrt{t}~
 {{\mu+{1\over 2}\sigma^2-r}\over \sigma}+{\cal O}(\sqrt{\delta t})~,
\end{eqnarray}
and
\begin{eqnarray}
 \langle X^2_t\rangle_{\bf Q}-\langle X_t\rangle_{\bf Q}^2=
 {\delta t\over t}
 \sum_{k=0}^{t/\delta t-1}\left[\langle \epsilon_{k\delta t}^2\rangle_{\bf Q}
 -\langle\epsilon_{k\delta t}\rangle_{\bf Q}^2\right]=4q(1-q)=1+{\cal O}
 (\delta t)~.
\end{eqnarray}
Thus, $\ln(S_t)$ is now normally distributed (w.r.t. the martingale measure ${\bf Q}$, that is) with mean $\ln(S_0)+
(r-{1\over 2}\sigma^2)
t$ and variance $\sigma^2 t$. This implies that
\begin{equation}
 S_t=S_0\exp\left(\sigma\sqrt{t}Z_t+\left[r-{1\over 2}\sigma^2\right]t\right)~,
\end{equation}
where $Z_t$ is normally distributed with mean zero and variance 1 under the
martingale measure ${\bf Q}$.

\subsection{Brownian Motion}

{}Consider the discrete process
\begin{equation}
 z_t\equiv\sqrt{\delta t} \sum_{k=0}^{t/\delta t-1}\epsilon_{k\delta t}~,
\end{equation}
where $\epsilon_t=\pm 1$, and $P(\epsilon_t)=1/2$. In the limit $\delta t
\rightarrow 0$ this is {\em Brownian motion}. The variable $z_t$ is normally
distributed with mean zero and variance $t$:
\begin{equation}
 P(z,t)={1\over \sqrt{2\pi t}}\exp\left(-{z^2\over 2t}\right)
\end{equation}
is the probability distribution for $z$ at time $t$.

{}The formal definition of Brownian motion is as follows:\\
\noindent The process $W=(W_t:t\geq 0)$ is a ${\bf P}$-Brownian motion if and
only if:\\
$\bullet$ $W_t$ is continuous, and $W_0=0$;\\
$\bullet$ under ${\bf P}$ the value of $W_t$ is distributed as a normal random
variable $N(0,t)$ of mean 0 and variance $t$;\\
$\bullet$ the increment $W_{s+t}-W_s$ is distributed as a normal $N(0,t)$
under ${\bf P}$, and is independent of ${\cal F}_s$, that is, of the history
of what the process did up to time $s$.

{}Let us ask the following question: what is the probability that starting at
$z=0$ the Brownian motion $z_t$ defined above hits $z_*$ by time $T$? Without loss of generality we can assume
that $z_*\geq 0$. Let us first consider the case where $z_*>0$. Then our
probability is
\begin{eqnarray}
 &&P(z_0=0~\&~\exists t_*\leq T:~z_{t_*}=z_*)=
 P(z_0=0~\&~\exists t_*\leq T:~z_{t_*}=z_*~\&~z_T\geq z_*)+\nonumber\\
 &&\,\,\,\,\,\,P(z_0=0~\&~\exists t_*\leq T:~z_{t_*}=z_*~\&~z_T\leq z_*)~.
\end{eqnarray}
Note, however, that
\begin{eqnarray}
 &&P(z_0=0~\&~\exists t_*\leq T:~z_{t_*}=z_*~\&~z_T\geq z_*)=\nonumber\\
 &&\,\,\,\,\,\,\,P(z_0=0~\&~\exists t_*\leq T:~z_{t_*}=z_*~\&~z_T\leq z_*)~.
\end{eqnarray}
Indeed, starting from $z_*$ at $t=t_*$ it is as probable that we end up with
$z_T\leq z_*$ as with $z_T\geq z_*$. Thus, we have
\begin{equation}
 P(z_0=0~\&~\exists t_*\leq T:~z_{t_*}=z_*)=
 2P(z_0=0~\&~\exists t_*\leq T:~z_{t_*}=z_*~\&~z_T\geq z_*)~.
\end{equation}
Next, note that
\begin{eqnarray}
 &&P(z_0=0~\&~\exists t_*\leq T:~z_{t_*}=z_*~\&~z_T\geq z_*)=\nonumber\\
 &&\,\,\,\,\,\,\,\ P(z_0=0~\&~z_T\geq z_*)=\int_{z_*}^\infty {dz\over\sqrt{2\pi T}}
 \exp\left(-{z^2\over 2T}\right)~,
\end{eqnarray}
so that
\begin{equation}
 P(z_0=0~\&~\exists t_*\leq T:~z_{t_*}=z_*)=2
 \int_{z_*}^\infty {dz\over\sqrt{2\pi T}}
 \exp\left(-{z^2\over 2T}\right)~.
\end{equation}
For $z_*\rightarrow 0$ this probability goes to 1.

\section{Stochastic Calculus}

{}A stochastic process $X$ is a continuous process $(X_t:t\geq 0)$ such that
\begin{equation}\label{stoch}
 X_t=X_0+\int_0^t \sigma_s dW_s +\int_0^t\mu_s ds~,
\end{equation}
where $\sigma$ and $\mu$ are random ${\cal F}$-previsible processes\footnote{A continuous ${\cal F}$-previsible process $\phi_s$ is defined as a process which at time $s$ is known given the filtration ${\cal F}_s$, {\em i.e.}, $\phi_s = \phi({\cal F}_s)$, so $\phi_s$ is a {\em functional} of the filtration ${\cal F}_s$.} such that
\begin{equation}
 \int_0^t\left[\sigma_s^2+|\mu_s|\right]ds
\end{equation}
is finite for all $t$ (with probability 1). In the differential form we have
\begin{equation}
 dX_t=\sigma_t dW_t+\mu_t dt~.
\end{equation}
Given a process $X$, there is only one pair of volatility $\sigma$ and drift
$\mu$ that satisfies (\ref{stoch}) for all $t$. (This uniqueness comes from
the {\em Doob-Meyer decomposition} of semimartingales.)

{}If $\sigma$ and $\mu$ depend on $W$ only via $X$ (that is, if
$\sigma_t=\sigma(X_t,t)$ and $\mu_t=\mu(X_t,t)$, where $\sigma(x,t)$ and
$\mu(x,t)$ are deterministic functions), we have
\begin{equation}
 dX_t=\sigma(X_t,t)dW_t+\mu(X_t,t)dt~,
\end{equation}
which is a {\em stochastic differential equation} (SDE).

\subsection{It{\^o} Calculus}

{}We can think of Brownian motion $W_t$ as a limit $\delta t\rightarrow 0$
of the process
\begin{equation}
 W_t\equiv \sqrt{\delta t}\sum_{k=0}^{t/\delta t-1}\epsilon_{k\delta t}~.
\end{equation}
Consider the increment
\begin{equation}
 \delta W_t\equiv W_{t+\delta t}-W_t=\epsilon_t\sqrt{\delta t}~.
\end{equation}
The continuous version of this is given by:
\begin{equation}
 dW_t=\epsilon_t (dt)^{1/2}~.
\end{equation}
This implies that
\begin{equation}
 (dW_t)^n=(\epsilon_t)^n (dt)^{n/2}~.
\end{equation}
In particular,
\begin{equation}
 (dW_t)^2=dt~.
\end{equation}
This implies that if
\begin{equation}
 dX_t=\sigma_t dW_t+\mu_t dt~,
\end{equation}
then
\begin{equation}
 (dX_t)^2=\sigma_t^2 dt +{\cal O}(dt^{3/2})~.
\end{equation}
This has important consequences.

{}Thus, consider a function $f(x,t)$. We will denote partial derivatives w.r.t.
$x$ via a prime:
\begin{equation}
 \partial_x f(x,t)\equiv f^\prime(x,t)~.
\end{equation}
Then we have (we keep only terms of order $dW_t$ and $dt$):
\begin{eqnarray}
 &&df(X_t,t)=f^\prime(X_t,t)dX_t+{1\over 2}f^{\prime\prime}(X_t,t)(dX_t)^2+
 \partial_t f(X_t,t)dt=\nonumber\\
 &&\,\,\,\,\,\,\,\sigma_t f^\prime(X_t,t)dW_t+\left[\mu_t f^\prime(X_t,t)+{1\over 2}
 \sigma_t^2 f^{\prime\prime}(X_t,t)+\partial_t f(X_t,t)\right]dt~.
\end{eqnarray}
As an example consider the function
\begin{equation}
 f(X_t)=\exp(X_t)~.
\end{equation}
Then we have
\begin{equation}
 df(X_t)=f(X_t)\left[\sigma_t dW_t+\left(\mu_t+{1\over 2}\sigma_t^2\right)dt
 \right]~.
\end{equation}
So a solution to the SDE
\begin{equation}\label{SDE1}
 dS_t=S_t\left[\sigma_t dW_t+{\widetilde \mu}_t dt\right]
\end{equation}
is given by
\begin{equation}\label{S1}
 S_t = S_0\exp\left[\int_0^t \sigma_s dW_s +\int_0^t\left({\widetilde\mu_t}-
 {1\over 2}\sigma_t^2\right)dt\right]~.
\end{equation}
Note the difference between the drift ${\widetilde \mu}_t$ in the SDE
(\ref{SDE1}) and in the exponent in (\ref{S1}), which is shifted by $\sigma_t^2/2$.

\subsection{Radon-Nikodym Process}

{}Two measures ${\bf P}$ and ${\bf Q}$ are equivalent if they operate on the
same sample space, and agree on what is possible.

{}Consider a binomial tree. The Radon-Nikodym process is defined as follows:
\begin{equation}
 \zeta_i\equiv {Q({\cal F}_i)\over P({\cal F}_i)}~.
\end{equation}
This process is a ${\bf P}$-martingale ($i\leq j$):
\begin{eqnarray}
 \langle\zeta_j\rangle_{{\bf P},{\bf F}_i}=&&\sum_{\epsilon_{i+1},\dots,
 \epsilon_j} P_{\epsilon_1,\dots,\epsilon_i}(\epsilon_{i+1},\dots,\epsilon_j)
 ~{Q(\epsilon_1,\dots,\epsilon_j)\over P(\epsilon_1,\dots,\epsilon_j)}=
 \nonumber\\
 &&\sum_{\epsilon_{i+1},\dots,
 \epsilon_j} P_{\epsilon_1,\dots,\epsilon_i}(\epsilon_{i+1},\dots,\epsilon_j)
 ~{Q(\epsilon_1,\dots,\epsilon_i) Q_{\epsilon_1,\dots,\epsilon_i}
 (\epsilon_{i+1},\dots,\epsilon_j)\over
 P(\epsilon_1,\dots,\epsilon_i) P_{\epsilon_1,\dots,\epsilon_i}
 (\epsilon_{i+1},\dots,\epsilon_j)}=
 \nonumber\\
 &&{Q(\epsilon_1,\dots,\epsilon_i) \over
 P(\epsilon_1,\dots,\epsilon_i)} \sum_{\epsilon_{i+1},\dots,
 \epsilon_j} Q_{\epsilon_1,\dots,\epsilon_i}
 (\epsilon_{i+1},\dots,\epsilon_j)=\nonumber\\
 &&{Q(\epsilon_1,\dots,\epsilon_i) \over
 P(\epsilon_1,\dots,\epsilon_i)}=\zeta_i~.
\end{eqnarray}
This, in particular, implies, that
\begin{equation}
 \zeta_i=\left\langle {d{\bf Q}\over d{\bf P}}
 \right\rangle_{{\bf P},{\cal F}_i}~,
\end{equation}
where
\begin{equation}
 {d{\bf Q}\over d{\bf P}}\equiv \zeta_T={Q({\cal F}_T)\over P({\cal F}_T)}
\end{equation}
is the Radon-Nikodym derivative for some horizon time $T$.

{}We can use the Radon-Nikodym process to compute expectations w.r.t. the
measure ${\bf Q}$. Thus, we have:
\begin{equation}
 \langle X_T\rangle_{\bf Q}=\left\langle {d{\bf Q}\over d{\bf P}}
 X_T \right\rangle_{\bf P}~.
\end{equation}
More generally, we have
\begin{equation}
 \langle X_j\rangle_{{\bf Q},{\cal F}_i}=\zeta_i^{-1}
 \langle \zeta_j X_j \rangle_{{\bf P},{\cal F}_i}~,~~~i\leq j\leq T~,
\end{equation}
which can be seen from
\begin{eqnarray}
 &&\langle\zeta_j X_j\rangle_{{\bf P},{\bf F}_i}=\sum_{\epsilon_{i+1},\dots,
 \epsilon_j} P_{\epsilon_1,\dots,\epsilon_i}(\epsilon_{i+1},\dots,\epsilon_j)
 ~{Q(\epsilon_1,\dots,\epsilon_j)\over P(\epsilon_1,\dots,\epsilon_j)}~X_j=
 \nonumber\\
 &&\,\,\,\,\,\,\,{Q(\epsilon_1,\dots,\epsilon_i) \over
 P(\epsilon_1,\dots,\epsilon_i)} \sum_{\epsilon_{i+1},\dots,
 \epsilon_j} Q_{\epsilon_1,\dots,\epsilon_i}
 (\epsilon_{i+1},\dots,\epsilon_j) X_j=\zeta_i
 \langle X_j\rangle_{{\bf Q},{\bf F}_i}~,
\end{eqnarray}
where $i\leq j\leq T$.

\subsection{Path Integral}

{}We can generalize the notions of the change of measure and Radon-Nikodym
process to continuous processes using {\em path integral}. Consider a
${\bf P}$-Brownian motion $W_t$ between $t=0$ and some horizon time $T$.
Let $x(t)$ be the values of $W_t$ (note that $x(0)=0$). We will
divide the time interval $[t_0,t_f]$, $0\leq t_0<t_f\leq T$, into $N$ intervals
$[t_{i-1},t_i]$, $t_N=t_f$, $t_i-t_{i-1}\equiv \Delta t_i>0$. Let the
corresponding values of $x(t)$ be $x_i\equiv x(t_i)$, $x_N\equiv x_f$,
$\Delta x_i\equiv
x_i-x_{i-1}$. Let ${\cal O}_t$, $0\leq t\leq T$, be a previsible process. That
is, ${\cal O}_t$ depends only on the path
${\cal F}_t=\{(x(s),s)|s\in [0,t]\}$:
\begin{equation}
 {\cal O}_t={\cal O}({\cal F}_t)~.
\end{equation}
The conditional expectation (here ${\cal F}_{t_0}=\{(x_*(s),s)|s\in [0,t_0],
x_*(0)=0, x_*(t_0)=x_0\}$, where $x_*(s)$ is fixed)
\begin{equation}
 \langle {\cal O}_{t_f} \rangle_{{\bf P},{\cal F}_{t_0}}
\end{equation}
can then be thought of as a $\Delta t_i\rightarrow 0$,
that is, $N\rightarrow \infty$, limit of the corresponding discrete expression:
\begin{equation}
 \langle {\cal O}_{t_f} \rangle_{{\bf P},{\cal F}_{t_0}}=
 \lim~\prod_{i=1}^N \int_{-\infty}^\infty {dx_i\over
 \sqrt{2\pi \Delta t_i}}\exp\left(-{(\Delta x_i)^2\over 2\Delta t_i}\right)~
 {\cal O}_{t_f,{\cal F}_{t_0}}~,
\end{equation}
where
\begin{equation}
 {\cal O}_{t_f,{\cal F}_{t_0}}={\cal O}({\cal F}_{t_0}\cup \{(x_1,t_1),
 \dots,(x_N,t_N)\})~.
\end{equation}
This limit is nothing but a Euclidean path integral
\begin{equation}
 \langle {\cal O}_{t_f} \rangle_{{\bf P},{\cal F}_{t_0}}=
 \int {\cal D}x~\exp(-S[x;t_0,t_f])~{\cal O}_{t_f,{\cal F}_{t_0}}~,
\end{equation}
where ${\cal D}x$ includes the properly normalized measure, and
\begin{equation}
 S[x;t_0,t_f] \equiv \int_{t_0}^{t_f} {{\dot x}^2(t)\over 2}~dt
\end{equation}
is the Euclidean action functional
for a free particle on ${\bf R}$ (dot in ${\dot x}(t)$ denotes time derivative).

{}To illustrate the above discussion, consider the following simple example.
Let
\begin{equation}
 {\cal  O}_t=\exp\left(\int_0^t \rho(s) dW_s\right)~,
\end{equation}
where $\rho(s)$ is a deterministic function. In the path integral we can
rewrite ${\cal O}_t$ as
\begin{equation}
 {\cal O}_t=\exp\left(\int_0^t \rho(s){\dot x}(s)ds\right)~.
\end{equation}
In particular, we have
\begin{eqnarray}
 &&{\cal O}_{t_f,{\cal F}_{t_0}}=\left.
 \exp\left(\int_0^{t_0} \rho(s){\dot x}_*(s)ds\right)
 \exp\left(\int_{t_0}^{t_f} \rho(s){\dot x}(s)ds\right)\right|_{x(t_0)=x_0}=
 \nonumber\\
 &&\,\,\,\,\,\,\,{\cal O}({\cal F}_{t_0})
 \left.\exp\left(\int_{t_0}^{t_f} \rho(s){\dot x}(s)ds\right)
 \right|_{x(t_0)=x_0}~.
\end{eqnarray}
The corresponding expectation is given by:
\begin{eqnarray}
 &&\langle {\cal O}_{t_f} \rangle_{{\bf P},
 {\cal F}_{t_0}}={\cal O}({\cal F}_{t_0})
 \int {\cal D}x~\left.\exp\left(-S[x;t_0,t_f]+
 \int_{t_0}^{t_f} \rho(s){\dot x}(s)ds\right)\right|_{x(t_0)=x_0}=\nonumber\\
 &&{\cal O}({\cal F}_{t_0})
 \exp\left(\int_{t_0}^{t_f} {\rho^2(s)\over 2}~ds\right)
 \int {\cal D}x~\left.\exp\left(-
 \int_{t_0}^{t_f} {({\dot x}(s)-\rho(s))^2\over 2}~ds\right)
 \right|_{x(t_0)=x_0}=\nonumber\\
 &&{\cal O}({\cal F}_{t_0})
 \exp\left(\int_{t_0}^{t_f} {\rho^2(s)\over 2}~ds\right)~.\label{exp.exp}
\end{eqnarray}
Here we have used the change of variable $x(t)=y(t)+\int_{t_0}^t \rho(s)ds$,
$t_0\leq t \leq t_f$, in the
path integral, and took into account that
\begin{equation}\label{path}
 \int {\cal D}y ~\left.\exp(-S[y;t_0,t_f])\right|_{y(t_0)=x_0}=
 \int {\cal D}y ~\exp(-S[y;t_0,t_f])=1~,
\end{equation}
which follows from our definition of the path integral. In particular, note
that the boundary condition $y(t_0)=x_0$ at the initial time $t_0$ is
immaterial -- the path integral (\ref{path}) is independent of $y(t_0)$. The change of the measure ${\cal D}x/{\cal D}y$ is also trivial -- see the derivation of (\ref{measure}).

{}Recall from the definition of the Brownian motion that $Z_{s,s+t} \equiv W_{s+t}-W_s$ is a normal $N(0,t)$ independent of ${\cal F}_s$.
In the path integral language this can be seen as follows. Let $z(r)\equiv y(s+r)-y(s)$, where $y(s)$ corresponds to $W_s$,. Then $z(0)=0$,
${\dot z}(r)={\dot y}(r+s)$, and
\begin{equation}
 \int_s^{s+t} {\dot y}^2(s^\prime)~ds^\prime=\int_0^t {\dot z}^2(r)~dr~.
\end{equation}
Thus, we have
\begin{eqnarray}
 \langle f({Z}_{s,s+t})\rangle_{{\bf P},{\cal F}_s}=&&\int{\cal D}y~
 \exp(-S[y;s,s+t])~\left. f(y(s+t)-y(s))\right|_{y(s)=x_*(s)}=\nonumber\\
 &&\int {\cal D}z \exp(-S[z;0,t])\left. f(z)\right|_{z(0)=0}=\nonumber\\
 &&\langle f(W_t)\rangle_{\bf P}~,
\end{eqnarray}
so $Z_{s,s+t}$ behaves the same way as the Brownian motion $W_t$ regardless of
the history ${\cal F}_s$. This is an example of what we mentioned above, that analytic computations are more streamlined in the continuous langauge, especially once we employ path integral, which makes things much simpler and more intuitive.

\subsection{Continuous Radon-Nikodym Process}

{}Suppose we want to change measure from ${\bf P}$ to ${\bf Q}$. We can
define the continuous Radon-Nikodym process
\begin{equation}
 \zeta_t\equiv {Q({\cal F}_t)\over P({\cal F}_t)}~.
\end{equation}
Then we have
\begin{eqnarray}
 &&\langle X_T\rangle_{{\bf Q}}=\left\langle {d{\bf Q}\over d{\bf P}} X_T
 \right\rangle_{\bf P}~,\\
 &&\langle X_t\rangle_{{\bf Q},{\cal F}_s}=\zeta_s^{-1}
   \langle \zeta_t X_t\rangle_{{\bf P},{\cal F}_s}~,\\
 &&\zeta_t=\left\langle{d{\bf Q}\over d{\bf P}}\right\rangle_{{\bf P},{\cal
 F}_t}~,\\
 &&{d{\bf Q}\over d{\bf P}}\equiv \zeta_T~,
\end{eqnarray}
which are continuous versions of the corresponding discrete statements.

\subsection{Cameron-Martin-Girsanov Theorem}

{}Let $W_t$ be a ${\bf P}$-Brownian motion, and let $\gamma_t$ be an
${\cal F}$-previsible process (we will impose a condition on $\gamma_t$ below).
Define a measure ${\bf Q}$ via
\begin{equation}
 {d{\bf Q}\over d{\bf P}}=\exp\left(-\int_0^T\gamma_s dW_s-
{1\over 2}\int_0^T \gamma_s^2~ds\right)~.
\end{equation}
This measure is equivalent to ${\bf P}$, and
\begin{equation}
 {\widetilde W}_t\equiv W_t+\int_0^t \gamma_s~ ds
\end{equation}
is a ${\bf Q}$-Brownian motion.

{}To see this, let us first compute the Radon-Nikodym process $\zeta_t$.
In fact, it is given by
\begin{equation}\label{zeta}
 \zeta_t=\exp\left(-\int_0^t\gamma_s dW_s-
 {1\over 2}\int_0^t \gamma_s^2~ds\right)~.
\end{equation}
This process is previsible:
\begin{equation}
 \zeta_t=\zeta({\cal F}_t)~.
\end{equation}
A quick way to see that $\zeta_t$ is given by (\ref{zeta}) is to use (\ref{exp.exp}), according to which, since $\gamma_s$ is previsible, we have
\begin{equation}
 \left\langle\exp\left(-\int_0^T\gamma_s dW_s\right)\right\rangle_{{\bf P}, {\cal F}_t} = \exp\left(-\int_0^t\gamma_s dW_s\right)\exp\left({1\over 2}\int_t^T \gamma_s^2~ds\right)~.
\end{equation}
However, in deriving (\ref{exp.exp}) we did not deal with the measure, so it is instructive to directly compute the expectation
(let $\gamma_s =\gamma^*_s$ and $x(s)=x_*(s)$, $s\in[0,t]$, for the
path ${\cal F}_t$) using the path integral:
\begin{eqnarray}
 &&\left\langle{d{\bf Q}\over d{\bf P}}\right\rangle_{{\bf P},{\cal
 F}_t}=\int {\cal D}x~\exp(-S[x;t,T])~
 \left({d{\bf Q}\over d{\bf P}}\right)_{{\cal F}_t}=\nonumber\\
 &&\zeta({\cal F}_t)\int {\cal D}x~\left.\exp\left(-S[x;t,T]-
 \int_t^T\gamma_s {\dot x}(s) ds -
 {1\over 2}\int_t^T \gamma_s^2~ds\right)\right|_{x(t)=x_*(t),~\gamma_t=
 \gamma^*_t}=\nonumber\\
 &&\zeta({\cal F}_t)\int {\cal D}x~\left.\exp\left(-
 {1\over 2}\int_t^T (\dot{x}(s)+\gamma_s)^2~ds\right)
 \right|_{x(t)=x_*(t),~\gamma_t=\gamma^*_t}=\nonumber\\
 &&\zeta({\cal F}_t)\int {\cal D}x~\left.\exp\left(-
 {1\over 2}\int_t^T \dot{y}^2(s)~ds\right)
 \right|_{y(t)=x_*(t)}~,\label{measure}
\end{eqnarray}
where $y(s)\equiv x(s)+\int_t^s \gamma_{s^\prime}~ds^\prime$, $t\leq s\leq
T$. To evaluate this last integral, we need to convert ${\cal D}x$ into
${\cal D}y$ with the appropriate measure. This measure, in fact, is trivial:
${\cal D}x={\cal D}y$. To see this, let us discretize our path integral.
Then we have ($t_0=t$, $t_N=T$)
\begin{equation}
 \Delta y_i\equiv y(t_i)-y(t_{i-1})=\Delta x_i+\gamma_{i-1} \Delta t_i~.
\end{equation}
We, therefore, have
\begin{eqnarray}
 &&\int {\cal D}x~\left.\exp\left(-
 {1\over 2}\int_t^T \dot{y}^2(s)~ds\right)
 \right|_{y(t)=x_*(t)}=\nonumber\\
 &&\lim~\prod_{i=1}^N \int_{-\infty}^\infty {dx_i\over
 \sqrt{2\pi\Delta t_i}}\left.\exp\left(-{1\over 2}\left[{\Delta x_i\over
 \Delta t_i}+
 \gamma_{i-1}\right]^2\Delta t_i
 \right)\right|_{x_0=x_*(t),~\gamma_0=\gamma^*_t}
 =\nonumber\\
 &&\lim~\prod_{i=1}^N \int_{-\infty}^\infty {d\Delta x_i\over
 \sqrt{2\pi\Delta t_i}}\left.\exp\left(-{1\over 2}\left[{\Delta x_i\over
 \Delta t_i}+
 \gamma_{i-1}\right]^2\Delta t_i
 \right)\right|_{x_0=x_*(t),~\gamma_0=\gamma^*_t}
 =\nonumber\\
 &&\lim~\prod_{i=1}^N \int_{-\infty}^\infty {d\Delta y_i\over
 \sqrt{2\pi\Delta t_i}}\left.\exp\left(-{(\Delta y_i)^2\over 2
 \Delta t_i}
 \right)\right|_{y_0=x_*(t)}=\nonumber\\
 &&\lim~\prod_{i=1}^N \int_{-\infty}^\infty {d y_i\over
 \sqrt{2\pi\Delta t_i}}\left.\exp\left(-{(\Delta y_i)^2\over 2
 \Delta t_i}
 \right)\right|_{y_0=x_*(t)}=\nonumber\\
 &&\int {\cal D}y~\left.\exp\left(-
 {1\over 2}\int_t^T \dot{y}^2(s)~ds\right)
 \right|_{y(t)=x_*(t)}=1~.
\end{eqnarray}
The key points in the above computation are the following. First, we can change
the integration variables from $x_i$ to $\Delta x_i=x_i-x_{i-1}$,
$i=1,\dots,N$. Note that
\begin{equation}
 x_i=x_0+\sum_{k=1}^i \Delta x_k~,
\end{equation}
so that
\begin{equation}
 {\partial x_i\over \partial\Delta x_j}=\theta_{ij}~,
\end{equation}
where $\theta_{ij}=0$ if $i<j$, and $\theta_{ij}=1$ if $i\geq j$. This implies
that the corresponding measure is trivial:
\begin{equation}
 \det(\theta_{ij})=1~.
\end{equation}
Next, consider the change of variables from $\Delta x_i$ to $\Delta y_i$. We
have
\begin{eqnarray}
 {\cal M}_{ij}\equiv
 {\partial\Delta y_i\over \partial \Delta x_j}=&&\delta_{ij}+
 \Delta t_i {\partial \gamma_{i-1}\over\partial\Delta x_j}=\nonumber\\
 &&\delta_{ij}+
 \Delta t_i \sum_{k=1}^N {\partial \gamma_{i-1}\over\partial x_k}~
 {\partial x_k\over \partial\Delta x_j}=\nonumber\\
 &&\delta_{ij}+
 \Delta t_i \sum_{k=j}^N {\partial \gamma_{i-1}\over\partial x_k}~.
\end{eqnarray}
Note, however, that $\gamma_{i-1}$ is independent of $x_k$ with $k\geq i$.
This implies that ${\cal M}_{ij}=0$ if $i<j$, and ${\cal M}_{ii}=1$ (that is,
${\cal M}_{ij}$ is a Jordanian matrix with unit diagonal elements). It then
follows that
\begin{equation}
 \det({\cal M}_{ij})=1~,
\end{equation}
so that the measure corresponding to the change of variables
from $\Delta x_i$ to $\Delta y_i$ is also trivial. Finally, we can change
variables from $\Delta y_i$ to $y_i$ also with a trivial measure.
Note that once we change variables from $x_i$ to $\Delta x_i$ the boundary
condition $x_0=x_*(t)$ becomes immaterial, and it remains such upon changing
variables from $\Delta x_i$ to $\Delta y_i$ to $y_i$. This also completes our proof of (\ref{exp.exp}).

{}Thus, we see that $\zeta_t$ is indeed given by (\ref{zeta}). This
implies that $\zeta_t$ is
a ${\bf P}$-martingale (see below), and the measures ${\bf Q}$ and ${\bf P}$
are equivalent. More precisely, we must impose a non-trivial condition on
$\gamma_t$. In particular, note that the SDE for $\zeta_t$ is given by:
\begin{equation}
 d\zeta_t=-\gamma_t\zeta_t dW_t~,
\end{equation}
so that the volatility of $\zeta_t$ is $-\gamma_t\zeta_t$, and the drift is
zero. So $\zeta_t$ is a stochastic process if (this is a technical condition)
\begin{equation}
 \int_0^t \gamma_s^2\zeta_s^2~ds
\end{equation}
is finite (with probability 1).

{}We can now show that ${\widetilde W}_t$ is a ${\bf Q}$-Brownian motion.
Clearly, ${\widetilde W}_t$ is continuous, and ${\widetilde W}_0=0$. Let
$f({\widetilde W}_t)$ be a deterministic function of ${\widetilde W}_t$.
Then we have ($y(s)\equiv x(s)+\int_0^s \gamma_{s^\prime}~ds^\prime$,
$s\in [0,t]$):
\begin{equation}
 \langle f({\widetilde W}_t)\rangle_{\bf Q}=
 \langle \zeta_t f({\widetilde W}_t)\rangle_{\bf P}=
 \int{\cal D}y~\exp(-S[y;0,t])~f(y)=\
 \langle f({W}_t)\rangle_{\bf P}~,
\end{equation}
which, in particular, implies that ${\widetilde W}_t$ is a normal $N(0,t)$
under ${\bf Q}$ just as $W_t$ is under the measure ${\bf P}$.

{}Next, let us define a process:
\begin{equation}
 {\widetilde Z}_{s,s+t}\equiv {\widetilde W}_{s+t}-{\widetilde W}_s=
 Z_{s,s+t}+\int_s^{s+t}\gamma_{s^\prime}~ds^\prime~,
\end{equation}
where
\begin{equation}
 {Z}_{s,s+t}\equiv {W}_{s+t}-{W}_s~.
\end{equation}
Then we have ($y(\tau)\equiv x(s)+\int_s^\tau \gamma_{s^\prime}~ds^\prime$,
$\tau\in[s,s+t]$):
\begin{eqnarray}
 \langle f({\widetilde Z}_{s,s+t})\rangle_{{\bf Q},{\cal F}_s}=&&
 \zeta_s^{-1}\langle \zeta_{s+t} f({\widetilde Z}_{s,s+t})
 \rangle_{{\bf P},{\cal F}_s}=\nonumber\\
 &&\int{\cal D}y~\exp(-S[y;s,s+t])~\left. f(y(s+t)-y(s))\right|_{y(s)=x_*(s)}
 =\nonumber\\
 &&\langle f({Z}_{s,s+t})\rangle_{{\bf P},{\cal F}_s}~,
 \end{eqnarray}
so that the process ${\widetilde Z}_{s,s+t}$ under ${\bf Q}$ behaves the same
way as the process ${Z}_{s,s+t}$ under ${\bf P}$. Thus, ${\widetilde W}_t$ is
indeed a ${\bf Q}$-Brownian motion.

\section{Continuous Martingales}

{}A stochastic process $M_t$ is a {\em martingale} w.r.t. a measure ${\bf P}$
if and only if for all $t\geq 0$ the expectation $\langle |M_t|\rangle_{\bf P}$
is finite, and
\begin{equation}
 \langle M_t\rangle_{{\bf P},{\cal F}_s}=M_s~,~~~0\leq s\leq t~.
\end{equation}
That is, a martingale is expected to be driftless.

{}Just as in the discrete case, we have the {\em tower law} for conditional
expectations:
\begin{equation}
 \left\langle\langle X_T\rangle_{{\bf P},{\cal F}_t}
 \right\rangle_{{\bf P},{\cal F}_s}=\langle X_T
 \rangle_{{\bf P},{\cal F}_s}~,~~~0\leq s\leq t\leq T~.
\end{equation}
This implies that the process
\begin{equation}
 N_t\equiv \langle X_T\rangle_{{\bf P},{\cal F}_t}
\end{equation}
is a ${\bf P}$-martingale provided
that $\langle |X_T|\rangle_{\bf P}$ is finite. The fact that this last
condition is necessary as well as sufficient can be seen as follows. First,
note that $N_T=X_T$. However, we must have finite
$\langle |N_T|\rangle_{\bf P}=\langle |X_T|\rangle_{\bf P}$. Next, note that
\begin{equation}
 |N_t|=\left|\langle X_T\rangle_{{\bf P},{\cal F}_t}\right|\leq
 \langle |X_T|\rangle_{{\bf P},{\cal F}_t}\leq
 \langle |X_T|\rangle_{\bf P}~.
\end{equation}
This then implies that $|N_t|$ is bounded by $\langle |X_T|\rangle_{\bf P}$,
and, therefore, so is its expectation.

{}Note that a ${\bf P}$-Brownian motion $W_t$ is a ${\bf P}$-martingale.
To check the first condition, recall that for any process $A_t$ we have
\begin{equation}
 \langle A_t\rangle^2_{\bf P}\leq \langle A_t^2\rangle_{\bf P}~.
\end{equation}
This implies that
\begin{equation}
 \langle |W_t|\rangle_{\bf P}\leq \sqrt{\langle |W_t|^2\rangle_{\bf P}}=
 \sqrt{t}~.
\end{equation}
Furthermore,
\begin{equation}
 \langle W_t\rangle_{{\bf P},{\cal F}_s}=
 \langle W_s\rangle_{{\bf P},{\cal F}_s}+
 \langle (W_t-W_s)\rangle_{{\bf P},{\cal F}_s}=W_s~.
\end{equation}
Thus, $W_t$ is indeed a ${\bf P}$-martingale.

\subsection{Driftlessness}

{}Next, consider a general stochastic process $X_t$:
\begin{equation}
 dX_t=\sigma_t dW_t+\mu_t dt~.
\end{equation}
Let us show that $X_t$ can be a ${\bf P}$-martingale only if
$\mu_t\equiv 0$. From the definition of a martingale we have
\begin{equation}
 0=\langle dX_t\rangle_{{\bf P},{\cal F}_t}=\langle \sigma_t dW_t
 \rangle_{{\bf P},{\cal F}_t}+\langle \mu_t dt
 \rangle_{{\bf P},{\cal F}_t}=\sigma_t\langle dW_t
 \rangle_{{\bf P},{\cal F}_t}+\mu_t dt=\mu_t dt~.
\end{equation}
Note that $W_t$ is previsible, but $dW_t$ is not.

{}Next, suppose that $X_t$ is a driftless stochastic process:
\begin{equation}
 dX_t=\sigma_t dW_t~.
\end{equation}
Then we have
\begin{equation}
 X_t=X_0+\int_0^t \sigma_t dW_t~.
\end{equation}
Note that $X_t$ is a previsible process: $X_t=X({\cal F}_t)$.
Let us compute the conditional expectation:
\begin{eqnarray}
 &&\langle X_t\rangle_{{\bf P},{\cal F}_s}=
 X({\cal F}_s)+\int {\cal D}x~\exp(-S[x;s,t])~\left.\int_s^t \sigma_{s^\prime}
 {\dot x}(s^\prime)ds^\prime\right|_{x(s)=x_*(s),~\sigma_s=\sigma^*_s}
 =\nonumber\\
 &&X({\cal F}_s)+
 \lim~
 \left[\prod_{i=1}^N \int_{-\infty}^\infty {dx_i\over\sqrt{2\pi\Delta t_i}}
 \exp\left(-{{\Delta x_i}^2\over \Delta t_i}\right)\right]\left.
 \sum_{k=1}^N\sigma_{k-1} \Delta x_k\right|_{x_0=x_*(s),~\sigma_0=\sigma^*_s}=
 \nonumber\\
 &&X({\cal F}_s)=X_s~.
\end{eqnarray}
Here we have taken into account that we can change integration variables from
$x_i$ to $\Delta x_i$ with a trivial measure. Then integration over the
$\Delta x_k$ variable makes the corresponding $k$th term in the sum vanish as
$\sigma_{k-1}$ is independent of $\Delta x_k$.

{}Thus, a driftless stochastic process is a martingale subject to the
condition that $\langle |X_t|\rangle_{\bf P}$ is finite. We can guarantee this
as follows. Let $Z_t\equiv X_t-X_0$, and
$\langle Z_t\rangle_{\bf P}=\langle X_t\rangle_{\bf P}-X_0=0$. On the other
hand,
\begin{equation}
 \langle |X_t|\rangle_{\bf P}\leq \sqrt{\langle X_t^2\rangle_{\bf P}}=
 \sqrt{\langle Z_t^2\rangle_{\bf P}+X_0^2}~,
\end{equation}
so that $\langle |X_t|\rangle_{\bf P}$ is finite if
$\langle Z_t^2\rangle_{\bf P}$ is finite. On the other hand,
\begin{equation}
 Z_t^2=\int_0^t d(Z_s)^2=2\int_0^t Z_s\sigma_s dW_s+\int_0^t \sigma^2_s ds~.
\end{equation}
Since $Z_s\sigma_s$ is a previsible process,
\begin{equation}
 \left\langle \int_0^t Z_s\sigma_s dW_s\right\rangle_{\bf P}=0~,
\end{equation}
so we have
\begin{equation}
 \langle Z_t^2\rangle_{\bf P}=\left\langle\int_0^t \sigma_s^2 ds\right\rangle~,
\end{equation}
and if the r.h.s. of this equation is finite, then so is
$\langle |X_t|\rangle_{\bf P}$. Here we note that this condition is
sufficient but not necessary.

{}Thus, consider the following SDE:
\begin{equation}
 dX_t=\sigma_t X_t dW_t~.
\end{equation}
The solution to this SDE is given by
\begin{equation}\label{expmar}
 X_t=X_0\exp\left(\int_0^t\sigma_s dW_s-{1\over 2}\int_0^t \sigma_s^2 ds
 \right)~.
\end{equation}
Note that $|X_t|={\rm sign}(X_0) X_t$, so that the requirement that
$\langle |X_t|\rangle_{\bf P}$ be finite is satisfied. Thus, (\ref{expmar})
is an exponential martingale.

{}Here the following remarks are in order. Consider a driftless process
\begin{equation}\label{SDErho}
 dX_t=\rho_t dW_t~.
\end{equation}
In general the condition on
$\langle |X_t|\rangle_{\bf P}$ is non-trivial. Let us formally
rewrite this SDE as follows:
\begin{equation}\label{SDEsigma}
 dX_t=\sigma_t X_t dW_t~,
\end{equation}
where $\rho_t\equiv \sigma_t X_t$. But
the exponential martingale (\ref{expmar})
satisfies the condition on
$\langle |X_t|\rangle_{\bf P}$ regardless of $\sigma_t$. The reason for this
apparent discrepancy is that not all SDEs of the form (\ref{SDErho}) can be
rewritten in the form (\ref{SDEsigma}). For instance, if $\rho_t\equiv\rho$
is constant, the solution to (\ref{SDErho}) is simply $X_t=\rho W_t+X_0$.
This process can take (with probability 1) both positive as well as
negative values,
while the exponential martingale (\ref{expmar}), which is the solution to
(\ref{SDEsigma}), would take only either positive or negative values (depending
on whether $X_0$ in (\ref{expmar}) is positive or negative).

{}On the other hand, suppose
in (\ref{SDEsigma}) we take
\begin{equation}
 \sigma_t={1\over{W_t+\beta}}
\end{equation}
with constant $\beta$. The formal solution to this SDE is then given by
\begin{equation}
 X_t=X_0\exp(Y_t)~
\end{equation}
where
\begin{equation}
 dY_t=\sigma_tdW_t-{1\over 2}\sigma_t^2 dt={dW_t\over{W_t+\beta}}-
 {1\over 2}{dt\over(W_t+\beta)^2}~.
\end{equation}
The solution to this SDE is given by (we are assuming the boundary
condition $Y_0=0$):
\begin{equation}
 Y_t=\ln|W_t+\beta|-\ln|\beta|~.
\end{equation}
Thus, we have
\begin{equation}
 X_t=X_0\left|{{W_t+\beta}\over\beta}\right|~.
\end{equation}
It is clear that this is {\em not} a martingale. The reason why this occurred
is the following. Note that the volatility of $X_t$ is given by
\begin{equation}
 \rho_t=\sigma_t X_t= {X_0\over|\beta|}~{\rm sign}(W_t+\beta)~,
\end{equation}
which is discontinuous (albeit the requirement that $\int_0^t \rho_s^2 ds$ be
finite is formally satisfied) and not previsible. This is why the above formal manipulations
did not yield a martingale.

\subsection{Martingale Representation Theorem}

{}From the above discussion we have the continuous version of the
{\em martingale representation theorem}:\\
\indent
Suppose $M_t$ is a ${\bf Q}$-martingale whose volatility $\sigma_t$ is
always non-vanishing (with probability 1). Then if $N_t$ is any other
${\bf Q}$-martingale, there exists an ${\cal F}$-previsible process $\phi$
such that $\int_0^t \phi_s^2\sigma_s^2 ds$ is finite (with probability 1),
and
\begin{equation}
N_t=N_0+\int_0^t \phi_s dM_s~.
\end{equation}
Further, $\phi_s$ is (essentially) unique.

{}This can be seen as follows. Since $M_t$ is a ${\bf Q}$-martingale, we have
\begin{equation}
 dM_t=\sigma_t dW_t~.
\end{equation}
Similarly, since $N_t$ is a ${\bf Q}$-martingale, we have
\begin{equation}
 N_t=\rho_t dW_t~,
\end{equation}
where $\rho_t$ is the volatility of $N_t$. Then the process $\phi$ is given
by $\phi_t=\rho_t/\sigma_t$,
which is well defined as $\sigma_t$ never vanishes. Moreover, since
$\int_0^t \rho_s^2 ds$ is finite, then
$\int_0^t \phi_s^2\sigma_s^2 ds$ is also finite.

\section{Continuous Hedging}

{}Suppose we have a stock $S_t$ and a cash bond $B_t$. We will assume that
the latter is deterministic. To replicate a claim $X$ at the maturity time
$T$, we have the following hedging strategy.

{}First, define a discounted stock process $Z_t\equiv B_t^{-1}S_t$. We need a
measure ${\bf Q}$ that makes $Z_t$ into a martingale with {\em positive}
volatility.

{}Next, define the process $E_t\equiv \langle B_T^{-1} X\rangle_{{\bf Q},
{\cal F}_t}$. This process is a ${\bf Q}$-martingale. It then follows from the
martingale representation theorem that there exists a previsible process
$\phi_t$ such that
\begin{equation}
 dE_t=\phi_t dZ_t~.
\end{equation}
Also, define the process
\begin{equation}
 \psi_t\equiv E_t-\phi_t Z_t~.
\end{equation}
This process is also previsible.

{}At time $t$ hold a portfolio $(\phi_t,\psi_t)$ consisting of $\phi_t$
units of stock $S$ and $\psi_t$ units of the cash bond $B$. The value of this
portfolio is
\begin{equation}
 V_t=\phi_t S_t+\psi_t B_t= B_t E_t~.
\end{equation}
Let us show that this portfolio is {\em self-financing}.

{}First, note that
\begin{equation}
 V_T=B_T E_T =X~,
\end{equation}
so at time $T$ it replicates the claim $X$. Furthermore,
\begin{eqnarray}
 dV_t=&&B_t dE_t+E_t dB_t=\nonumber\\
 &&\phi_t B_t dZ_t +(\psi_t+\phi_t Z_t)dB_t=\nonumber\\
 &&\phi_t dS_t+\psi_t dB_t~.
\end{eqnarray}
The price of the claim $X$ at time $t$, therefore, is given by
\begin{equation}
 V_t=B_t\langle B_T^{-1}X\rangle_{{\bf Q},{\bf F}_t}~.
\end{equation}
In particular, $V_0=B_0\langle B_T^{-1}X\rangle_{\bf Q}$.

\subsection{Change of Measure in the General One-Stock Model}

{}Let us consider a general one-stock model:
\begin{eqnarray}
 &&dB_t=r_t B_t dt~,\\
 &&dS_t=S_t\left[\sigma_t dW_t+\mu_t dt\right]~,
\end{eqnarray}
where $\sigma_t,\mu_t,r_t$ are general previsible processes.

{}The above equations have the following solutions:
\begin{eqnarray}
 &&B_t=B_0 \exp\left(\int_0^t r_s~ds\right)~,\\
 &&S_t=S_0\exp \left(\int_0^t \sigma_s dW_s+\int_0^t\left[\mu_s-{1\over 2}
 \sigma^2_s\right] ds\right)~.
\end{eqnarray}
The discounted stock process is given by:
\begin{equation}
 Z_t=B_t^{-1}S_t=
 S_0\exp \left(\int_0^t \sigma_s dW_s+\int_0^t\left[\mu_s-r_s-{1\over 2}
 \sigma^2_s\right] ds\right)~.
\end{equation}
We need to change the measure from ${\bf P}$ to ${\bf Q}$ so that $Z_t$ is
a martingale. Let us define
\begin{equation}
 {\widetilde W}_t\equiv W_t+\int_0^t \gamma_s~ds~,
\end{equation}
where
\begin{equation}
 \gamma_t\equiv {{\mu_t-r_t}\over \sigma_t}~.
\end{equation}
Then we have
\begin{equation}
 Z_t=B_t^{-1}S_t=
 S_0\exp \left(\int_0^t \sigma_s d{\widetilde W}_s-
 {1\over 2}\int_0^t
 \sigma^2_s ds\right)~.
\end{equation}
Note that $Z_t$ is an exponential ${\bf Q}$-martingale, where
the measure ${\bf Q}$ is such that ${\widetilde W}_t$ is a ${\bf Q}$-Brownian
motion. The corresponding Radon-Nikodym process is given by:
\begin{equation}
 \zeta_t=\exp\left(-\int_0^t\gamma_s dW_s-
 {1\over 2}\int_0^t\gamma_s^2 ~ds\right)~.
\end{equation}
Using $\zeta_t$ we can obtain ${\bf Q}$ from ${\bf P}$.

\subsection{Terminal Value Pricing}

{}Let us assume that $B_t$ (and, therefore, $r_t$) is a deterministic
function (independent of which particular history ${\cal F}_t$ the stock
follows up to time $t$). Also, let us assume that the log-volatility $\sigma_t$
is a deterministic function: $\sigma_t=\sigma(S_t,t)$. Then the following is
true.

{}Suppose the derivative $X$ is given by $f(S_T)$, where $f(x)$ is some
deterministic function. Then the value of the derivative at time $t$ is
given by $V(S_t,t)$, where
\begin{equation}
 V(z,t)\equiv B_t\langle B_T^{-1} f(S_T)\rangle_{{\bf Q},~S_t=z}~.
\end{equation}
The process $\phi_t$ is then given by
\begin{equation}\label{phiPDE}
 \phi_t=\left. \partial_z V(z,t)\right|_{z=S_t}~.
\end{equation}
This can be seen as follows.

{}First, note that
\begin{equation}
 dS_t=S_t\left[\sigma_t d{\widetilde W}_t+r_t dt\right]~.
\end{equation}
Next,
\begin{eqnarray}
 &&dV_t=dV(S_t,t)=\partial_z V(S_t,t)dS_t+{1\over 2}\partial^2_z V(S_t,t)
 (dS_t)^2 +\partial_t V(S_t,t)dt=\nonumber\\
 &&\,\,\,\,\,\,\,\sigma_t S_t\partial_z V(S_t,t)d{\widetilde W}_t+\nonumber\\
 &&\,\,\,\,\,\,\,\left[r_t S_t \partial_z V(S_t,t) + {1\over 2}\sigma_t^2 S_t^2
 \partial^2_z V(S_t,t)+\partial_t V(S_t,t)\right]dt~.
\end{eqnarray}
On the other hand,
\begin{eqnarray}
 dV_t=&&\phi_t dS_t+\psi_t dB_t=\nonumber\\
 &&\sigma_t S_t \phi_t d{\widetilde W}_t+r_t\left[\phi_t S_t+\psi_t B_t
 \right]dt=
 \nonumber\\
 &&\sigma_t S_t \phi_t d{\widetilde W}_t+r_t V_t dt~.
\end{eqnarray}
Comparing these two expressions we see that $\phi_t$ is indeed given by
(\ref{phiPDE}). Moreover, we have the following partial differential equation
(PDE) for $V(z,t)$:
\begin{equation}\label{pricePDE}
 r_t z \partial_z V(z,t) + {1\over 2}\sigma_t^2 z^2
 \partial^2_z V(z,t)+\partial_t V(z,t)-r_t V(z,t)=0
\end{equation}
with the boundary condition $V(z,T)=f(z)$. This PDE, which is called the {\em Black-Scholes equation}, gives another way of
solving the pricing problem. Using the so-called {\em Greeks}
\begin{eqnarray}
 &&\Theta\equiv {\partial V\over \partial t}~,\\
 &&\Delta \equiv {\partial V\over \partial S}~,\\
 &&\Gamma \equiv {\partial^2 V\over \partial S^2}~,\\
 &&\nu\equiv {\partial V\over \partial \sigma}~,\\
 &&\rho\equiv {\partial V\over \partial r}~,
\end{eqnarray}
where the last two definitions are given for the sake of completeness, we have
\begin{equation}
 \Theta + rS\Delta + {1\over 2}\sigma^2S^2\Gamma= rV~.
\end{equation}
The five Greeks above (and there are more) are called Theta, Delta, Gamma, Vega and Rho.

\subsection{A Different Formulation}

{}Suppose we have a general stock model
\begin{eqnarray}
 &&dB_t=r_t B_t dt~,\\
 &&dS_t=\sigma_t dW_t +\mu_t dt~.
\end{eqnarray}
The discounted stock process is given by $Z_t=B_t^{-1} S_t$, and we have
\begin{equation}
 dZ_t=B_t^{-1}\left[\sigma_t dW_t +\left(\mu_t-r_t S_t\right)dt\right]~.
\end{equation}
The shift $\gamma_t$ that will make $Z_t$ into a martingale is given by
\begin{equation}
 \gamma_t={{\mu_t-r_tS_t}\over\sigma_t}~.
\end{equation}
The corresponding change of measure, however, is not always possible
(we will discuss an example of this in the following). The
reason why is that the volatility $\sigma_t$ might sometimes be vanishing.
Suppose, however, that the volatility never vanishes. Then we can find the
martingale measure ${\bf Q}$:
\begin{eqnarray}
 &&dZ_t=B_t^{-1}\sigma_t d{\widetilde W}_t~,\\
 &&dS_t=\sigma_t d{\widetilde W}_t+r_t S_t dt~.
\end{eqnarray}
In the following we will assume that both $r_t$ and $\sigma_t$ are
deterministic.

{}We have:
\begin{equation}\label{priceex}
 V(z,t)\equiv B_t\langle B_T^{-1} f(S_T)\rangle_{{\bf Q},~S_t=z}~.
\end{equation}
We, therefore, have:
\begin{eqnarray}
 &&dV_t=dV(S_t,t)=\partial_z V(S_t,t) dS_t+{1\over 2}\partial^2_z V(S_t,t)
 (dS_t)^2+\partial_t V(S_t,t)dt=\nonumber\\
 &&\sigma_t \partial_z V(S_t,t)d{\widetilde W}_t+
 \left[r_t S_t \partial_z V(S_t,t) + {1\over 2}\sigma^2_t\partial^2_z V(S_t,t)
 +\partial_t V(S_t,t)\right]dt~.
\end{eqnarray}
On the other hand,
\begin{eqnarray}
 dV_t=&&\phi_t dS_t+\psi_tdB_t=\nonumber\\
 &&\sigma_t\phi_t d{\widetilde W_t}+r_t\left[\phi_t S_t+\psi_t B_t
 \right]dt=\nonumber\\
 &&\sigma_t\phi_t d{\widetilde W_t}+r_t V_t dt~.
\end{eqnarray}
We, therefore, have
\begin{eqnarray}
 &&\phi_t=\partial_z V(S_t,t)~,\\
 &&\psi_t=B_t^{-1}\left[V_t-\phi_t S_t\right]=
 B_t^{-1}\left[V(S_t,t)-S_t\partial_z V(S_t,t)\right]~,
\end{eqnarray}
and the following PDE for $V(z,t)$:
\begin{equation}\label{PDEpr}
 r_t S_t \partial_z V(S_t,t) + {1\over 2}\sigma^2_t\partial^2_z V(S_t,t)
 +\partial_t V(S_t,t)-r_t V(S_t,t)=0
\end{equation}
with the boundary condition $V(z,T)=f(z)$.

\subsection{An Instructive Example}

{}As an example consider the following stock model (we will assume that the
cash bond is constant, that is, we have zero interest rates):
\begin{equation}
 S_t=S_0+\alpha W_t^2-\beta t
\end{equation}
with constant $\alpha$ and $\beta$. The corresponding SDE is
\begin{equation}
 dS_t=2\alpha W_t dW_t+(\alpha-\beta)dt~,
\end{equation}
so that the volatility and the drift are given by:
\begin{eqnarray}
  &&\sigma_t=2\alpha W_t~,\\
  &&\mu_t=\alpha-\beta~.
\end{eqnarray}
The shift $\gamma_t$ is then
\begin{equation}
 \gamma_t={{\alpha-\beta}\over 2\alpha W_t}~,
\end{equation}
which is ill-defined at $t=0$. It is then not difficult to see that the
change of measure via the Radon-Nikodym process is not possible -- the
measures ${\bf Q}$ and ${\bf P}$ are not equivalent in this case.

{}Suppose we are lucky, and $\beta=\alpha$, so that $S_t$ is a martingale
to begin with. Still,
to hedge a generic claim we would need to use the martingale representation
theorem to determine the previsible process $\phi_t$. However, since $\sigma_t$
vanishes at $t=0$, for a generic claim this might not be possible.
Nonetheless, we can still try to hedge claims of the form
$X=f(S_T)$ using the PDE approach to pricing.

{}In the above example we have (restricting to times $t\leq T <S_0/\alpha$ so that $S_t > 0$):
\begin{eqnarray}
 &&S_t=S_0+\alpha(W_t^2-t)~,\\
 &&\sigma_t^2=4\alpha^2 W_t^2= 4\alpha (S_t-S_0+\alpha t)~.
\end{eqnarray}
The corresponding pricing PDE then reads:
\begin{equation}
 2\alpha\left[z-S_0+\alpha t\right]\partial^2_z V(z,t)+\partial_t V(z,t)=0
\end{equation}
with the boundary condition $V(z,T)=f(z)$.

{}This PDE can be simplified as follows. Let
\begin{equation}
 y\equiv z-S_0+\alpha t~,
\end{equation}
and $V(z,t)\equiv U(y,t)$.
Then we have:
\begin{equation}
 \alpha\left[2y\partial^2_y U(y,t)+\partial_y U(y,t)\right]+
 \partial_t U(y,t)=0
\end{equation}
with the boundary condition $U(y,T)=f(y+S_0-\alpha T)$.

{}Note that for the allowed stock values $y$ is non-negative. We can therefore
perform the following change of variables:
\begin{equation}
 y\equiv \alpha x^2~.
\end{equation}
Let $V(y,t)\equiv C(x,t)$. Then we have the following PDE:
\begin{equation}\label{PDEquad}
 {1\over 2}\partial_x^2 C(x,t)+\partial_t C(x,t)=0
\end{equation}
with the boundary condition $C(x,T)=f(\alpha x^2+S_0-\alpha T)$.

{}Note that this is nothing but the terminal value pricing in terms of $W_t$ --
the variable $x$ is simply the value of $W_t$. So in this example we might as
well price the option directly via ({\ref{priceex}) -- indeed, in this case
we know $S_t=S(W_t,t)$ explicitly. However, in general we might not have an
explicit solution of the SDE for $S_t$, in which case we can
use the pricing PDE (\ref{PDEpr}) (and, if necessary, solve it numerically).

{}Let us determine $V(z,t)$ in the above example directly via (\ref{priceex}).
We have (in this case ${\bf P}={\bf Q}$):
\begin{equation}
 V(z,t)=\langle f(S_T)\rangle_{{\bf P},~S_t=z}~.
\end{equation}
Note that
\begin{equation}
 W_T=W_t+Z_{t,t+(T-t)}~,
\end{equation}
where $Z_{t,t+(T-t)}$ is a normal $N(0,T-t)$, and is independent of
${\cal F}_t$. Let the values of $Z_{t,t+(T-t)}$ be $x$. Then we have
\begin{eqnarray}
 S_T=&&S_0+\alpha\left[(W_t+x)^2-T\right]=\nonumber\\
 &&S_t+\alpha\left[x^2+2W_tx-(T-t)\right]=\nonumber\\
 &&S_t+\alpha\left[x^2+2\epsilon x\sqrt{{S_t-S_0+\alpha t}\over\alpha}
 -(T-t)\right]~,
\end{eqnarray}
where $\epsilon=\pm 1$ gives two values of $W_t$ corresponding to a given
$S_t$.

{}The pricing function $V(z,t)$ is given by:
\begin{eqnarray}
 &&V(z,t)=\nonumber\\
 &&\int_{-\infty}^\infty {dx\over\sqrt{2\pi(T-t)}}~
 \exp\left(-{x^2\over 2(T-t)}\right) \times \nonumber\\
 &&\,\,\,\,\,\,\, \times f\left(z+
 \alpha\left[x^2+2\epsilon x\sqrt{{z-S_0+\alpha t}\over\alpha}
 -(T-t)\right]\right)=\nonumber\\
 &&\int_{-\infty}^\infty {dy\over\sqrt{2\pi}}~
 \exp\left(-{y^2\over 2}\right) \times \nonumber \\
 &&\,\,\,\,\,\,\, \times f\left(z+
 \alpha(T-t)\left[y^2+2\epsilon y\sqrt{{z-S_0+\alpha t}\over\alpha(T-t)}
 -1\right]\right)~.
\end{eqnarray}
Note that $V(z,t)$ is the same for both $\epsilon=\pm 1$. Also, $V(z,T)=f(z)$
as it should be.

{}We can obtain the same result from the PDE (\ref{PDEquad}). The solution
to this PDE with the appropriate boundary condition is given by:
\begin{equation}
 C(x,t)=\int_{-\infty}^\infty dx^\prime K(x^\prime-x,T-t)f\left(
 \alpha(x^\prime)^2+S_0-\alpha T\right)~,
\end{equation}
where $K(y,\tau)$ is the solution to the PDE
\begin{equation}
 {1\over 2}\partial_y^2 K(y,\tau)=\partial_\tau K(y,\tau)
\end{equation}
with the boundary condition $K(y,0)=\delta(y)$. This solution is given by:
\begin{equation}
 K(y,\tau)={1\over\sqrt{2\pi\tau}}~\exp\left(-{y^2\over 2\tau}\right)~.
\end{equation}
We, therefore, have
\begin{eqnarray}
 C(x,t)=\int_{-\infty}^\infty {dx^\prime\over\sqrt{2\pi(T-t)}}~
  \exp\left(-{(x^\prime-x)^2\over 2(T-t)}\right)~f\left(
 \alpha(x^\prime)^2+S_0-\alpha T\right)~.
\end{eqnarray}
It is not difficult to see that this is the same as $V(z,t)$ we obtained above
once we go back from $x$ to $z$ via $z=\alpha x^2+S_0-\alpha t$.

\subsection{The Heat Kernel Method}

{}In the previous subsection we solved a pricing PDE using the {\em heat
kernel method}. It can also be used in the general case. Thus, let us go back
to the general pricing PDE (\ref{pricePDE}):
\begin{equation}
 r_t z \partial_z V(z,t) + {1\over 2}\sigma_t^2 z^2
 \partial^2_z V(z,t)+\partial_t V(z,t)-r_t V(z,t)=0
\end{equation}
with the boundary condition $V(z,T)=f(z)$. Let us simplify this equation as
follows. Let
\begin{equation}
 V(z,t)\equiv \exp\left(-\int_t^T r_s ~ds\right)~U(z,t)~.
\end{equation}
The PDE for $U(z,t)$ is given by:
\begin{equation}
 r_t z \partial_z U(z,t) + {1\over 2}\sigma_t^2 z^2
 \partial^2_z U(z,t)+\partial_t U(z,t)=0
\end{equation}
with the boundary condition $U(z,T)=f(z)$. Next, let us change variables
from $(z,t)$ to $(y,t)$, where
\begin{equation}
 y\equiv \exp\left(\int_t^T r_s ~ds\right)z~.
\end{equation}
Let $U(z,t)=Y(y,t)$. Then we have:
\begin{eqnarray}
 &&\partial_t U=\partial_t Y+\partial_y Y\partial_t y=
 \partial_t Y-r_t\exp\left(\int_t^T r_s ~ds\right)\partial_y Y~,\\
 &&\partial_z U=\exp\left(\int_t^T r_s ~ds\right)\partial_y Y~,\\
 &&\partial_z^2 U=\exp\left(2\int_t^T r_s ~ds\right)\partial_y^2 Y~.
\end{eqnarray}
The PDE for $Y(y,t)$ is given by (this is the {\em diffusion equation}):
\begin{equation}
 {1\over 2}D(y,t)~
 \partial^2_y Y(y,t)+\partial_t Y(y,t)=0
\end{equation}
with the boundary condition $U(y,T)=f(y)$. Here
\begin{equation}
 D(y,t)\equiv y^2 \sigma^2_t=y^2\sigma^2(z,t)=y^2\sigma^2
 \left(\exp\left(-\int_t^T r_s ~ds\right)y,t\right)~.
\end{equation}
The solution to the PDE for $Y(y,t)$ is given by:
\begin{equation}
 Y(y,t)=\int_{-\infty}^\infty dy^\prime ~K(y^\prime,y;T-t)~f(y^\prime)~.
\end{equation}
The {\em heat kernel} $K(x^\prime,x;\tau)$ is the solution to the equation
\begin{equation}
 {1\over 2} D(x,T-\tau)\partial_x^2 K(x^\prime,x;\tau)=\partial_\tau
 K(x^\prime,x;\tau)
\end{equation}
with the boundary condition $K(x^\prime,x;0)=\delta(x^\prime-x)$.
In terms of this heat kernel
we can write the pricing function $V(z,t)$ as follows:
\begin{equation}
 V(z,t)=\exp\left(-\int_t^T r_s ~ds\right)\int_{-\infty}^\infty dy^\prime~
 K\left(y^\prime,\exp\left(\int_t^T r_s ~ds\right)z;T-t\right)~f(y^\prime)~.
\end{equation}
We can subsequently use $V(z,t)$ to compute the processes $\phi_t$ and
$\psi_t$, and hedge the derivative $f(S_T)$.

\section{European Options: Call, Put and Binary}

{}A call option is a right (but not obligation) to buy a stock at the maturity
time $T$ for the strike price $k$ agreed on at time $t=0$. So the claim for the
call option is given by:
\begin{equation}
 f^c(S_T,k)=(S_T-k)^+~.
\end{equation}
The price of the
call option is given by:
\begin{equation}
 V^c_t(k)=B_t\langle B_T^{-1}f^c(S_T,k)\rangle_{{\bf Q},{\cal F}_t}=
 B_t\langle B_T^{-1}(S_T-k)^+\rangle_{{\bf Q},{\cal F}_t}~.
\end{equation}
Here $(x)^+=x$ if $x>0$, and $(x)^+=0$ if $x\leq 0$.

{}A put option is a right (but not obligation) to sell a stock at the maturity
time $T$ for the strike price $k$ agreed on at time $t=0$. So the claim for the
put option is given by:
\begin{equation}
 f^p(S_T,k)=(k-S_T)^+~.
\end{equation}
The price of the
put option is given by:
\begin{equation}
 V^p_t(k)=B_t\langle B_T^{-1}f^p(S_T,k)\rangle_{{\bf Q},{\cal F}_t}=
 B_t\langle B_T^{-1}(k-S_T)^+\rangle_{{\bf Q},{\cal F}_t}~.
\end{equation}
Note that $f^c(S_T,k)-f^p(S_T,k)=S_T-k$. Consequently, we have
\begin{eqnarray}
 V^c_t(k)-V^p_t(k)=&&B_t\langle B_T^{-1}(S_T-k)\rangle_{{\bf Q},{\cal F}_t}=
 \nonumber\\
 &&S_t-B_t B_T^{-1} k=V^f_t(k)~,
\end{eqnarray}
where $V^f_t(k)$ is the price of the forward with a strike price $k$. This
result is called the {\em put-call parity}.

{}A binary (digital) option is a derivative which pays \$1 at the maturity time
$T$ if a stock grows over the strike price $k$
agreed on at time $t=0$. So the claim for the
binary option is given by:
\begin{equation}
 f^b(S_T,k)=\theta(S_T-k)~,
\end{equation}
where $\theta(x)$ is the Heavyside step function.
The price of the
binary option is given by:
\begin{equation}
 V^b_t(k)=B_t\langle B_T^{-1}f^b(S_T,k)\rangle_{{\bf Q},{\cal F}_t}=
 B_t\langle B_T^{-1}\theta(S_T-k)\rangle_{{\bf Q},{\cal F}_t}~.
\end{equation}
Note that
\begin{equation}
 {d\over dx}~(x)^+=\theta(x)~.
\end{equation}
This implies that
\begin{equation}
 V^b_t(k)=-{\partial\over\partial k}~V^c_t(k)=B_t^{-1}B_T-
 {\partial\over\partial k}~V^p_t(k)~.
\end{equation}
Thus, we can determine the price of the binary option from the spectrum
of the prices of the call (put) options.

\section{The Black-Scholes Model}

{}The Black-Scholes model is given by:
\begin{eqnarray}
 &&B_t=\exp(rt)~,\\
 &&S_t=S_0\exp(\sigma W_t+\mu t)~,
\end{eqnarray}
where $r,\sigma,\mu$ are constant.

{}The first step is to define the discounted stock process:
\begin{equation}
 Z_t\equiv B_t^{-1} S_t=S_0\exp\left(\sigma W_t+[\mu-r] t\right)~.
\end{equation}
The SDE for $Z_t$ is given by:
\begin{equation}
 dZ_t=Z_t\left[\sigma dW_t+\left(\mu-r+{1\over 2}\sigma^2 \right)dt\right]~.
\end{equation}
We can make $Z_t$ into a martingale via the following change of variable:
\begin{equation}
 {\widetilde W}_t=W_t+\gamma t=W_t+{1\over \sigma}\left(\mu-r+{1\over 2}\sigma^2 \right)t~.
\end{equation}
Note that ${\widetilde W}_t$ is a ${\bf Q}$-Brownian motion, where the measure
${\bf Q}$ is related to the original measure ${\bf P}$ via the Radon-Nikodym
process
\begin{equation}
 \zeta_t=\exp\left(-\gamma W_t-{1\over 2}\gamma^2 t\right)~.
\end{equation}
Also, note that $dZ_t=\sigma Z_t d{\widetilde W}_t$.

{}The next step is to take a claim $X=X_T$, and construct the process
\begin{equation}
 E_t=\langle B_T^{-1} X\rangle_{{\bf Q},{\cal F}_t}=\exp(-rT)
 \langle X\rangle_{{\bf Q},{\cal F}_t}~.
\end{equation}
This process is a ${\bf Q}$-martingale. We can therefore define
\begin{equation}
 \phi_t\equiv {dE_t\over dZ_t}~,
\end{equation}
which is a previsible process.

{}Finally, the self financing portfolio $(\phi_t,\psi_t)$ consists of
holding $\phi_t$ units of stock and $\psi_t$ units of the cash bond at time
$t$, where
\begin{equation}
 \psi_t=E_t-\phi_t Z_t~.
\end{equation}
The price of this portfolio is given by
\begin{equation}
 V_t=\phi_t S_t+\psi_t B_t=B_t E_t~.
\end{equation}
Thus, the price of the claim $X$ at time $t$ is:
\begin{equation}
 V_t=\exp(-r[T-t])\langle X\rangle_{{\bf Q},{\cal F}_t}~.
\end{equation}
We can use this formula to value various derivatives in the Black-Scholes
model.

\subsection{Call Option}

{}For the call option we have $X=f^c(S_T,k)=(S_T-k)^+$. To compute the pricing
function
\begin{equation}
 V^c(z,t,k)=\exp(-r[T-t])\langle f^c(S_T,k)\rangle_{{\bf Q},~S_t=z}~,
\end{equation}
let us rewrite $S_T$ as follows:
\begin{equation}
 S_T=S_0\exp\left(\sigma{\widetilde W}_T+\left[r-{1\over 2}\sigma^2\right]T
 \right)=S_t\exp\left(\sigma x+\left[r-{1\over 2}\sigma^2\right][T-t]
 \right)~,
\end{equation}
where $x$ stands for the values of the process $W_T-W_t$, which is a normal
$N(0,T-t)$, and is independent of ${\cal F}_t$.
Then we have:
\begin{eqnarray}
 &&V^c(z,t,k)=\nonumber\\
 &&e^{-r[T-t]}\int_{-\infty}^\infty {dx\over\sqrt{2\pi(T-t)}}~
 \exp\left(-{x^2\over 2(T-t)}\right) \times\nonumber\\
 &&\,\,\,\,\,\,\,\times\left(z\exp\left(\sigma x+\left[r-{1\over 2}\sigma^2\right][T-t]
 \right)-k\right)^+=\nonumber\\
 &&\int_{x^*}^\infty {dx\over\sqrt{2\pi(T-t)}}~
 \exp\left(-{x^2\over 2(T-t)}\right) \times\nonumber\\
 &&\,\,\,\,\,\,\,\times\left(z\exp\left(\sigma x-{1\over 2}\sigma^2[T-t]
 \right)-ke^{-r[T-t]}\right)~,
\end{eqnarray}
where
\begin{equation}
 x_*={1\over \sigma}
 \left[\ln\left({k\over z}\right)-
 \left[r-{1\over 2}\sigma^2\right][T-t]\right]~.
\end{equation}
We have:
\begin{eqnarray}
 V^c(z,t,k)=&&z \int_{x^*-\sigma (T-t)}^\infty {dx\over\sqrt{2\pi(T-t)}}~
 \exp\left(-{x^2\over 2(T-t)}\right)-\nonumber\\
 &&k e^{-r[T-t]}\int_{x^*}^\infty {dx\over\sqrt{2\pi(T-t)}}~
 \exp\left(-{x^2\over 2(T-t)}\right)~.
\end{eqnarray}
Let
\begin{equation}
 \Phi(y)\equiv \int_{-\infty}^y {dy^\prime\over\sqrt{2\pi}}~
 \exp\left(-{(y^\prime)^2 \over 2}\right)~.
\end{equation}
Then
\begin{eqnarray}
 &&V^c(z,t,k)=z\Phi\left({{\ln\left({z\over k}\right)+
 \left[r+{1\over 2}\sigma^2\right][T-t]}\over \sigma\sqrt{T-t}}\right)-\nonumber\\
 &&\,\,\,\,\,\,\,k e^{-r[T-t]}\Phi\left({{\ln\left({z\over k}\right)+
 \left[r-{1\over 2}\sigma^2\right][T-t]}\over \sigma\sqrt{T-t}}\right)~.
\end{eqnarray}
This is the Black-Scholes formula for pricing a European call option.

\subsection{Put Option}

{}We can use the put-call parity to price a put option with a strike $k$:
\begin{equation}
 V^p(z,t,k)=V^c(z,t,k)-z+k e^{-r[T-t]}~.
\end{equation}
Let us introduce the function
\begin{equation}
 {\widetilde \Phi}(y)\equiv 1-\Phi(y)=\int_y^\infty
 {dy^\prime\over\sqrt{2\pi}}~
 \exp\left(-{(y^\prime)^2 \over 2}\right)~.
\end{equation}
Then we have:
\begin{eqnarray}
 &&V^p(z,t,k)=k e^{-r[T-t]}{\widetilde \Phi}\left({{\ln\left({z\over k}\right)+
 \left[r-{1\over 2}\sigma^2\right][T-t]}\over \sigma\sqrt{T-t}}\right)-\nonumber\\
 &&\,\,\,\,\,\,\,z{\widetilde \Phi}\left({{\ln\left({z\over k}\right)+
 \left[r+{1\over 2}\sigma^2\right][T-t]}\over \sigma\sqrt{T-t}}\right)~.
\end{eqnarray}
This is the Black-Scholes formula for pricing a European put option.

\subsection{Binary Option}

{}We can price a binary option with a strike $k$ either directly or
using the relation between the binary and call (put) prices. The result is
\begin{equation}
 V^b(z,t,k)=e^{-r[T-t]}\Phi\left({{\ln\left({z\over k}\right)+
 \left[r-{1\over 2}\sigma^2\right][T-t]}\over \sigma\sqrt{T-t}}\right)~.
\end{equation}
This is the Black-Scholes formula for pricing a European binary option.

\section{Hedging in the Black-Scholes Model}

{}In this section we discuss explicit hedges for European options in the
Black-Scholes model. Since these options are of the form $X=f(S_T)$, we can use
the pricing function $V(z,t)$ to compute $\phi_t$ and $\psi_t$.
Thus, we have
\begin{eqnarray}
 &&V_t=V(S_t,t)~,\\
 &&\phi_t=\partial_z V(S_t,t)~,\\
 &&\psi_t=B_t^{-1}\left[V_t-\phi_t S_t\right]=\exp(-rt)\left[V(S_t,t)-
 S_t\partial_z V(S_t,t)\right]~.
\end{eqnarray}
These formulas are all we need to hedge a European option in the Black-Scholes
model.

\subsection{Call Option}

{}For the call option we have:
\begin{eqnarray}
 &&V^c(z,t,k)=z\Phi\left({{\ln\left({z\over k}\right)+
 \left[r+{1\over 2}\sigma^2\right][T-t]}\over \sigma\sqrt{T-t}}\right)-\nonumber\\
 &&\,\,\,\,\,\,\,k e^{-r[T-t]}\Phi\left({{\ln\left({z\over k}\right)+
 \left[r-{1\over 2}\sigma^2\right][T-t]}\over \sigma\sqrt{T-t}}\right)~.
\end{eqnarray}
This gives
\begin{eqnarray}
 &&\phi_t=\Phi\left({{\ln\left({S_t\over k}\right)+
 \left[r+{1\over 2}\sigma^2\right][T-t]}\over \sigma\sqrt{T-t}}\right)~,\\
 &&\psi_t=-ke^{-rT}\Phi\left({{\ln\left({S_t\over k}\right)+
 \left[r-{1\over 2}\sigma^2\right][T-t]}\over \sigma\sqrt{T-t}}\right)~.
\end{eqnarray}
Note that the cash bond is always in the borrowing (albeit $B_t\psi_t$ is
bounded by the exercise price $k$). Also, note that at $t=T$ we have
\begin{eqnarray}
 &&\phi_T=\theta(S_T-k)~,\\
 &&\psi_T=-ke^{-rT}\theta(S_T-k)~.
\end{eqnarray}
So if $S_T>k$, we have one unit of stock (which is
worth $S_T$), and we are short $k\exp(-rT)$ units of the cash bond
(which is worth $-k$). We deliver the stock to the call option holder, receive
$k$ dollars for the transaction, and break even. On the other hand, if $S_T<k$,
we are holding no stock or cash bond, neither do we have any obligations, so we
also break even (unless the option holder decides to exercise the call option
and buy the stock for $k$ dollars, in which case we have a surplus of $k-S_T$
dollars at time $T$ -- this is because the option holder did not exercise the
option optimally). Finally, if $S_T=k$, some care is needed as the
step-function is discontinuous. We will address this point in detail when
we discuss the hedge for the binary option.

\subsection{Put Option}

{}For the put option we have:
\begin{eqnarray}
 &&V^p(z,t,k)=k e^{-r[T-t]}{\widetilde \Phi}\left({{\ln\left({z\over k}\right)+
 \left[r-{1\over 2}\sigma^2\right][T-t]}\over \sigma\sqrt{T-t}}\right)-\nonumber\\
 &&\,\,\,\,\,\,\,z{\widetilde \Phi}\left({{\ln\left({z\over k}\right)+
 \left[r+{1\over 2}\sigma^2\right][T-t]}\over \sigma\sqrt{T-t}}\right)~.
\end{eqnarray}
This gives
\begin{eqnarray}
 &&\phi_t=-{\widetilde \Phi}\left({{\ln\left({S_t\over k}\right)+
 \left[r+{1\over 2}\sigma^2\right][T-t]}\over \sigma\sqrt{T-t}}\right)~,\\
 &&\psi_t=ke^{-rT}{\widetilde \Phi}\left({{\ln\left({S_t\over k}\right)+
 \left[r-{1\over 2}\sigma^2\right][T-t]}\over \sigma\sqrt{T-t}}\right)~.
\end{eqnarray}
Note that the stock holding is always short.
Also, note that at $t=T$ we have
\begin{eqnarray}
 &&\phi_T=\theta(S_T-k)-1~,\\
 &&\psi_T=ke^{-rT}\left[1-\theta(S_T-k)\right]~.
\end{eqnarray}
So if $S_T<k$, we are short one unit of stock
(which is worth $-S_T$), and we are holding $k\exp(-rT)$ units of the cash bond
(which is worth $k$). If the put option holder decides to exercise the option
and sell us one unit of stock, we receive that one unit of stock, pay the
option holder $k$ dollars, and break even. Similarly, we break even if
$S_T > k$ (provided that the option holder exercises the option optimally,
or else we end up with a surplus). Once again, for $S_T=k$ some additional
care is needed -- see below.

\subsection{Binary Option}

{}For the binary option we have:
\begin{equation}
 V^b(z,t,k)=e^{-r[T-t]}\Phi\left({{\ln\left({z\over k}\right)+
 \left[r-{1\over 2}\sigma^2\right][T-t]}\over \sigma\sqrt{T-t}}\right)~.
\end{equation}
This gives
\begin{eqnarray}
 &&\phi_t= \left[\sqrt{2\pi(T-t)}\sigma S_t\exp(r[T-t])\right]^{-1}\times\nonumber\\
 &&\,\,\,\,\,\,\,\times\exp\left(-{{\left[\ln\left({S_t\over k}\right)+
 \left[r-{1\over 2}\sigma^2\right](T-t)\right]^2}
 \over 2\sigma^2(T-t)}\right)~,\\
 &&\psi_t=e^{-rT}\left[\Phi\left({{\ln\left({S_t\over k}\right)+
 \left[r-{1\over 2}\sigma^2\right][T-t]}\over \sigma\sqrt{T-t}}\right)-\right.
 \nonumber\\
 &&\,\,\,\,\,\,\,\left.\left[\sqrt{2\pi(T-t)}\sigma\right]^{-1}
 \exp\left(-{{\left[\ln\left({S_t\over k}\right)+
 \left[r-{1\over 2}\sigma^2\right](T-t)\right]^2}
 \over 2\sigma^2(T-t)}\right)
 \right]~.
\end{eqnarray}
This hedge has an interesting behavior as $t\rightarrow T$. Suppose $S_T-k\not
=0$. Then we have
\begin{eqnarray}
 &&\phi_T=0~,\\
 &&\psi_T=e^{-rT}\theta(S_T-k)~.
\end{eqnarray}
Thus, we are holding no stock. If $S_T>k$, then we are holding $\exp(-rT)$
units of the cash bond (which is worth \$1), and we break even if the option
holder decides to exercise the option. If
$S_T<k$, we are holding no cash bond either, but we have no obligation in this
case, so we also break even.

{}Suppose, however, $S_T=k$. Then some care is needed. Recall that
\begin{equation}
 S_t=S_0\exp\left(\sigma {\widetilde W}_t+
 \left[r-{1\over 2}\sigma^2\right]t\right)~.
\end{equation}
So in this case
\begin{equation}
 k=S_0\exp\left(\sigma {\widetilde W}_T+
 \left[r-{1\over 2}\sigma^2\right]T\right)~,
\end{equation}
and
\begin{equation}
 \ln\left({S_t\over k}\right)+
 \left[r-{1\over 2}\sigma^2\right](T-t)=\sigma[{\widetilde W}_t-
 {\widetilde W}_T]~.
\end{equation}
Let $t=T-\delta t$. Then ${\widetilde W}_T-{\widetilde W}_t$
is itself a Brownian motion with
variance $\delta t$. For small $\delta t$ we have
\begin{equation}
 {\widetilde W}_T-{\widetilde W}_t=\epsilon_t\sqrt{\delta t}~,
\end{equation}
and
\begin{equation}
 \ln\left({S_t\over k}\right)+
 \left[r-{1\over 2}\sigma^2\right](T-t)=-\sigma\epsilon_t\sqrt{\delta t}~,
\end{equation}
where $\epsilon_t=\pm 1$. This implies that, as $t\rightarrow T$, we have
\begin{eqnarray}
 &&\Phi\left({{\ln\left({S_t\over k}\right)+
 \left[r-{1\over 2}\sigma^2\right][T-t]}\over \sigma\sqrt{T-t}}\right)
 \rightarrow \Phi(-\epsilon_t)~,\\
 &&\left[\sqrt{2\pi(T-t)}\sigma\right]^{-1}
 \exp\left(-{{\left[\ln\left({S_t\over k}\right)+
 \left[r-{1\over 2}\sigma^2\right](T-t)\right]^2}
 \over 2\sigma^2(T-t)}\right)
 \rightarrow \nonumber\\
 &&\,\,\,\,\,\,\,\rightarrow{\exp\left(-{1\over 2}\right)\over\sqrt{2\pi(T-t)}}~.
\end{eqnarray}
This implies that the value of $\theta(S_T-k)$ in the hedges for
the call and put options is either $\Phi(+1)$ or $\Phi(-1)$ for $S_T=k$, that
is, it is random. This, however, does not pose a problem as this value is
{\em previsible}. In the case of the binary option, however, for $S_T=k$ to
hedge we would need to borrow more and more cash bond and buy more and more
stock as $t\rightarrow T$. This is, however, an idealized model, and in
practice, where we do have transaction costs, this singular behavior is
smoothed out -- without going into details, let us simply observe that, for one thing, buying more and more stock becomes prohibitive in the presence of transactions costs.

\section{Price, Time and Volatility Dependence}

{}In this section we discuss how various option prices depend on the strike
price $k$, maturity time $T$ and volatility $\sigma$. Let $F=S_0\exp(rT)$ be
the forward price at $t=0$.

\subsection{Call Option}

{}The price of the call option is given by:
\begin{equation}
 V^c=e^{-rT}\left[F\Phi\left({\ln\left({F\over k}\right)\over
 \sigma\sqrt{T}} + {1\over 2}\sigma\sqrt{T}\right)-
 k\Phi\left({\ln\left({F\over k}\right)\over
 \sigma\sqrt{T}} -{1\over 2}\sigma\sqrt{T}\right)\right]~.
\end{equation}
Suppose $\ln(F/k)<0$, and
$|\ln(F/k)|\gg\sigma\sqrt{T}$. Then the option is out of the money and
unlikely to recover by the maturity time $T$. In this case $V^c$ is small.
On the other hand, if $\ln(F/k)\gg\sigma\sqrt{T}$, then the option loses most
of its optionality, and essentially becomes a forward struck at price $k$
for time $T$, whose value is $S_0-k\exp(-rT)$.

{}The maturity $T$ dependence goes as follows. For small $T$ the chances of
anything substantial happening get smaller, and the option value gets closer
and closer to the claim value taken at the current price: $(S_0-k)^+$.
On the other hand, as $T$ grows the option price also grows. The reason why is
that at time $T$ we must deliver one unit of stock if the option is in the
money, and the uncertainty in $S_T$ grows with $T$. In fact,
for large $T$ the option price approaches $S_0$, and the corresponding hedge
involves buying one unit of stock at time $t=0$ -- indeed, this is the
only way to guarantee that we will be able to deliver the stock at time $T$
for large $T$, even if the stock price becomes very large, which is not
unlikely as $T$ is large (as we get closer to the maturity time $T$, however,
our hedge is previsibly dictated by the stock movements).
It is important to note that this
is true even if the interest rate is vanishing. The reason why is that
the call option issuer has an obligation to deliver a volatile instrument
(that is, a stock) if the option is in the money at time $T$.

{}All else being equal, the option is worth more the more volatile the
stock is. If $\sigma$ is very small, the option resembles a riskless bond, and
is worth $\left(S_0-ke^{-rT}\right)^+$, which is the value of the corresponding
forward if the option is in the money, and zero otherwise. If $\sigma$ is very
large, then the option is worth $S_0$.

{}It is instructive to study the volatility dependence when
$\sigma\sqrt{T}\ll 1$ (this is relevant in the case of bonds with volatile
interest rates). It is clear that the value of the option is almost
independent of $\sigma$ if $|\ln(F/k)|\gg \sigma\sqrt{T}$, that is, if the
strike price is too different from the forward price. On the other hand,
suppose $|\ln(F/k)|
{\ \lower-1.2pt\vbox{\hbox{\rlap{$<$}\lower5pt\vbox{\hbox{$\sim$}}}}\ }
\sigma\sqrt{T}$. Let $\kappa\equiv \ln(F/k)/\sigma\sqrt{T}$. Note that
$|\kappa|
{\ \lower-1.2pt\vbox{\hbox{\rlap{$<$}\lower5pt\vbox{\hbox{$\sim$}}}}\ } 1$.
We have:
\begin{eqnarray}
 V^c\approx &&e^{-rT}\left[(F-k)\Phi(\kappa) +(F+k)
 \Phi^\prime(\kappa){\sigma\sqrt{T}\over 2}\right]\approx\nonumber\\
 &&k e^{-rT}\left[\kappa \Phi(\kappa)+\Phi^\prime(\kappa)\right]
 \sigma\sqrt{T}~.
\end{eqnarray}
In particular, for $\kappa=0$, that is, when the strike price $k$ is exactly
equal the forward price $F$, we have
\begin{equation}
 V^c\approx{ke^{-rT}\over\sqrt{2\pi}}~\sigma\sqrt{T}~.
\end{equation}
Note that the value of the option grows linearly with $\sigma$.

\subsection{Put Option}

{}The price of the put option is given by:
\begin{equation}
 V^p=e^{-rT}\left[
 k{\widetilde \Phi}\left({\ln\left({F\over k}\right)\over
 \sigma\sqrt{T}} -{1\over 2}\sigma\sqrt{T}\right)
 -F{\widetilde \Phi}\left({\ln\left({F\over k}\right)\over
 \sigma\sqrt{T}} + {1\over 2}\sigma\sqrt{T}\right)\right]~.
\end{equation}
Suppose $\ln(F/k)\gg\sigma\sqrt{T}$. Then the option is out of the money and
unlikely to recover by the maturity time $T$. In this case $V^p$ is small.
On the other hand, if $\ln(F/k)<0$ and $|\ln(F/k)|\gg\sigma\sqrt{T}$,
then the option loses most
of its optionality, and essentially is equivalent to a {\em short} holding
of a forward struck at price $k$
for time $T$. The value of this holding is $k\exp(-rT)-S_0$. Note that these
facts can also be deduced from the put-call parity.

{}The maturity $T$ dependence goes as follows. For small $T$ the chances of
anything substantial happening get smaller, and the option value gets closer
and closer to the claim value taken at the current price: $(k-S_0)^+$.
The large $T$ behavior is obscured in the case of a non-zero interest rate.
Indeed, the cost now of price $k$ for large $T$ goes to zero if $r>0$, so that
the price of the option goes to zero at large $T$ in this case. Let us,
therefore, consider the $r=0$ case. Then the put option price grows with
$T$ just as in the case of the call option (in fact, the put-call parity
tells us that $V^p=V^c-S_0+k$). In fact,
in the large $T$ limit $V^p$ approaches $k$. This is because
in the case of the
put option the option issuer must guarantee $k$ dollars at the maturity
even if the stock (which we receive at time $T$ if the option is in the
money) goes very low, which is not unlikely as $T$ is large. In fact, the
corresponding hedge consists of buying $k$ units of the cash bond at $t=0$
(the hedge is previsibly determined as we get closer to time $T$) -- indeed,
this is the only way we can guarantee that we will be able to make a payment
of $k$ dollars at time $T$ even if by then the stock price goes down to zero.
Thus, the important point here is that the option issuer is receiving a
volatile instrument (that is, a stock) if the option is in the money at time
$T$.

{}All else being equal, the option is worth more the more volatile the
stock is. If $\sigma$ is very small, the option resembles a riskless bond, and
is worth $\left(ke^{-rT}-S_0\right)^+$, which is the value of a {\em short}
holding of the corresponding
forward if the option is in the money, and zero otherwise. If $\sigma$ is very
large, then the option is worth $ke^{-rT}$.

{}Let us study the volatility dependence when
$\sigma\sqrt{T}\ll 1$. Thus, we have:
\begin{eqnarray}
 V^p\approx &&e^{-rT}\left[(k-F)
 {\widetilde \Phi}(\kappa) -(k+F)
 {\widetilde \Phi}^\prime (\kappa)
 {\sigma\sqrt{T}\over 2}\right]\approx\nonumber\\
 &&-ke^{-rT}\left[\kappa{\widetilde \Phi}(\kappa)+{\widetilde \Phi}^\prime
 (\kappa)\right]\sigma\sqrt{T}~.
\end{eqnarray}
In particular, for $\kappa=0$ we have
\begin{equation}
 V^p\approx {ke^{-rT}\over\sqrt{2\pi}}~\sigma\sqrt{T}~.
\end{equation}
As in the call option case, the value of the put option grows linearly
with $\sigma$.

\subsection{Binary Option}

{}The price of the binary option is given by:
\begin{equation}
 V^b=e^{-rT} \Phi\left({\ln\left({F\over k}\right)\over
 \sigma\sqrt{T}} -{1\over 2}\sigma\sqrt{T}\right)~.
\end{equation}
Suppose $\ln(F/k)<0$ and $|\ln(F/k)|\gg\sigma\sqrt{T}$.
Then the option is out of the money and
unlikely to recover by the maturity time $T$. In this case $V^b$ is small.
On the other hand, if $\ln(F/k)\gg\sigma\sqrt{T}$,
then the option loses most of its optionality, and essentially becomes a
riskless zero-coupon bond with face value \$1. The $t=0$ value of this
bond is $\exp(-rT)$.

{}The maturity $T$ dependence goes as follows.  For small $T$ the chances of
anything substantial happening get smaller, and the option value gets closer
and closer to the claim value taken at the current price: $\theta(S_0-k)$.
Once again, the large $T$ behavior is obscured in the case of a non-zero
interest rate. Let us, therefore, consider the $r=0$ case. Then
as $T$ grows the option price gets smaller if the
option is in the money at $t=0$ as the chances that the option
ends up out of the money grow with $T$. On the other hand,
if the option is out of the
money at $t=0$, the option price at first grows with $T$
as the chances that it ends up in the money grow with $T$. Eventually, however,
that is, as $T$ gets larger and larger, the price goes back
to zero. The reason for this is that the option issuer in this case
only has an obligation to deliver \$1 at time $T$ if the option is in the
money, that is, at the maturity time the transaction only involves a
non-volatile instrument -- a cash bond of a fixed amount. Then if at $t=0$
the option is out of the money, even if at some later time $t_*$ it ends up
in the money, as more and more time lapses, the chances that it ends up
out of the money grow, so the option price gets smaller. So for large $T$
the binary option price goes to zero regardless of the starting point.

{}The volatility dependence is the same as the $T$ dependence in the case of
$r=0$ -- indeed, in this case $\sigma$ and $T$ appear in the combination
$\sigma\sqrt{T}$. It is then clear that, all else being equal, this volatility
dependence remains the same even if $r>0$. Thus, if $\sigma$ is very small,
then the option resembles a riskless bond, and is worth
$e^{-rT}\theta\left(S_0-ke^{-rT}\right)$, which is $e^{-rT}$ if the option
is in the money, and zero otherwise. If $\sigma$ is very large, then the option
price goes to zero.

{}Let us study the volatility dependence when $\sigma\sqrt{T}\ll 1$. We have:
\begin{equation}
 V^b\approx e^{-rT}\left[\Phi(\kappa)-\Phi^\prime(\kappa)
 {\sigma\sqrt{T}\over 2}\right]~.
\end{equation}
In particular, for $\kappa=0$ we have
\begin{equation}
 V^b\approx {e^{-rT}\over 2}\left[1-{\sigma\sqrt{T}\over\sqrt{2\pi}}
 \right]~.
\end{equation}
Note that the option value decreases linearly with $\sigma$ in this case,
which is consistent with our discussion above.

\subsection{American Options}

{}An example of an American option is a call option which allows the option
holder to purchase the stock for the strike price $k$ at any time $\tau$,
which is called the {\em stopping time},
between $t=0$ and the expiration time $T$. The option issuer does not know in
advance what $\tau$ the investor will use, so the price of this option is
given by:
\begin{equation}
 V={\rm max}_{\,\tau}
 \left(\left\langle e^{-r\tau}(S_\tau-k)^+\right\rangle_{\bf Q}
 \right)~.
\end{equation}
That is, the option issuer must charge the value maximized over all possible
stopping strategies.

{}In general, if the option purchaser has a set of options $A$, and receives
a payoff $X_a$ at time $\tau_a\leq T$
after choosing $a\in A$, the option issuer
should charge
\begin{equation}
 V={\rm max}_{\,a\in A}
 \left(\left\langle e^{-r\tau_a} X_a\right\rangle_{\bf Q}\right)
\end{equation}
for the option. If the purchaser does not exercise the option optimally,
then the hedge will produce a surplus (for the issuer)
by the expiration date $T$.

\section{Upper and Lower Bounds on Option Prices}

{}For a European or an American call option the price $V^c$ should not be
greater than that of the stock $S_0$. Suppose $V^c>S_0$
for a European call option with maturity $T$. Then at time $t=0$
we sell the call option, and take a long position in one unit of stock plus
$V^c-S_0$ worth of the cash bond. If at maturity $t=T$ the stock is above
the strike, $S_T>k$, we deliver the stock, receive the payment of $k$ dollars,
plus we have $(V^c-S_0)e^{rT}$ dollars from the cash bond. So we end up making
a profit. On the other hand, if $S_T<k$, then we are left with one unit of
stock, which is worth $S_T$, plus $(V^c-S_0)e^{rT}$ dollars from the cash bond.
Thus, either way we make a profit, hence arbitrage. The above argument also
holds for an American call option where instead of $T$ we use the stopping
time $\tau$.

{}For a European or an American put option the price $V^p$ should not be
more than the strike price $k$. Suppose $V^p>k$. Then at time $t=0$ we sell
the call option and buy $V^p$ dollars worth of the cash bond. At the stopping
time $\tau$ the latter is worth $V^pe^{r\tau}\geq k$, so even if the option
is exercised, we end up making a profit. For a European put option the bound
is actually more severe, in particular, we must have $V^p\leq ke^{-rT}$.

{}For a European call option the price $V^c$ should not be lower than
$S_0-ke^{-rT}$. Suppose $V^c<S_0-ke^{-rT}$. Then we sell one unit of stock,
buy the call option, and hold $S_0-V^c$ worth of the cash bond. The latter
grows to $(S_0-V^c)e^{rT}>k$ at time $T$. If the option is in the money,
we receive one unit of stock for $k$ dollars (so we no longer have a short
position in the stock), so we make a profit. If $S_T<k$, we still make a
profit as our cash bond is worth more than $k$, hence arbitrage.

{}For a European put option the price $V^p$ should not be lower than
$ke^{-rT}-S_0$. Suppose $V^p<ke^{-rT}-S_0$. Then at time $t=0$ we buy one unit
of stock as well as the put option by shorting $V^p+S_0$ worth of the cash
bond. The latter position is worth $-(V^p+S_0)e^{rT}>-k$ at time $T$.
If $S_T<k$, then we exercise the option, and end up receiving a payment of
$k$ dollars, so we make a profit. On the other hand, if $S_T\geq k$, then
we still make a profit by selling the stock, hence arbitrage.

\subsection{Early Exercise}

{}Suppose an American call option is deep in the money at
time $t=0$. Then it is never optimal to exercise the option prior to
the expiration time $T$ if the investor plans to keep the stock
for the rest of the life of the option. The reason is that the later the
option is exercised, the less the worth of the strike price at the initial
time $t=0$. If the investor believes that the stock is overpriced, then
it might be tempting to exercise the option early, and sell the stock.
The profit from this would be $S_\tau-k$. However, even a better profit
is made by selling the option itself. Indeed, the price of the option
$V^c_\tau\geq S_\tau-ke^{-r(T-\tau)}>S_\tau-k$ for $\tau<T$. Alternatively, the
investor can keep the option, short the stock, and take a long position in
$S_\tau$ worth of the cash bond. By time $T$ this portfolio is worth
at least $S_\tau e^{r(T-\tau)}-k$. This is the case if $S_T\geq k$. If
$S_T<k$, then the investor makes a better profit $S_\tau e^{r(T-\tau)}-S_T$.

{}Since an American call option should not be exercised early, it then follows
that it is worth the same as the corresponding European call option.

{}In the case of an American put option the situation is somewhat different.
If it is deep in the money, then it should be exercised early. Thus, suppose
the stock price is almost zero. Then it is better to exercise early as the
stock price cannot go negative, and it is better to receive $k$ dollars now
than later.

{}Since there are circumstance such that an American put option should be
exercised early, it is always worth more than the corresponding European
put option.

\section{Equities and Dividends}

{}An equity is a stock that makes periodic cash payments (that is,
dividend payments) to the stock holder. The simplest model would be an equity
with continuous dividends. Thus, let the stock price $S_t$ and the cash
bond follow the Black-Scholes model. The dividend payment in time $dt$ starting
at time $t$ is $\rho S_t dt$, where $\rho$ is the dividend rate.

{}The stock itself is not tradable in this model as we must also
take into account the dividend payments up to time $t$. We, therefore, need
to find a new process corresponding to a tradable.
Let us consider the following simple portfolio strategy. Let us
instantaneously reinvest
all the dividends by buying more stock. Starting with one unit of
stock at time $t=0$, at time $t$ we would then have $\exp(\rho t)$ units of
stock, which is worth
\begin{equation}
 {\widetilde S}_t=\exp(\rho t)S_t=S_0\exp\left(\sigma W_t+[\mu+\rho]t\right)~.
\end{equation}
This {\em auxiliary} process corresponds to a tradable quantity. We can treat
${\widetilde S}_t$ as an {\em effective} stock process to hedge claims for
$S_t$.

{}Thus, consider a portfolio $(\phi_t,\psi_t)$ consisting of the $\phi_t$ units
of stock $S_t$ and $\psi_t$ units of the cash bond $B_t=\exp(rt)$. This
portfolio is equivalent to the portfolio $({\widetilde \phi}_t,\psi_t)$
consisting of ${\widetilde \phi}_t$ units of the reinvested stock ${\widetilde S}_t$ and
$\psi_t$ units of the cash bond $B_t$, where ${\widetilde \phi}_t=\exp(-\rho t)
\phi_t$. The self-financing equation reads:
\begin{eqnarray}
 dV_t=&&{\widetilde\phi}_t d{\widetilde S}_t+\psi_t dB_t=\nonumber\\
      &&\phi_t dS_t+\psi_t dB_t+\rho\phi_t S_t dt~.
\end{eqnarray}
Note that in terms of the tilded quantities we have a self-financing property
as expected, while in terms of the original quantities we do not as $S_t$ is
not tradable.

{}Now we proceed in the standard way. The discounted effective stock is
${\widetilde Z}_t=B_t^{-1} {\widetilde S}_t$, whose SDE is
\begin{equation}
 d{\widetilde Z}_t={\widetilde Z}_t\left[\sigma dW_t+
 \left(\mu+\rho+{1\over 2}\sigma^2-r\right)dt\right]~.
\end{equation}
We must find a measure ${\bf Q}$
that makes ${\widetilde Z}_t$ into a martingale.
In particular, ${\widetilde W}_t=W_t+\gamma t$ is a ${\bf Q}$-Brownian motion.
The corresponding shift is given by:
\begin{equation}
 \gamma={{\mu+\rho+{1\over 2}\sigma^2-r}\over \sigma}~.
\end{equation}
Thus, we have $d{\widetilde Z}_t=\sigma {\widetilde Z}_t d{\widetilde W}_t$.

{}To construct a hedging strategy, we introduce the process $E_t=\langle
B_T^{-1} X\rangle_{{\bf Q},{\cal F}_t}$. Then the process ${\widetilde
\phi}_t$ is determined from the equation $dE_t={\widetilde \phi}_t d{\widetilde Z}_t$,
while $\psi_t=E_t-{\widetilde \phi}_t{\widetilde Z}_t$. The self-financing
portfolio then consists of holding $\phi_t=\exp(\rho t){\widetilde \phi}_t$
units of stock $S_t$ and $\psi_t$ units of the cash bond $B_t$.

{}What about the derivative price? Note that under the measure ${\bf Q}$ we
have
\begin{equation}
 S_t=S_0\exp\left(\sigma {\widetilde W}_t+\left[r-\rho-
{1\over 2}\sigma^2\right]t\right)~.
\end{equation}
Thus, the effect of the dividends is to replace $r$ with $r-\rho$. This
tells us that all the options in the above equity model can be obtained from
the corresponding options in the vanilla Black-Scholes model via the
substitution $r\rightarrow r-\rho$. In particular, note that the forward
price is now given by $F=S_0\exp([r-\rho]T)$, which is the forward price we should use in the corresponding Black-Scholes formulas for various option prices (including call, put and binary).

\subsection{An Example}

{}Consider the following 5-year contract (so $T=5$).
The UK {\small FTSE} stock index
$S_t$ pays out 90\% of the ratio of the terminal and initial values of
{\small FTSE}, or it pays 130\% if otherwise it would be less, or 180\% if
otherwise it would be more. The data is $\mu=7\%$, $\sigma=15\%$, $\rho=
4\%$, $r=6.5\%$. The {\small FTSE} index is composed of 100 different stocks,
so their separate dividend payments approximate a continuously paying stream.

{}Assuming $S_0=1$, the claim $X$ is
\begin{equation}
 X={\rm min}\left\{{\rm max}\left\{1.3,.9S_T\right\},1.8\right\}~.
\end{equation}
This claim can be rewritten using the identities
\begin{eqnarray}
 &&{\rm max}\{a,b\}=(a-b)^+ + b~,\\
 &&{\rm min}\{a,b\}=a-(a-b)^+~.
\end{eqnarray}
Thus, we have
\begin{eqnarray}
 X=&&{\rm min}\left\{{\rm max}\left\{1.3,.9S_T\right\},1.8\right\}=\nonumber\\
 &&{\rm min}\left\{(.9S_T-1.3)^+ +1.3,1.8\right\}=\nonumber\\
 &&(.9S_T-1.3)^+ +1.3-\left((.9S_T-1.3)^+ +1.3-1.8\right)^+=\nonumber\\
 &&1.3+(.9S_T-1.3)^+ -\left((.9S_T-1.3)^+ -.5\right)^+=\nonumber\\
 &&1.3 +(.9S_T-1.3)^+ -(.9S_T-1.8)^+=\nonumber\\
 &&1.3 +.9\left[(S_T-1.44)^+ -(S_T-2)^+\right]~.
\end{eqnarray}
That is, $X$ is actually the difference of two {\small FTSE} calls plus some
cash. The calls can be evaluated with the Black-Scholes formula, where for the
forward price we use $F=\exp([r-\rho]T)$.

\subsection{Periodic Dividends}

{}Suppose at deterministic times $T_i$ the equity pays a dividend of a fraction
$\rho$ of the stock price which was current just before the dividend was paid.
The stock price process is modeled as
\begin{equation}
 S_t=S_0(1-\rho)^{n[t]} \exp(\sigma W_t+\mu t)~,
\end{equation}
where $n[t]\equiv{\rm max}\{i:T_i\leq t\}$ is the number of dividend payments
made by time $t$. As usual, we also have a cash bond $B_t=\exp(rt)$.
Note that the stock process $S_t$ is discontinuous.

{}As in the case of the continuous dividends, we introduce the auxiliary
process
\begin{equation}
 {\widetilde S}_t=[1-\rho]^{-n[t]} S_t=S_0\exp(\sigma W_t+\mu t)~,
\end{equation}
which would correspond to reinvesting the dividends back into the stock. We can
now hedge as before in terms of ${\widetilde S}_t$ and $B_t$. The corresponding
portfolio is $({\widetilde \phi}_t,\psi_t)$, which corresponds to the
portfolio $(\phi_t,\psi_t)$ of the actual stock $S_t$ and the cash bond $B_t$,
where $\phi_t=(1-\rho)^{-n[t]}{\widetilde \phi}_t$. Under the martingale
measure ${\bf Q}$ we have
\begin{equation}
 S_t=S_0(1-\rho)^{n[t]}\exp\left(\sigma {\widetilde W}_t+\left[r-{1\over 2}
 \sigma^2\right]t\right)~.
\end{equation}
So all the option prices can be computed using the corresponding Black-Scholes
formulas with the forward price given by $F=S_0(1-\rho)^{n[T]}\exp(rT)$.

\section{Multiple Stock Models}

{}In many cases it is important to model movements of multiple securities
which are intertwined in a non-trivial way. Let us consider a model containing
$n$ stocks $S^i_t$ that depend on $n$ independent Brownian motions $dW^i_t$,
$i=1,\dots,n$:
\begin{eqnarray}
 &&dB_t=r_t B_t dt~,\\
 &&dS^i_t=S^i_t dY^i_t~,
\end{eqnarray}
where the stochastic processes $Y^i_t$ have the following SDEs:
\begin{equation}
 dY_t^i=\sum_{j=1}^n \sigma^{ij}_t dW^j_t+\mu^i_t dt~.
\end{equation}
Here $\Sigma_t\equiv(\sigma^{ij}_t)$
is the volatility matrix, and $\mu^i_t$ are the drifts.

{}Note that
\begin{equation}
 \langle dW^i_t dW^j_t\rangle_{\bf P}=\delta^{ij} dt~.
\end{equation}
This implies that
\begin{equation}
 \langle dY^i_t dY^j_t\rangle_{\bf P}=M^{ij} dt~,
\end{equation}
where
\begin{equation}
 M^{ij}_t=\sum_{k=1}^n \sigma^{ik}_t\sigma^{jk}_t~,
\end{equation}
or in the matrix form
\begin{equation}
 M_t=\Sigma_t \Sigma^T_t~,
\end{equation}
where superscript $T$ denotes transposition. Note that $M$ is a symmetric
matrix $M^{ij}_t=M^{ji}_t$ with positive semi-definite determinant:
\begin{equation}
 \det(M_t)={\det}^2(\Sigma_t)\geq 0~.
\end{equation}
$M^{ij}$ is the {\em covariance matrix}.

{}Let us define:
\begin{eqnarray}
 (\sigma^i_t)^2\equiv M^{ii}_t~,\\
 \rho^{ij}_t\equiv {M^{ij}_t\over{\sigma^i_t\sigma^j_t}}~.
\end{eqnarray}
Note that $\rho^{ii}_t\equiv 1$.
Here $\sigma^i_t$ are the volatilities for the processes
$Y^i_t$ (so they are the log-volatilities for the stock processes $S^i_t$),
while $\rho^{ij}_t$ ($i\not=j$) is the correlation between the process $Y^i_t$ and
the process $Y^j_t$. {\em I.e.}, $\rho^{ij}_t$ is the {\em correlation matrix}. Note that if any of
$\sigma^i_t$ are zero, then
$M_t$ (and, therefore, $\Sigma_t$) has vanishing determinant:
\begin{equation}
 \sigma^i_t=\sum_{k=1}^n \left(\sigma^{ik}_t\right)^2~,
\end{equation}
so that if $\sigma^i_t=0$, then $\sigma^{ik}=0$, $k=1,\dots,n$, and
$\det(\Sigma_t)=0$. As we will see in a moment, we will need to assume that
$\det(\Sigma_t)\not=0$. Then it follows that all $\sigma^i_t\not=0$, and
the matrix $\rho^{ij}_t$ is well defined.

{}The solution to the above SDEs is given by:
\begin{eqnarray}
 B_t=&&\exp\left(\int_0^t r_s~ds\right)~,\\
 S^i_t=&&S^i_0\exp\left(\sum_{j=1}^n \int_0^t\sigma^{ij}_s dW^j_s +
 \int_0^t\left[\mu^i_s-{1\over 2}\sum_{j=1}^n\left(\sigma^{ij}_s\right)^2
 \right]ds\right)=\nonumber\\
 &&S^i_0\exp\left(\sum_{j=1}^n \int_0^t\sigma^{ij}_s dW^j_s +
 \int_0^t\left[\mu^i_s-{1\over 2}\left(\sigma^i_s\right)^2
 \right]ds\right)~.
\end{eqnarray}
This shows that $\sigma^i_t$ are indeed log-volatilities of $S^i_t$.

{}Next, we need to find a new measure ${\bf Q}$ under which {\em all} the
discounted stock prices $Z^i_t=B_t^{-1}S^i_t$ become ${\bf Q}$-martingales
simultaneously. Let
\begin{equation}
 {\widetilde W}^i_t\equiv W^i_t+\int_0^t \gamma^i_s~ds~.
\end{equation}
The corresponding Radon-Nikodym process is
\begin{equation}
 \zeta_t=\prod_{i=1}^n \zeta^i_t~,
\end{equation}
where $\zeta^i_t$ are individual Radon-Nikodym processes. The discounted
stock processes have the following SDEs:
\begin{equation}
 dZ^i_t=Z^i_t\left[\sum_{j=1}^n\sigma^{ij}_t d{\widetilde W^j}_t+
 \left(\mu^i_t-r_t-\sum_{j=1}^n\sigma^{ij}_t\gamma^j_t\right)dt\right]~.
\end{equation}
To make the drift terms vanish simultaneously, we must make sure that the
matrix equation
\begin{equation}
 \sum_{j=1}^n\sigma^{ij}_t\gamma^j_t=\mu^i_t-r_t
\end{equation}
has a solution for $\gamma^i_t$. This is guaranteed if the matrix
$\sigma^{ij}_t$ is invertible. Then the matrix $M^{ij}_t$ is also invertible.
Let $M^{-1}_t$
be the inverse of $M_t$ (note that since $M_t$ is a symmetric matrix,
its left and right inverse matrices coincide). Then we have
\begin{equation}
 \gamma^i_t=\sum_{j,k=1}^n \left(M^{-1}_t\right)^{jk}\sigma^{ji}_t
 \left[\mu^k_t-r_t\right]~.
\end{equation}
The shift $\gamma^i_t$ is referred to as the {\em market price of risk} for
the stock $S^i_t$.

{}To construct replicating strategies, we proceed as follows. We introduce a
${\bf Q}$-martingale $E_t\equiv\langle B_T^{-1}
X\rangle_{{\bf Q},{\cal F}_t}$. The
$n$-factor martingale representation theorem then implies that there exist
previsible processes $\phi^i_t$ such that
\begin{equation}
 E_t=E_0+\sum_{i=1}^n \int_0^t \phi^i_t dZ^i_t
\end{equation}
as long as the matrix $\Sigma_t$ is invertible.
Indeed, since $E_t$ is a ${\bf Q}$-martingale, we have
\begin{equation}
 dE_t=\sum_{j=1}^n Z^j_t \lambda^j_t d{\widetilde W}^j_t
\end{equation}
for some previsible processes $\lambda^i_t$.
On the other hand,
\begin{equation}
 \sum_{i=1}^n \phi^i_t dZ^i_t=\sum_{i,j=1}^n Z^i_t \phi^i_t \sigma^{ij}_t
 d{\widetilde W}^j_t~.
\end{equation}
And since $\Sigma_t$ is invertible, the matrix equation
\begin{equation}
 \sum_{i=1}^n Z^i_t \phi^i_t \sigma^{ij}_t=Z^j_t \lambda^j_t
\end{equation}
has a solution for $Z^i_t\phi^i_t$. Since $Z^i_t$ is non-vanishing, then we also
have a solution for $\phi^i_t$, which is previsible as $Z^i_t$ and $\lambda^i_t$ are previsible.

{}The hedging portfolio
then is $(\phi^1_t,\dots,\phi^n_t,\psi_t)$,
where $\psi_t=E_t-\sum_{i=1}^n\phi^i_t
Z^i_t$, so the value of the portfolio is $V_t=B_t E_t$. We then have
\begin{equation}
 dV_t=\sum_{i=1}^n \phi^i_t dS^i_t +\psi_t dB_t~,
\end{equation}
that is, the portfolio is self-financing.

\subsection{The Degenerate Case}

{}Let us consider a situation where we have $N$ stocks $S^I_t$, $I=1,\dots,N$,
but fewer Brownian motions $W^i_t$, $i=1,\dots,n$, $n<N$.
Here we would like to discuss this case in detail.

{}Thus, we have
\begin{equation}
 dS^I_t=S^I_t dY^I_t~,
\end{equation}
where
\begin{equation}
 dY^I_t=\sum_{i=1}^n\sigma^{Ii}_t dW^i_t +\mu^I_t dt~.
\end{equation}
That is,
\begin{equation}
 S^I_t=S^I_0\exp\left(\sum_{i=1}^n \int_0^t\sigma^{Ii}_s
 dW^i_s +
 \int_0^t\left[\mu^I_s-{1\over 2}\sum_{i=1}^n\left(\sigma^{Ii}_s
 \right)^2
 \right]ds\right)~.
\end{equation}
In the following we will assume that the matrix $\Sigma_t\equiv(\sigma^{ij})$
is invertible (see below).

{}The discounted
stock processes have the following SDEs:
\begin{equation}
 dZ^I_t=Z^I_t\left[\sum_{i=1}^n\sigma^{Ii}_t d{\widetilde W^i}_t+
 \left(\mu^I_t-r_t-\sum_{i=1}^n\sigma^{Ii}_t\gamma^i_t\right)dt\right]~,
\end{equation}
where
\begin{equation}
 {\widetilde W}^i_t=W^i_t+\int_0^t\gamma^i_s~ds~.
\end{equation}
To make the drift terms vanish simultaneously, we must make sure that
\begin{equation}
 \sum_{i=1}^n\sigma^{Ii}_t\gamma^i_t=\mu^I_t-r_t~.
\end{equation}
Thus, we have more equations than unknowns. That is, this system is
overconstrained, and this imposes non-trivial conditions on the drifts.
In particular, we have
\begin{eqnarray}
 &&\sum_{j=1}^n\sigma^{ij}_t\gamma^j_t=\mu^i_t-r_t~,\\
 &&\sum_{i=1}^n\sigma^{\alpha i}\gamma^i=\mu^\alpha_t-r_t~,
\end{eqnarray}
where $I=(i,\alpha)$, $\alpha=n+1,\dots,N$. This implies that
\begin{equation}
 \gamma^i_t=\sum_{j,k=1}^n \left(M^{-1}_t\right)^{jk}\sigma^{ji}_t
 \left[\mu^k_t-r_t\right]~,
\end{equation}
and we have the following conditions on the drifts:
\begin{equation}
 \mu^\alpha_t=r_t+\sum_{i,j,k=1}^n\sigma^{\alpha i}
 \left(M^{-1}_t\right)^{jk}\sigma^{ji}_t
 \left[\mu^k_t-r_t\right]~.
\end{equation}
Note that these conditions come from the requirement that there exist a
martingale measure ${\bf Q}$, which is the requirement that there be no
arbitrage.

{}This fact has an important implication. In particular, with the above
restrictions on $\mu^\alpha_t$ only $n$ out of the original $N$ processes
are independent. Thus, note that
\begin{equation}
 dY^i_t=\sum_{j=1}^n\sigma^{ij}_t dW^j_t+\mu^i_t dt~.
\end{equation}
This implies that
\begin{equation}
 dW^i_t=\sum_{j,k=1}^n \left(M^{-1}_t\right)^{jk}\sigma^{ji}_t
 \left[dY^k_t-\mu^k_t dt\right]~.
\end{equation}
On the other hand,
\begin{eqnarray}
 dY^\alpha_t=
 &&\sum_{i=1}^n\sigma^{\alpha i}_t dW^i_t+\mu^\alpha_t dt=\nonumber\\
 &&\sum_{i,j,k=1}^n \sigma^{\alpha i}_t
 \left(M^{-1}_t\right)^{jk}\sigma^{ji}_t
 \left[dY^k_t-\mu^k_t dt\right] +\mu^\alpha_t dt=\nonumber\\
 &&\sum_{i,j,k=1}^n \sigma^{\alpha i}_t
 \left(M^{-1}_t\right)^{jk}\sigma^{ji}_t \left[dY^k_t-r_t dt\right]+
 r_t dt~.
\end{eqnarray}
That is, once we specify the cash bond, the processes $Y^\alpha_t$ are
determined via the processes $Y^i_t$. In particular, we have only $(n+1)$
independent
(including the cash bond) tradables in this market, and
not $(N+1)$ independent
tradables. The implication of the above discussion is that we
can still hedge all the claims in this market using the independent $(n+1)$
tradables.

{}Finally, let us note that if we have fewer stocks than Brownian motions
that they depend on, we will not be able to hedge. Another way of phrasing
this is that in this case the market is not {\em complete}, in particular,
we have more then one martingale measure, so that we do not have unique
prices for claims as the system is underconstrained.

\subsection{Arbitrage-free Complete Models}

{}The above discussion illustrates the general result due to Harrison and
Pliska. Thus, suppose we have a market of securities and a cash bond. Then:\\
$\bullet$ the market is arbitrage free if and only if there is at least
one equivalent martingale measure (EMM) ${\bf Q}$;\\
$\bullet$ if so, the market is complete if and only if there is exactly one
such EMM ${\bf Q}$ and no other.

{}Thus, a market is arbitrage free if there is no {\em guaranteed}
way of making riskless profits. An arbitrage opportunity would be a
(self-financing) trading strategy which starts at zero value and terminates
with a positive value at some definite date $T$.

{}For simplicity let us assume that we have one stock $S_t$ and the cash bond
$B_t$. Thus, suppose there exists a measure ${\bf Q}$ such that it makes the
discounted stock process $Z_t=B_t^{-1} S_t$ into a martingale. Let us consider
a self-financing portfolio $(\phi_t,\psi_t)$, whose value
\begin{equation}
 V_t=\phi_t S_t +\psi_t B_t
\end{equation}
satisfies the self-financing equation
\begin{equation}
 dV_t=\phi_t dS_t +\psi_t dB_t~.
\end{equation}
The discounted value of this portfolio, that is, $E_t\equiv B_t^{-1} V_t$,
then satisfies the following SDE:
\begin{eqnarray}
 dE_t=&&-V_t B_t^{-2} dB_t+B_t^{-1} dV_t=\nonumber\\
 &&-\left[\phi_t S_t+\psi_t B_t\right] B_t^{-2}dB_t+B_t^{-1}\left[
 \phi_t dS_t+\psi_t dB_t\right]=\nonumber\\
 &&\phi_t\left[-B_t^{-2}S_t dB_t+B_t^{-1}dS_t\right]=\nonumber\\
 &&\phi_t dZ_t~.
\end{eqnarray}
And since $Z_t$ is a ${\bf Q}$-martingale, then so is $E_t$.

{}Now, suppose our strategy starts from zero value ($V_0=0$), and finishes
with a non-negative payoff ($V_T\geq 0$). We have
\begin{equation}
 \langle E_T\rangle_{\bf Q}=E_0=B_0^{-1} V_0=0~.
\end{equation}
However, since $V_T\geq 0$, then $E_T\geq 0$ (since $B_T>0$). But the
${\bf Q}$-expectation of $E_T$ is zero, so $E_T=0$, and this implies that
$V_T=0$ as well. That is, a self-financing strategy cannot make something
from nothing if there exists a martingale measure ${\bf Q}$. No free lunch!

{}Next, let us see how completeness, that is, being able to hedge
any possible derivative claim with a self-financing portfolio,
implies uniqueness of the martingale measure.
Thus, suppose that we can hedge any claim, but we have two different
martingale measures ${\bf Q}$ and ${\bf Q}^\prime$. Let $I_A$ be the
{\em indicator function}, which takes value 1 if the event $A$ has happened in
the history ${\cal F}_T$, and zero otherwise. Consider a claim
$X_T=B_T I_A$. This is a valid claim, so we should be able to hedge it
according to our assumption. Our discounted stock process $Z_t$ is both
a ${\bf Q}$- and ${\bf Q}^\prime$-martingale, and, therefore, so is
the discounted value $E_t$ of our self-financing portfolio. This then implies
that (note that $E_T=B_T^{-1}X_T=I_A$)
\begin{eqnarray}
 &&E_0=\langle E_T\rangle_{\bf Q}=\langle I_A \rangle_{\bf Q}={\bf Q}(A)~,\\
 &&E_0=\langle E_T\rangle_{{\bf Q}^\prime}=
 \langle I_A \rangle_{{\bf Q}^\prime}={\bf Q}^\prime(A)~.
\end{eqnarray}
That is, for an arbitrary event $A$ we have ${\bf Q}(A)={\bf Q}^\prime(A)$, so
that the two measures ${\bf Q}$ and ${\bf Q}^\prime$ are actually identical.
So hedging indeed implies a unique EMM.

\section{Numeraires}

{}The {\em numeraire} is usually chosen to be the cash bond, but it can be chosen
to be any tradable instrument available. In particular, the numeraire can have
volatility. Thus, let us consider a market with
$n$ Brownian motions $W^i_t$, $i=1,\dots,n$. Let $S^I_t$, $I=1,\dots,N$
be the stocks, where $N\geq n$, and let $B_t$ be the numeraire. We have
\begin{eqnarray}
 &&dB_t=B_t\left[\sum_{i=1}^n \rho^i_t dW^i_t +r_t dt\right]~,\\
 &&dS^I_t=S^I_t\left[\sum_{i=1}^n\sigma^{Ii}_t dW^i_t +\mu^I_t dt\right]~.
\end{eqnarray}
The discounted stock processes have the following SDEs:
\begin{equation}
 dZ^I_t=Z^I_t\left[\sum_{i=1}^n \left(\sigma^{Ii}_t-\rho^i_t\right)
 dW^i_t+
 \left(\mu^I_t-r_t\right)dt\right]~.
\end{equation}
Let ${\widehat\sigma}^{Ii}_t=\sigma^{Ii}_t-\rho^i_t$, and
${\widehat\mu}^I_t\equiv\mu^I_t-r_t$. Then we have
\begin{equation}
 dZ^I_t=Z^I_t\left[\sum_{i=1}^n {\widehat \sigma}^{Ii}_t
 d{\widetilde W^i}_t+
 \left({\widehat \mu}^I_t-\sum_{i=1}^n {\widehat \sigma}^{Ii}_t
 \gamma^i_t\right) dt\right]~,
\end{equation}
where
\begin{equation}
 {\widetilde W}^i_t=W^i_t+\int_0^t\gamma^i_s~ds~.
\end{equation}
To make the drift terms vanish simultaneously, we must make sure that
\begin{equation}
 \sum_{i=1}^n{\widehat \sigma}^{Ii}_t\gamma^i_t={\widehat \mu}^I_t~.
\end{equation}
That is,
\begin{eqnarray}
 &&\sum_{j=1}^n{\widehat \sigma}^{ij}_t\gamma^j_t={\widehat \mu}^i_t~,\\
 &&\sum_{i=1}^n{\widehat \sigma}^{\alpha i}\gamma^i=
 {\widehat \mu}^\alpha_t~,
\end{eqnarray}
where $I=(i,\alpha)$, $\alpha=n+1,\dots,N$.

{}Assuming that the matrix
${\widehat \Sigma}_t\equiv({\widehat\sigma}^{ij}_t)$ is invertible, we have
(${\widehat M}_t\equiv{\widehat\Sigma}_t{\widehat\Sigma}_t^T$):
\begin{equation}
 \gamma^i_t=\sum_{j,k=1}^n \left({\widehat M}^{-1}_t\right)^{jk}
 {\widehat \sigma}^{ji}_t {\widehat \mu}^k_t~,
\end{equation}
and we have the following conditions on the drifts:
\begin{equation}
 {\widehat \mu}^\alpha=\sum_{i,j,k=1}^n{\widehat \sigma}^{\alpha i}
 \left({\widehat M}^{-1}_t\right)^{jk}{\widehat \sigma}^{ji}_t
 {\widehat \mu}^k_t~.
\end{equation}
Note that these conditions, which are non-trivial if $N>n$,
come from the requirement that there exist a
martingale measure ${\bf Q}$.

{}To construct replicating strategies, we proceed as follows. We introduce a
${\bf Q}$-martingale $E_t\equiv\langle B_T^{-1}
X\rangle_{{\bf Q},{\cal F}_t}$. The
$n$-factor martingale representation theorem then implies that there exist
previsible processes $\phi^i_t$ such that
\begin{equation}
 E_t=E_0+\sum_{i=1}^n \int_0^t \phi^i_t dZ^i_t
\end{equation}
as long as the matrix ${\widehat \Sigma}_t$ is invertible.
Indeed, since $E_t$ is a ${\bf Q}$-martingale, we have
\begin{equation}
 dE_t=\sum_{j=1}^n Z^j_t \lambda^j_t d{\widetilde W}^j_t
\end{equation}
for some previsible processes $\lambda^i_t$.
On the other hand,
\begin{equation}
 \sum_{i=1}^n \phi^i_t dZ^i_t=\sum_{i,j=1}^n Z^i_t \phi^i_t
 {\widehat \sigma}^{ij}_t
 d{\widetilde W}^j_t~.
\end{equation}
And since ${\widehat \Sigma}_t$ is invertible, the matrix equation
\begin{equation}
 \sum_{i=1}^n Z^i_t \phi^i_t {\widehat \sigma}^{ij}_t=Z^j_t \lambda^j_t
\end{equation}
has a solution for $\phi^i_t$.

{}The hedging portfolio
then is $(\phi^1_t,\dots,\phi^n_t,\psi_t)$, which corresponds to holding
$\phi^i_t$ units of the stocks $S^i_t$, $i=1,\dots,n$, and $\psi_t$ units of
the numeraire $B_t$,
where $\psi_t=E_t-\sum_{i=1}^n\phi^i_t
Z^i_t$, so the value of the portfolio is
\begin{equation}
 V_t=B_t E_t=\sum_{i=1}^n\phi^i_t S^i_t +\psi_t B_t~.
\end{equation}
We then have
\begin{eqnarray}
 dV_t=&&d(B_tE_t)=B_t dE_t+E_t dB_t +B_t \sum_{j=1}^n Z^j_t\lambda^j_t
 \rho^j_t dt=\nonumber\\
 &&B_t\sum_{j=1}^n Z^j_t\lambda^j_t d{\widetilde W}^j_t +
 \left(\psi_t +\sum_{i=1}^n\phi^i_t Z^i_t\right)dB_t +
 B_t \sum_{j=1}^n Z^j_t\lambda^j_t
 \rho^j_t dt=\nonumber\\
 &&B_t\sum_{i,j=1}^n Z^i_t\phi^i_t{\widehat \sigma}^{ij}_t
 \left[d{\widetilde W}^j_t
 +\rho^j_t dt\right]+
 \left(\psi_t +\sum_{i=1}^n\phi^i_t Z^i_t\right)dB_t=\nonumber\\
 &&\sum_{i=1}^n\phi^i_t \left[B_t dZ^i_t+B_t \sum_{j=1}^n Z^i_t
 {\widehat\sigma}^{ij}_t\rho^j_t+Z^i_t dB_t\right]+\psi_t dB_t=\nonumber\\
 &&\sum_{i=1}^n \phi^i_t d\left(B_t Z^i_t\right) +\psi_t dB_t=\nonumber\\
 &&\sum_{i=1}^n \phi^i_t dS^i_t +\psi_t dB_t~.
\end{eqnarray}
Here we have taken into account that
\begin{equation}
 dB_t=B_t\left[\sum_{i=1}^n \rho^i_t d{\widetilde W}^i_t+\left(r_t-
 \sum_{i=1}^n \rho^i_t\gamma^i_t\right)dt\right]~.
\end{equation}
Thus, as we see, the portfolio is self-financing even though the numeraire
is volatile.

\subsection{Change of Numeraire}

{}Suppose we have stocks $S^i_t$ plus two other securities $B_t$ and $C_t$
either of which can be a numeraire. If we choose $B_t$ as the numeraire, then
we need to find a measure ${\bf Q}$ such that $B_t^{-1}S^i_t$ and
$B_t^{-1}C_t$ are martingales. On the other hand, if we choose $C_t$ as the
numeraire, then we need to find a measure ${\bf Q}^C$ such that $C_t^{-1}S^i_t$
and $C_t^{-1}B_t$ are martingales.

{}Let $\zeta_t$ be the Radon-Nikodym process
\begin{equation}
 \zeta_t=\left\langle {d{\bf Q}^C\over d{\bf Q}}\right\rangle_{{\bf Q},
 {\cal F}_t}~.
\end{equation}
Then for any process $X_t$ we have:
\begin{equation}
 \zeta_s\langle X_t\rangle_{{\bf Q}^C,{\cal F}_s}=
 \langle\zeta_t X_t\rangle_{{\bf Q},{\cal F}_s}~.
\end{equation}
Thus, if $X_t$ is a ${\bf Q}^C$-martingale, then
\begin{equation}
 \zeta_s X_s=\langle\zeta_t X_t\rangle_{{\bf Q},{\cal F}_s}~.
\end{equation}
That is, $\zeta_t X_t$ is a ${\bf Q}$-martingale. The process $\zeta_t$ that
satisfies this property is given by:
\begin{equation}
 \zeta_t=B_t^{-1}C_t~.
\end{equation}
Indeed, the canonical ${\bf Q}$-martingales are $1,B_t^{-1}C_t,B_t^{-1}S^i_t$,
and the corresponding canonical ${\bf Q}^C$-martingales are $C_t^{-1}B_t,1,
C_t^{-1}S^i_t$.

{}Let us compute the price of a claim $X=X_T$ under the measure ${\bf Q}^C$:
\begin{eqnarray}
 V^C_t=&&C_t\langle C_T^{-1} X\rangle_{{\bf Q}^C,{\cal F}_t}=\nonumber\\
 &&C_t\zeta_t^{-1}\langle \zeta_T C_T^{-1} X\rangle_{{\bf Q},{\cal F}_t}=
 \nonumber\\
 &&B_t\langle B_T^{-1} X\rangle_{{\bf Q},{\cal F}_t}=V_t~,
\end{eqnarray}
where $V_t$ is the price of the claim $X$ under the measure ${\bf Q}$. Thus,
the prices under ${\bf Q}$ and ${\bf Q}^C$ agree, that is, the prices are
independent of the choice of the numeraire.

\section{Foreign Exchange}

{}Consider the Black-Scholes foreign currency model. Let $B_t$ be the dollar cash
bond, $D_t$ be the sterling cash bond, and $C_t$ be the dollar worth of one
pound. Then the model is
\begin{eqnarray}
 &&B_t=\exp(rt)~,\\
 &&D_t=\exp(ut)~,\\
 &&C_t=C_0\exp(\sigma W_t+\mu t)~,
\end{eqnarray}
where the dollar interest rate $r$, the sterling interest rate $u$, as
well as the log-volatility $\sigma$ and the log-drift $\mu$ for the exchange
rate are all constant.

{}Let us consider this model from the viewpoint of the dollar investor.
The dollar cash bond is tradable. Since the sterling cash is not dollar
tradable (this is because there is non-zero sterling interest rate $u$),
$C_t$, which is the dollar worth of one pound, is not tradable either. On the
other hand, the sterling cash bond $D_t$ is not dollar tradable as it is the
price of a tradable instrument (the sterling cash bond), but it is a sterling
price. There is, however, a dollar tradable we can construct. It is given by
\begin{equation}
 S_t=D_t C_t~.
\end{equation}
This is the dollar price of the sterling cash bond, so it is dollar tradable.
Note that this tradable is volatile, and has the same behavior as a US stock.

{}The discounted process is now $Z_t=B_t^{-1}S_t=B_t^{-1}D_t C_t$. That is,
\begin{equation}
 C_t=B_t D_t^{-1} Z_t=\exp([r-u]t) Z_t=\exp({\widehat r}t) Z_t~.
\end{equation}
The quantity ${\widehat r}\equiv r-u$ can be thought of as the effective
dollar interest rate. Then the price $F_t$
of a sterling forward contract (that is,
the price at time $t$ for trading sterling at a future date $T$) is given by
\begin{equation}
 F_t=C_t\exp\left({\widehat r}[T-t]\right)=C_t\exp\left([r-u][T-t]\right)~.
\end{equation}
All option prices are then given by the corresponding Black-Scholes formulas
with the forward price given by $F_t$.

\section{The Interest Rate Market}

{}We can regard a promise of a dollar at the maturity time $T$ as an asset,
which has some worth at time $t$ before $T$. This asset is called a {\em
discount bond}. Let its price at time $0\leq t\leq T$ be $P(t,T)$.
Then $P(T,T)=1$. A discount bond behaves like a stock, but it has this boundary
condition at the maturity.

{}The {\em yield} of a discount bond is given by:
\begin{equation}
 R(t,T)=-{\ln(P(t,T))\over {T-t}}~.
\end{equation}
This has the meaning of an average interest rate over the period of time
$T-t$.

{}The {\em instantaneous rate}, or {\em short rate}, is given by:
\begin{equation}
 r_t=R(t,t)~.
\end{equation}
This is the rate of instantaneous borrowing.

{}The {\em forward rate} is given by:
\begin{equation}
 f(t,T)=-\partial_T\ln(P(t,T))~.
\end{equation}
This has the meaning of the forward rate of instantaneous borrowing at time
$T$.

{}We have the following relations:
\begin{eqnarray}
 &&f(t,T)=R(t,T)+(T-t)\partial_T R(t,T)~,\\
 &&r_t=f(t,t)~,\\
 &&P(t,T)=\exp\left(-\int_t^T f(t,u) du\right)~.
\end{eqnarray}
The latter gives the discount bond price in terms of the forward rate.

\subsection{The Heath-Jarrow-Morton (HJM) Model}

{}In the HJM model the forward rate for each maturity $T$ is a stochastic
process:
\begin{equation}
 f(t,T)=f(0,T)+\int_0^t\sigma(s,T)dW_s+\int_0^t\alpha(s,T)ds~,~~~0\leq t\leq
 T~,
\end{equation}
or in the differential form
\begin{equation}
 d_t f(t,T)=\sigma(t,T)dW_t+\alpha(t,T)dt~,
\end{equation}
where the volatilities $\sigma(t,T)$ and drifts $\alpha(t,T)$ are previsible
processes. The formula
\begin{eqnarray}
 &&P(t,T)=\exp\left(-\int_t^T f(0,u)du-\int_0^t\left(\int_t^T \sigma(s,u)du
 \right)dW_s\right.-\nonumber\\
 &&\,\,\,\,\,\,\,\left.\int_0^t\left(\int_t^T\alpha(s,u)du\right)ds\right)
\end{eqnarray}
then gives the price of the discount bond.

{}To hedge claims, we need a cash product. The simplest cash product is
an account, or a cash bond, formed by starting with \$1 at $t=0$ and
reinvesting continuously at the instantaneous rate $r_t$:
\begin{eqnarray}
 &&dB_t=r_t B_t dt~,\\
 &&B_t=\exp\left(\int_0^t r_s~ds\right)~.
\end{eqnarray}
Since
\begin{equation}
 r_t=f(t,t)=f(0,t)+\int_0^t\sigma(s,t)dW_s+\int_0^t\alpha(s,t)ds~,
\end{equation}
we have
\begin{eqnarray}
 &&B_t=\nonumber\\
 &&\exp\left(\int_0^t f(0,u)du +\int_0^t du\int_0^u\sigma(s,u)dW_s+
 \int_0^t du\int_0^u \alpha(s,u) ds\right)=\nonumber\\
 &&\exp\left(\int_0^t f(0,u)du +\int_0^t \left(\int_s^t\sigma(s,u)du\right)
 dW_s+\right.\nonumber\\
 &&\,\,\,\,\,\,\,\left.\int_0^t \left(\int_s^t \alpha(s,u)du\right) ds\right)~.
\end{eqnarray}
Note that we have changed the order of integration in the last two terms.

{}The discounted asset price is given by:
\begin{eqnarray}
 &&Z(t,T)=B_t^{-1}P(t,T)=\nonumber\\
 &&\exp\left(\int_0^t\Sigma(s,T)dW_s-\int_0^t f(0,u)du-
 \int_0^t\left(\int_s^T\alpha(s,u)du\right)ds\right)~,
\end{eqnarray}
where
\begin{equation}
 \Sigma(t,T)\equiv-\int_t^T\sigma(t,u)du
\end{equation}
plays the role of the log-volatility of $P(t,T)$.

{}Next, we need to change the measure from ${\bf P}$ to ${\bf Q}$
so that $Z(t,T)$ becomes a ${\bf Q}$-martingale.
Since
\begin{equation}
 d_t Z(t,T)=Z(t,T)\left(\Sigma(t,T)dW_t+\left[{1\over 2}
 \Sigma^2(t,T)-\int_t^T\alpha(t,u)du\right]dt\right)~,
\end{equation}
the corresponding shift $\gamma_t$ is given by
\begin{equation}
 \gamma_t={1\over 2}\Sigma(t,T)-{1\over\Sigma(t,T)}\int_t^T\alpha(t,u)du~.
\end{equation}
We have $d_t Z(t,T)=\Sigma(t,T)Z(t,T)d{\widetilde W}_t$, where
\begin{equation}
 {\widetilde W}_t=W_t+\int_0^t \gamma_s~ds
\end{equation}
is a ${\bf Q}$-Brownian motion. Note that the SDE for $P(t,T)$ is
given by
\begin{eqnarray}
 d_t P(t,T)=&&B_t d_t Z(t,T)+Z(t,T)dB_t=\nonumber\\
 &&P(t,T)\left[\Sigma(t,T)d{\widetilde W}_t+r_t dt\right]
\end{eqnarray}
under the martingale measure.

{}The rest (that is, the hedging, self-financing portfolios and
pricing) is as usual. Thus, the price of a claim $X=X_T$ is given by
\begin{eqnarray}
 V_t=&&B_t\langle B_T^{-1} X\rangle_{{\bf Q},{\cal F}_t}=\nonumber\\
     &&\left\langle e^{-\int_t^T r_s~ds} X\right\rangle_{{\bf Q},{\cal F}_t}~.
\end{eqnarray}
In particular, the price $P(t,S)$ of the $S$-bond is the same as the price
of the claim $X_S=\$1$:
\begin{equation}
 P(t,S)=\left\langle e^{-\int_t^S r_s~ds} \right\rangle_{{\bf Q},{\cal F}_t}~,
 ~~~t\leq S\leq T~.
\end{equation}
Note that this is nothing but a path integral of an exponential operator.

{}This has an important implication. Thus, for the discounted $S$-bond we have:
\begin{equation}
 Z(t,S)=B_t^{-1} P(t,S)=\langle B_S^{-1}\rangle_{{\bf Q},{\bf F}_t}~.
\end{equation}
This implies that $Z(t,S)$ is a ${\bf Q}$-martingale for all $S$. That is, a
single shift $\gamma_t$ should make all discounted $S$-bonds into
${\bf Q}$-martingales,
{\em i.e.,} the market price of risk for all $S$-bonds must be
the same. It then follows that $\gamma_t$ should be independent of the maturity
$T$. Recall that we have
\begin{equation}
 \int_t^T\alpha(t,u)du={1\over 2}\Sigma^2(t,T)-\Sigma(t,T)\gamma_t~,~~~0\leq t
 \leq T~.
\end{equation}
Differentiating w.r.t. $T$ we obtain:
\begin{equation}
 \alpha(t,T)=\sigma(t,T)\left[\gamma_t-\Sigma(t,T)\right]~,
\end{equation}
so the $T$-dependence of the ${\bf P}$-drifts $\alpha(t,T)$ cannot be
arbitrary.

{}Note that under the risk-neutral measure ${\bf Q}$ we have
\begin{eqnarray}
 &&d_t f(t,T)=\sigma(t,T)\left[d{\widetilde W}_t-\Sigma(t,T)dt\right]~,\\
 &&r_t=f(0,t)+\int_0^t\sigma(s,t)d{\widetilde W}_s-
 \int_0^t\sigma(s,t)\Sigma(s,t)ds~.
\end{eqnarray}
That is, these expressions no longer depend on the ${\bf P}$-drifts, but
only on the volatilities.

\subsection{Multi-factor HJM Models}

{}The drawback of the single-factor HJM model is that the correlation of
a $T$-bond and an $S$-bond is exactly 1. This can be lifted by considering
a multi-factor HJM model:
\begin{equation}
 f(t,T)=f(0,T)+\sum_{i=1}^n\int_0^t\sigma^i(s,T)dW^i_s+
 \int_0^t\alpha(s,T)ds~,
\end{equation}
where $W^i_t$ are independent Brownian motions. Note that the correlation
between the $T$-bond and the $S$-bond
\begin{equation}
 {{\sum_{i=1}^n\sigma^i(t,T)\sigma^i(t,S)}\over\sqrt{\sum_{i,j=1}^n
 \left[\sigma^i(t,T)\right]^2\left[\sigma^j(t,S)\right]^2}}
\end{equation}
is now generally different from 1.

{}As in the single factor model, to find a martingale measure ${\bf Q}$,
we have a restriction on the drift:
\begin{equation}
 \alpha(t,T)=\sum_{i=1}^n\sigma^i(t,T)\left[\gamma^i_t-\Sigma^i(t,T)\right]~,
\end{equation}
where $\Sigma^i(t,T)\equiv-\int_t^T\sigma^i(t,u)du$, and $\gamma^i_t$ are
independent of $T$.

{}Since we have $n$ independent Brownian motions, to hedge a claim we need
a portfolio consisting of $n$ separate instruments (plus the cash bond). Here
we can choose whichever $n$ instruments we like, and the answer will always
be the same. Thus, consider hedging the claim $X=X_T$ with discount bonds
$P(t,T^i)$, $i=1,\dots,n$. We must make sure that all $T^i>T$.

{}The value of a self-financing strategy $(\phi^1_t,\dots,\phi^n_t,
\psi_t)$ is then
\begin{equation}
 V_t=\sum_{i=1}^n \phi^i_t P(t,T^i)+\psi_t B_t~.
\end{equation}
Its discounted value $E_t=B_t^{-1} V_t$ has the SDE
\begin{equation}
 dE_t=\sum_{i=1}^n\phi^i_t dZ(t,T^i)~.
\end{equation}
To read off $\phi^i_t$ from this equation (note that
$E_t=\langle B_T^{-1}X\rangle_{{\bf Q},{\cal F}_t}$),
we must make sure that the
volatility matrix
\begin{equation}
 A_t=(A^{ij}_t)\equiv \left(\Sigma^i(t,T^j)\right)
\end{equation}
is non-singular. The rest goes as usual.

\section{Short-rate Models}

{}A short-rate model posits a risk-neutral measure ${\bf Q}$ and a short-rate
process $r_t$. The cash bond process is then given by
\begin{equation}
 B_t=\exp\left(\int_0^t r_s~ds\right)~,
\end{equation}
while the bond price is given by
\begin{equation}
 P(t,T)=\left\langle\exp\left(-\int_t^T r_s~ds\right)\right\rangle_{{\bf Q},
 {\cal F}_t}~.
\end{equation}
The price at time $t$ of a claim $X=X_T$ is
\begin{equation}
 V_t=\left\langle\exp\left(-\int_t^T r_s~ds\right)X\right\rangle_{{\bf Q},
 {\cal F}_t}~.
\end{equation}
One then works with a parametrized family of processes, which typically are
Markovian (but need not be), and chooses the parameters to best fit the market.

{}It is clear that the HJM models are short-rate models. Let us, however,
show that short-rate models are HJM models. Let us focus on the case
where $r_t$ is a Markov process. Then we have
\begin{equation}
 dr_t=\rho(r_t,t)dW_t+\nu(r_t,t) dt~,
\end{equation}
where we have chosen $\rho(y,t)$ and $\nu(y,t)$
to be deterministic functions.

{}Let
\begin{equation}
 V(x,t,T)\equiv
 \left\langle\exp\left(-\int_t^T r_s~ds\right)\right\rangle_{{\bf Q},
 ~r_t=x}~.
\end{equation}
This is nothing but the pricing function for the claim $X_T=1$. In particular,
$V(r_t,t,T)=P(t,T)$, and $V(x,T,T)=1$.
This pricing function satisfies a pricing PDE.
To derive this PDE, let us use the fact that the discounted bond process
$Z(t,T)=B_t^{-1}P(t,T)=B_t^{-1}V(r_t,t,T)$ must be a martingale under the
risk-neutral measure ${\bf Q}$. Thus, we have:
\begin{eqnarray}
 &&d_t Z(t,T)=B_t^{-1} \left[d_t V(r_t,t,T)-r_t V(r_t,t,T)\right]=\nonumber\\
 &&B_t^{-1}\Big[\rho(r_t,t)\partial_x V(r_t,t,T) dW_t+ \Big(\nu(r_t,t)\partial_x V(r_t,t,T)+\nonumber\\
 &&\,\,\,\,\,\,\,{1\over 2}\rho^2(r_t,t)\partial^2_x
 V(r_t,t,T)+\partial_t V(r_t,t,T) - r_t V(r_t,t,T)
 \Big) dt \Big]~.
\end{eqnarray}
The requirement that the drift term vanish then implies that
\begin{equation}
 \nu(r_t,t) \partial_x V(r_t,t,T)+ \partial_t
 V(r_t,t,T)+{1\over 2}\rho^2(r_t,t) \partial^2_x V(r_t,t,T)-r_t V(r_t,t, T)=0
\end{equation}
with the boundary condition $V(x,T,T)=1$.

{}Next, let
\begin{equation}
 g(x,t,T)\equiv -\ln(V(x,t,T))~.
\end{equation}
Note that $P(t,T)=\exp(-g(r_t,t,T))$, and
\begin{equation}
 f(t,T)=-\partial_T\ln(P(t,T))=\partial_T g(r_t,t,T)~.
\end{equation}
We therefore have
\begin{eqnarray}
 &&d_t f(t,T)=\rho(r_t,t)\partial_x\partial_T g(r_t,t,T)dW_t+\Big[\nu(r_t,t) \partial_x\partial_T g(r_t,t,T) +\nonumber\\
 &&\,\,\,\,\,\,\, \partial_t\partial_T
 g(r_t,t,T)+{1\over 2}\rho^2(r_t,t)
 \partial^2_x\partial_T g(r_t,t,T)\Big]dt~.
\end{eqnarray}
Since ${\bf Q}$ is the risk-neutral measure, we must have
\begin{equation}
 d_t f(t,T)=\sigma(t,T)dW_t-\sigma(t,T)\Sigma(t,T) dt~.
\end{equation}
This gives
\begin{eqnarray}
 &&\sigma(t,T)=\rho(r_t,t)\partial_x\partial_T g(r_t,t,T)~,\\
 &&\Sigma(t,T)=-\rho(r_t,t)\partial_x g(r_t,t,T)~.
\end{eqnarray}
Matching the drift terms requires that
\begin{eqnarray}\label{condg}
 &&\nu(r_t,t)\partial_x\partial_T g(r_t,t,T)+ \partial_t\partial_T
 g(r_t,t,T)+{1\over 2}\rho^2(r_t,t)\partial^2_x\partial_T g(r_t,t,T)=
 \nonumber\\
 &&\rho^2(r_t,t)\partial_x\partial_T g(r_t,t,T)\partial_x g(r_t,t,T)~.
\end{eqnarray}
This condition is indeed satisfied; in terms of $g(x,t,T)$
the pricing PDE for $V(x,t,T)$ reads:
\begin{equation}
 \nu(r_t,t)
 \partial_x g(r_t,t,T)+ \partial_t
 g(r_t,t,T)+{1\over 2}\rho^2(r_t,t) \left[\partial^2_x g(r_t,t,T)
 -\left(\partial_x g(r_t,t,T)\right)^2\right]+r_t=0~.
\end{equation}
Differentiating this equation w.r.t. $T$ we obtain (\ref{condg}).
This shows that short-rate models
are indeed HJM models.

{}Note that in a sense the choice of the drift $\nu_t$ is not particularly
important in the short-rate models -- indeed, $W_t$ is a ${\bf Q}$-Brownian
motion, but we can depart from the martingale measure ${\bf Q}$ to a ``real
world'' measure ${\bf P}$ via $W_t=W^\prime_t+\gamma_t$, where $\gamma_t$ is
an arbitrary previsible process. Under this new measure we have
\begin{equation}
 dr_t=\rho(r_t,t)dW^\prime_t +\nu^\prime_t dt~,
\end{equation}
where
\begin{equation}
 \nu^\prime_t=\nu(r_t,t)+\rho(r_t,t)\gamma_t
\end{equation}
can be an arbitrary previsible process under the measure ${\bf P}$.

\subsection{The Ho and Lee Model}

{}The {\em Ho and Lee} model is given by:
\begin{equation}
 dr_t=\rho(t) dW_t+\nu(t) dt~.
\end{equation}
That is, neither $\rho$ nor $\nu$ depend on $r_t$, but only on time $t$.

{}The function $g(x,t,T)$ can be computed as follows. Note that
\begin{eqnarray}
 \int_t^T r_s~ds=&&\left.sr_s\right|_t^T-\int_t^T sdr_s=\nonumber\\
 &&\left.sr_s\right|_t^T-T\int_t^T dr_s+\int_t^T (T-s)dr_s=\nonumber\\
 &&(T-t)r_t+\int_t^T (T-s)dr_s~.
\end{eqnarray}
This implies that
\begin{equation}
 V(x,t,T)=\exp\left(-x[T-t]\right)\left\langle\exp\left(-\int_t^T
 (T-s)dr_s\right)\right\rangle_{{\bf Q},~r_t=x}~.
\end{equation}
The expectation can be readily computed using the path integral
techniques:
\begin{eqnarray}
 &&\left\langle\exp\left(-\int_t^T
 (T-s)dr_s\right)\right\rangle_{{\bf Q},~r_t=x}=\exp\left(-\int_t^T(T-s)\nu(s)ds \right)\times\nonumber\\
 &&\,\,\,\,\,\,\,\times
 \int{\cal D}z~\exp(-S[z;t,T])\left.\exp\left(-\int_t^T
 (T-s)\rho(s){\dot z}(s)ds\right)\right|_{r_t=x}=\nonumber\\
 &&\exp\left(-\int_t^T(T-s)\nu(s)ds+{1\over 2}\int_t^T(T-s)^2\rho^2(s)ds\right)\times\nonumber\\
 &&\,\,\,\,\,\,\,\times \int{\cal D}y~\left.\exp(-S[y;t,T])\right|_{r_t=x}=\nonumber\\
 &&\exp\left(-\int_t^T(T-s)\nu(s)ds+{1\over 2}\int_t^T(T-s)^2\rho^2(s)ds\right)
\end{eqnarray}
{}From this it follows that
\begin{equation}
 g(x,t,T)=x(T-t)-{1\over 2}\int_t^T(T-s)^2\rho^2(s)ds+\int_t^T(T-s)\nu(s)ds~.
\end{equation}
We, therefore, have
\begin{eqnarray}
 &&\sigma(t,T)=\rho(t)\partial_x\partial_T g(r_t,t,T)=\rho(t)~,\\
 &&\Sigma(t,T)=-\rho(t)\partial_x g(r_t,t,T)=-\rho(t)(T-t)~,\\
 &&d_t f(t,T)=\rho(t)dW_t+\rho^2(t) (T-t)dt~,\\
 &&f(0,T)=\partial_T g(r_0,0,T)=r_0-\int_0^T (T-s)\rho^2(s)ds+
 \int_0^T \nu(s)ds~.
\end{eqnarray}
Note that the volatility surface $\sigma(t,T)$ is independent of $T$ in this
model.

\subsection{The Vasicek/Hull-White Model}

{}The {\em Vasicek} model is given by:
\begin{equation}
 dr_t=\rho(t) dW_t+\left[\nu(t)-\alpha(t)r_t\right] dt~.
\end{equation}
The corresponding equation for $g(x,t,T)$ is given by:
\begin{eqnarray}
 &&\left[\nu(t)-\alpha(t)x\right]\partial_x g(x,t,T)+\partial_t g(x,t,T)+\nonumber\\
 &&\,\,\,\,\,\,\,{1\over 2}\rho^2(t)\left[\partial^2_x g(x,t,T)-
 \left(\partial_x g(x,t,T)\right)^2\right]+x=0~.
\end{eqnarray}
The solution is given by:
\begin{equation}
 g(x,t,T)=x\eta(t,T)+\int_t^T\eta(s,T)\nu(s)ds-{1\over 2}
 \int_t^T\eta^2(s,T)\rho^2(s)ds~,
\end{equation}
where
\begin{equation}
 \eta(t,T)\equiv\int_t^T\beta(t,u)du~,
\end{equation}
and
\begin{equation}
 \beta(t,T)\equiv\exp\left(-\int_t^T\alpha(s)ds\right)~.
\end{equation}
Thus, we have
\begin{eqnarray}
 &&\sigma(t,T)=\rho(t)\beta(t,T)~,\\
 &&\Sigma(t,T)=-\rho(t)\eta(t,T)~,\\
 &&d_t f(t,T)=\rho(t)\beta(t,T)dW_t+\rho^2(t)\beta(t,T)\eta(t,T)dt~,\\
 &&f(0,T)=r_0\beta(0,T)+\int_0^T\beta(s,T)\nu(s)ds-\nonumber\\
 &&\,\,\,\,\,\,\,\int_0^T\beta(s,T)\eta(s,T)\rho^2(s)ds~.
\end{eqnarray}
Note that in this model the volatility surface depends on $T$.

{}Consider the case where $\rho,\nu,\alpha$ are all constant. Then
we have
\begin{equation}
 dr_t=\rho dW_t+\left[\nu-\alpha r_t\right]dt~.
\end{equation}
Note that the drift term pushes $r_t$ upward if $r_t$ is below $\nu/\alpha$,
and it pushes it downward if $r_t$ is above $\nu/\alpha$.
Moreover, the magnitude of the drift is proportional to the distance away
from this mean $\nu/\alpha$.
Processes with such ``mean-reverting'' behavior are known as
{\em Ornstein-Uhlenbeck processes}.

{}Let $r_t\equiv {\widehat r}_t\exp(-\alpha t)$. Then we have
\begin{equation}
 d{\widehat r}_t=\exp(\alpha t)\left[\rho dW_t+\nu dt\right]~.
\end{equation}
This has the solution
\begin{equation}
 r_t=\exp(-\alpha t)r_0+{\nu\over\alpha}\left[1-\exp(-\alpha t)\right]
 +\rho\exp(-\alpha t)\int_0^t\exp(\alpha s)dW_s~.
\end{equation}
Note that the mean of this process is
\begin{equation}
 \langle r_t\rangle=\exp(-\alpha t)r_0+
 {\nu\over\alpha}\left[1-\exp(-\alpha t)\right]~,
\end{equation}
which converges to $\nu/\alpha$ as $t$ gets large. The variance can be
computed as follows. Let
\begin{equation}
 r_t=\langle r_t\rangle +\xi_t~,
\end{equation}
where
\begin{equation}
 \xi_t\equiv\rho\exp(-\alpha t)\int_0^t\exp(\alpha s)dW_s~.
\end{equation}
Then
\begin{equation}
 v(t)\equiv\langle r_t^2\rangle-\left(\langle r_t\rangle\right)^2=
 \langle\xi_t^2\rangle~.
\end{equation}
Note that
\begin{equation}
 d\xi_t=\rho dW_t-\alpha\xi_t dt~.
\end{equation}
We therefore have:
\begin{eqnarray}
 d \langle \xi_t^2 \rangle=&&\langle d \xi_t^2\rangle=\nonumber\\
 &&2 \langle \xi_t d\xi_t\rangle +\langle (d\xi_t)^2\rangle=\nonumber\\
 &&2\rho \langle \xi_t dW_t\rangle+\left(\rho^2-2\alpha \langle \xi_t^2
 \rangle\right)
 dt=\nonumber\\
 &&\left(\rho^2-2\alpha \langle \xi_t^2\rangle\right)
 dt~.
\end{eqnarray}
That is,
\begin{equation}
 {dv(t)\over dt}=\rho^2-2\alpha v(t)~.
\end{equation}
The solution to this equation is given by
\begin{equation}
 v(t)=\rho^2~{{1-\exp(-2\alpha t)}\over 2\alpha}~,
\end{equation}
where we have taken into account the initial condition $v(0)=0$. Thus, the
variance converges to $\rho^2/2\alpha$ as $t$ gets large. Note that even
though the distribution for $r_t$ converges, the process $r_t$ itself does
not.

\subsection{The Cox-Ingersoll-Ross Model}

{}In the Ho and Lee as well as Vasicek models the short rate $r_t$ can
occasionally become negative. There are various ways to rectify this.

{}The {\em Cox-Ingersoll-Ross} model is given by:
\begin{equation}
 dr_t=\sqrt{r_t}\rho(t) dW_t+\left[\nu(t)-\alpha(t)r_t\right] dt~.
\end{equation}
The drift term is mean-reverting, while the volatility term is set up in such
a way that it gets smaller as $r_t$ approaches zero allowing the drift
term to dominate and stop $r_t$ from going below zero. In fact, as long as
$\nu(t)\geq \rho^2(t)/2$, this process actually stays strictly positive.
Such processes are called {\em autoregressive}.

{}Let $B(t,T)$ be the solution of the Riccati equation
\begin{equation}
 \partial_t B(t,T)={1\over 2}\rho^2(t) B^2(t,T)+\alpha(t)B(t,T)-1
\end{equation}
with the boundary condition $B(T,T)=1$. Then we have
\begin{equation}
 g(x,t,T)=xB(t,T)+\int_t^T\nu(s)B(s,T)ds~.
\end{equation}
Indeed, this satisfies the corresponding equation:
\begin{eqnarray}
 &&\left[\nu(t)-x\alpha(t)\right]\partial_x g(x,t,T)+\partial_t g(x,t,T)+\nonumber\\
 &&\,\,\,\,\,\,\,{1\over 2}x\rho^2(t)\left[\partial^2_x g(x,t,T)-\left(\partial_x
 g(x,t,T)\right)^2\right]+x=0~.
\end{eqnarray}
Let
\begin{equation}
 D(t,T)\equiv \partial_T B(t,T)~.
\end{equation}
Then we have:
\begin{eqnarray}
 &&\sigma(t,T)=\sqrt{r_t}\rho(t)D(t,T)~,\\
 &&\Sigma(t,T)=-\sqrt{r_t}\rho(t)B(t,T)~,\\
 &&d_t f(t,T)=\sqrt{r_t}\rho(t)D(t,T) dW_t +r_t\rho^2(t)D(t,T)B(t,T)dt~,\\
 &&f(0,T)=r_0 D(0,T)+\int_0^T D(s,T)\nu(s)ds~.
\end{eqnarray}
Note that the volatility surface in this model depends on the process $r_t$.

\subsection{The Black-Karasinski Model}

{}The {\em Black-Karasinski} model is given by:
\begin{eqnarray}
 &&r_t=\exp(X_t)~,\\
 &&dX_t=\rho(t)dW_t+\left[\nu(t)-\alpha(t)X_t\right]dt~.
\end{eqnarray}
This is another way of ensuring that $r_t$ stays positive. Note that
the process $X_t$ follows the Vasicek model.

\section{Interest Rate Products}

{}The simplest interest rate product is a forward contract. In a forward
contract at the current time $t$ we agree to make a payment $k$ at a future
time $T_1$, and in return receive \$1 at a later time $T_2$. According to
the pricing formula, we have
\begin{equation}
 V_t=B_t\langle B_{T_2}^{-1}\rangle_{{\cal F}_t}-B_t\langle B_{T_1}^{-1} k
 \rangle_{{\cal F}_t}=P(t,T_2)-kP(t,T_1)~,
\end{equation}
so the value of $k$ that gives this contract a nil value $V_t=0$ is given
by
\begin{equation}
 k={P(t,T_2)\over P(t,T_1)}~.
\end{equation}
We then hedge ourselves as follows. At time $t$ we buy $k$ units of the
$T_1$-bond, and sell one unit of the $T_2$-bond. The initial value of this
deal is zero, so it requires no investment. At time $T_1$ we receive $k$
dollars from the maturing $T_1$-bonds, which matches the payment $k$ we have
to make according to the forward contract. At time $T_2$ the dollar we receive
then covers the short $T_2$-bond. This is a {\em static} hedge.

\subsection{Forward Measures}

{}For the following applications it will be useful to define the notion
of the forward measures in the interest rate markets. Thus, in the interest
rate models it is often popular to use a $T$-bond as the numeraire. The
martingale measure for this numeraire is called the $T$-forward measure
${\bf P}_T$, and it makes the forward rate $f(t,T)$ into a
${\bf P}_T$-martingale.

{}The new numeraire is the $T$-bond normalized to have unit value at $t=0$:
$C_t=P(t,T)/P(0,T)$. The corresponding Radon-Nikodym process is
\begin{equation}
 \zeta_t={C_t\over B_t}={P(t,T)\over P(0,T)B_t}~.
\end{equation}
The forward price set at time $t$ for purchasing $X$ at time $T$ is given
by its current value $V_t$ scaled up by the return on a $T$-bond:
\begin{eqnarray}
 F_t=&&P^{-1}(t,T)V_t=P^{-1}B_t\langle B_T^{-1} X\rangle_{{\bf Q},{\cal F}_t}=
 \nonumber\\
 &&\langle X\rangle_{{\bf P}_T,{\cal F}_t}~,
\end{eqnarray}
so $F_t$ is a ${\bf P}_T$-martingale.

{}Note that we have:
\begin{equation}
 V_t=P(t,T)\langle X\rangle_{{\bf P}_T,{\cal F}_t}~,
\end{equation}
so the price of the claim $X$ at time $t$ is the conditional
${\bf P}_T$-expectation
of $X$ up to time $t$ (the forward price of $X$) discounted by the time value
of money (the $T$-bond).

{}Note that
\begin{equation}
 \zeta_t={Z(t,T)\over P(0,T)}~,
\end{equation}
where $Z(t,T)$ is the discounted $T$-bond process: $Z(t,T)=B_t^{-1}P(t,T)$.
This implies that
\begin{equation}
 d\zeta_t=\zeta_t\sum_{i=1}^n \Sigma^i(t,T)d{\widetilde W}^i_t~,
\end{equation}
where ${\widetilde W}^i_t$ are the ${\bf Q}$-Brownian motions. This then
implies that
\begin{equation}
 {\widehat W}^i_t={\widetilde W}^i_t-\int_0^t \Sigma^i(s,T)ds
\end{equation}
are the corresponding ${\bf P}_T$-Brownian motions.

{}Also, note that for the forward rate we have
\begin{equation}
 d_t f(t,T)=\sum_{i=1}^n \sigma^i(t,T)\left[d{\widetilde W}^i_t-
 \Sigma^i(t,T)dt\right]=\sum_{i=1}^n \sigma^i(t,T) d{\widehat W}^i_t~,
\end{equation}
so that $f(t,T)$ is a ${\bf P}_T$-martingale.
This, in particular, implies that
\begin{equation}
 f(t,T)=\langle f(T,T)\rangle_{{\bf P}_T,{\cal F}_t}=
 \langle r_T \rangle_{{\bf P}_T,{\cal F}_t}~,
\end{equation}
so $f(t,T)$ is the forward rate for $r_T$.

\subsection{Multiple Payment Contracts}

{}Most interest rate products do not just make a single payment $X$ at time
$T$. Instead, the contract specifies a sequence of payments $X_i$ made at a
sequence of times $T_i$, $i=1,\dots,n$.

{}To price such a contract, we can treat each payment separately:
\begin{equation}
 V_i(t)=B_t\langle B_{T_i}^{-1} X_i\rangle_{{\bf Q},{\cal F}_t}=
 P(t,T_i)\langle X_i\rangle_{{\bf P}_{T_i},{\cal F}_t}~.
\end{equation}
Note that in this case the forward measure, if used, would have to be changed
for each $i$.

{}Alternatively, we can roll up the payments into a {\em savings account}
as we receive them, and keep them until the last payment date $T$. That is,
as each payment is made, we use them to buy a $T$-bond, or invest it into the
bank account process $B_t$ until time $T$. Thus, in the former case the
payoff is a single payment at time $T$:
\begin{equation}
 X=\sum_{i=1}^n {X_i\over P(T_i,T)}~,
\end{equation}
and its worth at time $t$ is
\begin{equation}
 V_t=B_t\langle B_{T}^{-1} X\rangle_{{\bf Q},{\cal F}_t}=
 P(t,T)\langle X\rangle_{{\bf P}_{T},{\cal F}_t}~.
\end{equation}
In this case we need only one forward measure ${\bf P}_T$.

\subsection{Bonds with Coupons}

{}In practice zero-coupon bonds are not popular products, especially at the
long end. Instead, a bond usually pays not only its principal at maturity
$T$, but also makes smaller regular coupon payments of a fixed amount $c$
until then.

{}Thus, suppose a bond makes $n$ regular payments at an uncompounded rate $k$
at times $T_i=T_0+i\delta$, $i=1,2,\dots$, and also pays off a dollar at the
maturity time $T$.
The amount of the actual coupon payment is $k\delta$,
where $\delta$ is the payment period. This income stream is equivalent to
owning one $T$-bond plus $k\delta$ units of each $T_i$-bond, $i=1,\dots,
n$, where $n=I(T)-1$ is the total number of payments
before maturity time $T$, and
$I(t)\equiv {\rm min}(i:~t<T_i)$.
The price of the coupon bond at time $t$ then is
\begin{equation}
 P_c (t,T)=P(t,T)+k\delta\sum_{i=I(t)}^n P(t,T_i)~.
\end{equation}
At time $t=T_0$ we have
\begin{equation}
 P_c (T_0,T)=P(T_0,T)+k\delta\sum_{i=1}^n P(T_0,T_i)~.
\end{equation}
If we desire the coupon bond to start with its face value ($P_c(T_0,T)=1$),
then
\begin{equation}
 k={{1-P(T_0,T)}\over\delta{\sum_{i=1}^n P(T_0,T_i)}}
\end{equation}
is the corresponding coupon rate.

\subsection{Floating Rate Bonds}

{}A bond might also have {\em floating} coupon payments. Thus, consider a
bond that pays \$1 at the maturity time $T$, and also makes payments at times
$T_i=T_0+i\delta$, $i=1,2,\dots,$
of varying amounts. The amount of payment made
at time $T_i$ is determined by the {\small LIBOR} rate ({\em London Interbank Offer Rate}) at time $T_{i-1}$:
\begin{equation}
 L(T_{i-1})={1\over \delta}\left[{1\over P(T_{i-1},T_i)}-1\right]~.
\end{equation}
The actual coupon payment is
\begin{equation}
 X_i=\delta L(T_{i-1})={1\over P(T_{i-1},T_i)}-1~,
\end{equation}
which is the amount of interest we would get by buying a dollar's worth of
$T_i$-bond at time $T_{i-1}$. The value of this payment at time $T_0$ is
\begin{eqnarray}
 V_i(T_0)=&&B_{T_0}\langle B_{T_i}^{-1} X_i\rangle_{{\bf Q},{\cal F}_{T_0}}=
 \nonumber\\
 &&B_{T_0}\langle B_{T_i}^{-1} P^{-1}(T_{i-1},T_i)
 \rangle_{{\bf Q},{\cal F}_{T_0}}-
 B_{T_0}\langle B_{T_i}^{-1}\rangle_{{\bf Q},{\cal F}_{T_0}}=\nonumber\\
 &&B_{T_0}\left\langle\langle B_{T_i}^{-1} P^{-1}(T_{i-1},T_i)
 \rangle_{{\bf Q},{\cal F}_{T_{i-1}}}\right\rangle_{{\bf Q},{\cal F}_{T_0}}-
 B_{T_0}\langle B_{T_i}^{-1}\rangle_{{\bf Q},{\cal F}_{T_0}}~,
\end{eqnarray}
where in the last line we are using the tower law. Note that
\begin{eqnarray}
 \langle B_{T_i}^{-1} P^{-1}(T_{i-1},T_i)
 \rangle_{{\bf Q},{\cal F}_{T_{i-1}}}=&&P^{-1}(T_{i-1},T_i)
 \langle B_{T_i}^{-1}
 \rangle_{{\bf Q},{\cal F}_{T_{i-1}}}=\nonumber\\
 &&P^{-1}(T_{i-1},T_i)B^{-1}_{T_{i-1}}\left(B_{T_{i-1}}\langle B_{T_i}^{-1}
 \rangle_{{\bf Q},{\cal F}_{T_{i-1}}}\right)=\nonumber\\
 &&P^{-1}(T_{i-1},T_i)B^{-1}_{T_{i-1}}\left(P(T_{i-1},T_i)\right)=\nonumber\\
 &&B^{-1}_{T_{i-1}}~.
\end{eqnarray}
We therefore have
\begin{equation}
 V_i(T_0)=B_{T_0}\langle B_{T_{i-1}}^{-1}\rangle_{{\bf Q},{\cal F}_{T_0}}-
 B_{T_0}\langle B_{T_i}^{-1}\rangle_{{\bf Q},{\cal F}_{T_0}}=
 P(T_0,T_{i-1})-P(T_0,T_i)~.
\end{equation}
The total value of the variable coupon bond is given by:
\begin{eqnarray}
 V_0=&&P(T_0,T)+\sum_{i=1}^n V_i(T_0)=\nonumber\\
 &&P(T_0,T)+\sum_{i=1}^n\left[P(T_0,T_{i-1})
 -P(T_0,T_i)\right]=\nonumber\\
 &&P(T_0,T)+\left[P(T_0,T_0)-P(T_0,T_n)\right]=\nonumber\\
 &&1+\left[P(T_0,T)-P(T_0,T_n)\right]~.
\end{eqnarray}
If the maturity time $T$ coincides with the last coupon payment $T_n$, then
we have $V_0=1$. That is, the initial value of the variable coupon bond
is its face value. This is because this bond is equivalent to the following
sequence of trades. At time $T_0$ take a dollar and buy $T_1$-bonds with it.
At time $T_1$ take the interest from the $T_1$-bonds as the $T_1$-coupon,
and buy $T_2$ bonds with the leftover dollar principal. Repeat until we are
left with a dollar at time $T_n$. This has exactly the same cash flows as the
variable coupon bond, so the initial prices must match.

\subsection{Swaps}

{}Swaps are popular contracts that exchange a stream of varying payments for
a stream of fixed payments or {\em vice versa}. That is, we swap a floating
interest rate for a fixed one. In practice only the net difference is exchanged
at each payment date.

{}Consider a swap where we receive a stream of fixed rate
payments in exchange for
floating rate payments. This swap is simply a portfolio which is long a fixed
coupon bond and short a variable coupon bond. The former is worth
\begin{equation}
 P(T_0,T)+k\delta\sum_{i=1}^n P(T_0,T_i)~,
\end{equation}
while the latter costs
\begin{equation}
 P(T_0,T)+1-P(T_0,T_n)~.
\end{equation}
The fixed rate needed to give the swap initial null value then is
\begin{equation}
 k={{1-P(T_0,T_n)}\over \delta\sum_{i=1}^n P(T_0,T_i)}~.
\end{equation}
Note that this rate does not depend on the maturity $T$.

{}Suppose we would like to enter into a forward swap agreement. The value
of the swap at time $T_0$ is
\begin{equation}
 X=P(T_0,T_n)+k\delta \sum_{i=1}^n P(T_0,T_i)-1~.
\end{equation}
The price of $X$ at time $t$ before $T_0$ is
\begin{equation}
 V_t=B_t\langle B_{T_0}^{-1} X\rangle_{{\bf Q},{\cal F}_t}=
 P(t,T_n)+k\delta\sum_{i=1}^n P(t,T_i)-P(t,T_0)~,
\end{equation}
where we have taken into account that
\begin{equation}
 \langle B^{-1}_{T_0}P(T_0,T_i)\rangle_{{\bf Q},{\cal F}_t}=
 B_t^{-1} P(t,T_i)
\end{equation}
as $Z(t,T_i)=B_t^{-1} P(t,T_i)$ is a ${\bf Q}$-martingale.

{}Thus, the forward fixed rate $k$ needed to give the forward swap initial
null value at time $t$ is
\begin{equation}
 k={{P(t,T_0)-P(t,T_n)}\over\delta\sum_{i=1}^n P(t,T_i)}=
 {{1-F_t(T_0,T_n)}\over\delta\sum_{i=1}^n F_t(T_0,T_i)}~,
\end{equation}
where
\begin{equation}
 F_t(T_0,T_i)\equiv {P(t,T_i)\over P(t,T_0)}
\end{equation}
is the forward price at time $t$ for purchasing a $T_i$-bond at time $T_0$.

\subsection{Bond Options}

{}Consider a European
call option on a $T$-bond with the strike price $k$ and the
exercise date $\tau$. Its worth at time $t<\tau$ is
\begin{equation}
 V_t=B_t \langle B_\tau^{-1}\left(P(\tau,T)-k\right)^+\rangle_{{\bf Q},{\cal
 F}_t}~.
\end{equation}
Let us consider the Ho and Lee model:
\begin{equation}
 d_t f(t,T)=\rho dW_t +\rho^2 (T-t) dt
\end{equation}
with constant $\rho$. We then have
\begin{equation}
 f(t,T)=f(0,T)+ \rho W_t+{1\over 2}\rho^2 t(2T-t)~,
\end{equation}
and
\begin{eqnarray}
 P(t,T)=&&\exp\left(-\int_t^T f(t,u) du\right)=\nonumber\\
 &&\exp\left(-\int_t^T f(0,u) du -\rho(T-t)W_t-
 {1\over 2}\rho^2 tT(T-t)\right)~.
\end{eqnarray}
Also,
\begin{equation}
 r_t=f(t,t)=f(0,t)+\rho W_t +{1\over 2}\rho^2 t^2~,
\end{equation}
so we have
\begin{eqnarray}
 B_t =&&\exp\left(\int_0^t r_s ds \right)=\nonumber\\
 &&\exp\left(\int_0^t f(0,u)du +\rho
 \int_0^t W_s ds +{1\over 6} \rho^2 t^3\right)~.
\end{eqnarray}
Since we have $\int_0^t W_s ds$ in $B_t$, it is simpler to use the
$\tau$-forward measure instead of the measure ${\bf Q}$:
\begin{equation}
 V_t=P(t,\tau)\langle\left(P(\tau,T)-k\right)^+\rangle_{{\bf P}_\tau,{\cal
 F}_t}~.
\end{equation}
Note that
\begin{eqnarray}
 {\widehat W}_t=&&W_t-\int_0^t \Sigma(s,\tau)ds=\nonumber\\
 &&W_t+\rho\int_0^t (\tau-s)ds=\nonumber\\
 &&W_t+\rho\left(\tau t -{1\over 2}t^2\right)
\end{eqnarray}
is a ${\bf P}_\tau$-Brownian motion. (Moreover, $f(t,\tau)$ is a
${\bf P}_\tau$-martingale: $d_tf(t,\tau)=\rho d{\widehat W}_t$.)

{}Note that
\begin{eqnarray}
 P(\tau,T)=&&F_t(\tau,T)\exp\Big(-\rho(T-\tau)\left[W_\tau-
 W_t\right]-\nonumber\\
 &&{1\over 2}\rho^2 \left[\tau T(T-\tau)+t\tau(\tau-t)-
 tT(T-t)\right]\Big)=\nonumber\\
 &&F_t(\tau,T)\exp\left(-\rho(T-\tau)\left[{\widehat W}_\tau-
 {\widehat W}_t\right]-{1\over 2}\rho^2 (T-\tau)^2 (\tau-t)
 \right)=\nonumber\\
 &&F_t(\tau,T)\exp\left(-{\overline \sigma}\left[{\widehat W}_\tau-
 {\widehat W}_t\right]-{1\over 2}{\overline \sigma}^2 (\tau-t)
 \right)~,
\end{eqnarray}
where $F_t(\tau,T)=P(t,T)/P(t,\tau)$ is the forward price at time $t$
of purchasing a $T$-bond at time $\tau$, and
${\overline \sigma}\equiv \rho(T-\tau)$ is the {\em term volatility}.
Note that the process
${\widehat W}_\tau-{\widehat W}_t$ is a normal $N(0,\tau-t)$, and is
independent
of ${\cal F}_t$. It is then clear that the price of the call option is given
by the corresponding Black-Scholes formula
\begin{eqnarray}
 &&V_t=P(t,\tau)\left[F_t(\tau,T)\Phi\left({\ln\left({F_t(\tau,T)\over k}\right)
 \over
 {\overline \sigma}\sqrt{\tau-t}}+{{\overline \sigma}\sqrt{\tau-t}\over 2}
 \right)\right.\nonumber\\
 &&\,\,\,\,\,\,\,\left.-k\Phi\left({\ln\left({F_t(\tau,T)\over k}\right)
 \over
 {\overline \sigma}\sqrt{\tau-t}}-{{\overline \sigma}\sqrt{\tau-t}\over 2}
 \right)
 \right]~,
\end{eqnarray}
The reason why the Black-Scholes formula works in this model is that the
latter is actually log-normal.

\subsection{Bond Options in the Vasicek Model}

{}The most general single-factor model with log-normal bond prices is the
Vasicek model:
\begin{equation}
 dr_t=\rho(t) dW_t+\left[\nu(t)-\alpha(t)r_t\right] dt~.
\end{equation}
Let
\begin{eqnarray}
 &&\beta(t,T)\equiv\exp\left(-\int_t^T\alpha(s)ds\right)~,\\
 &&\eta(t,T)\equiv\int_t^T\beta(t,u)du~.
\end{eqnarray}
Then we have
\begin{eqnarray}
 &&\sigma(t,T)=\rho(t)\beta(t,T)~,\\
 &&\Sigma(t,T)=-\rho(t)\eta(t,T)~,\\
 &&d_t f(t,T)=\rho(t)\beta(t,T)dW_t+\rho^2(t)\beta(t,T)\eta(t,T)dt~.
\end{eqnarray}
That is,
\begin{equation}
 f(t,T)=f(0,T)+\int_0^t \rho(s)\beta(s,T)dW_s+
 \int_0^t\rho^2(s)\beta(s,T)\eta(s,T)ds~.
\end{equation}
We therefore have:
\begin{eqnarray}
 &&-\ln\left(P(t,T)\right)=\int_t^T f(0,u) du+\nonumber\\
 &&\int_t^T \left(\int_0^t
 \rho(s) \beta(s,u) dW_s\right) du+\int_t^T \left(
 \int_0^t\rho^2(s)\beta(s,u)\eta(s,u)ds\right)du=\nonumber\\
 &&\int_t^T f(0,u) du+\int_0^t
 \rho(s) \left[\eta(s,T)-\eta(s,t)\right] dW_s+\nonumber\\
 &&{1\over 2}\int_0^t\rho^2(s)\left[\eta^2(s,T)-\eta^2(s,t)\right]ds~.
\end{eqnarray}
This gives:
\begin{eqnarray}
 &&{P(\tau,T)\over F_t(\tau,T)}=\exp\left(-\int_t^\tau \rho(s)\left[
 \eta(s,T)-\eta(s,\tau)\right]dW_s -\right.\nonumber\\
 &&\,\,\,\,\,\,\,\left.{1\over 2}
 \int_t^\tau\rho^2(s)\left[\eta^2(s,T)-\eta^2(s,\tau)\right]ds
 \right)~.
\end{eqnarray}
Let us now go to the $\tau$-forward measure. The corresponding Brownian
motion is ${\widehat W}_t$:
\begin{equation}
 d{\widehat W}_t=dW_t-\Sigma(t,\tau)dt=dW_t+\rho(t)\eta(t,\tau)dt~.
\end{equation}
We, therefore, have:
\begin{eqnarray}
 &&{P(\tau,T)\over F_t(\tau,T)}=\exp\left(-\int_t^\tau \rho(s)\left[
 \eta(s,T)-\eta(s,\tau)\right]d{\widehat W}_s -\right.\nonumber\\
 &&\,\,\,\,\,\,\,\left.{1\over 2}
 \int_t^\tau\rho^2(s)\left[\eta(s,T)-\eta(s,\tau)\right]^2 ds
 \right)~.
\end{eqnarray}
Note that
\begin{equation}
 \eta(s,T)-\eta(s,\tau)=\int_\tau^T \beta(s,u)du~.
\end{equation}
Also, note that the process
\begin{equation}
 \zeta(t,\tau,T)\equiv -\int_t^\tau \rho(s)\left[
 \eta(s,T)-\eta(s,\tau)\right]d{\widehat W}_s
\end{equation}
is independent of ${\cal F}_t$. Moreover, for any real $\theta$ we have
\begin{eqnarray}
 &&\langle \exp(\theta
 \zeta(t,\tau,T))\rangle_{{\bf P}_\tau,{\cal F}_t}=\nonumber\\
 &&\int{\cal D}x~\left.\exp\left(-S[x;t,\tau]-\theta\int_t^\tau\rho(s)\left[
 \eta(s,T)-\eta(s,\tau)\right]{\dot x}(s)ds
 \right)\right|_{x(t)=x_*(t)}=
 \nonumber\\
 &&
 \exp\left({\theta^2\over 2}
 \int_t^\tau\rho^2(s)\left[\eta(s,T)-\eta(s,\tau)\right]^2
  ds\right)~,
\end{eqnarray}
which implies that $\zeta(t,\tau,T)$ is a normal $N(0,v(t,\tau,T))$
with the variance given by
\begin{equation}
 v(t,\tau,T)=\int_t^\tau\rho^2(s)\left[\eta(s,T)-\eta(s,\tau)\right]^2
  ds~.
\end{equation}
This then immediately implies that the price of the call option is given by:
\begin{eqnarray}
 &&V_t=P(t,\tau)\langle(P(\tau,T)-k)^+\rangle_{{\bf P}_\tau,{\cal F}_t}=\nonumber\\
 &&\,\,\,\,\,\,\,P(t,\tau)\left[F_t(\tau,T)\Phi\left({\ln\left({F_t(\tau,T)\over k}\right)
 \over
 \sqrt{v(t,\tau,T)}}+{\sqrt{v(t,\tau,T)}\over 2}\right)\right.\nonumber\\
 &&\,\,\,\,\,\,\,\left.-k\Phi\left({\ln\left({F_t(\tau,T)\over k}\right)
 \over
 \sqrt{v(t,\tau,T)}}-{\sqrt{v(t,\tau,T)}\over 2}\right)\right]~,
\end{eqnarray}
which, once again, is a Black-Scholes-like formula.

{}For constant $\rho$ and $\alpha$ we have:
\begin{eqnarray}
 &&\beta(s,u)=\exp(-\alpha(u-s))~,\\
 &&\eta(s,T)-\eta(s,\tau)=\int_\tau^T \beta(s,u)du={e^{\alpha s}\over \alpha}
 \left[\exp(-\alpha\tau)-\exp(-\alpha T)\right]~,\\
 &&v(t,\tau,T)={\rho^2\over\alpha^2}
 \left[\exp(-\alpha\tau)-\exp(-\alpha T)\right]^2\int_t^\tau \exp(2\alpha s)
 ds=\nonumber\\
 &&{\rho^2\over2\alpha^3}\left[1-\exp(-\alpha [T-\tau])\right]^2
 \left[1-\exp(-2\alpha [\tau-t])\right]~,
\end{eqnarray}
so that the dependence on $T-\tau$ and $\tau-t$ factorizes.

\subsection{Options on Coupon Bonds}

{}Suppose we have a bond with coupons:
\begin{equation}
 P_c(t,T)=P(t,T)+\kappa\delta\sum_{i=I(t)}^n P(t,T_i)~.
\end{equation}
Here we use $\kappa$ for the uncompounded rate for coupons to distinguish it from the strike price $k$.
Suppose the zero-coupon bonds follow a single-factor short-rate model with
deterministic $\rho(r_t,t)$ and $\nu(r_t,t)$. Then each zero-coupon
bond price can be viewed as a deterministic function $P(t,T)=V(r_t,t,T)$. In
this case we can price a call option on the coupon bond using the trick due to
Jamshidian.

{}The function $V(r_t,t,T)$ monotonically decreases with increasing $r_t$. This
means that $P_c(t,T)$ is also a decreasing function of $r_t$. Let $k$ be the
strike price for the call option. Then there exists $r_*$ such that
\begin{equation}
 \left.P_c(t,T)\right|_{r_t=r_*}=k~.
\end{equation}
Let $k_\tau\equiv V(r_*,t,\tau)$. Then we have
\begin{equation}
 (P_c(t,T)-k)^+=(P(t,T)-k_T)^+ +\kappa\delta \sum_{i=I(t)}^n (P(t,T_i)-k_{T_i})^+.
\end{equation}
We can therefore price a call option on a coupon bond using the corresponding
call options on the zero-coupon bonds of various maturities.

\subsection{Caps and Floors}

{}Suppose we are borrowing at a floating rate and want to ensure that it does
not go above a fixed rate $k$. The {\em cap} contract pays us the difference
between the {\small LIBOR} rate and the fixed rate at each payment time $T_i$:
\begin{equation}
 \delta(L(T_{i-1})-k)^+~,
\end{equation}
where
\begin{equation}
 L(T_i)={1\over \delta}\left(P^{-1}(T_{i-1},T_i)-1\right)~.
\end{equation}
An individual payment is called a {\em caplet}. If we can price caplets, then
we can also price the cap.

{}The caplet claim is
\begin{equation}
 X_{T_i}=\left(P^{-1}(T_{i-1},T_i)-1-k\delta\right)^+=
 K^{-1}P^{-1}(T_{i-1},T_i)(K-P(T_{i-1},T_i))^+~,
\end{equation}
where $K\equiv (1+\delta k)^{-1}$.
The price of the caplet at time $t$ is given by:
\begin{eqnarray}
 V_t=&&B_t\langle B_{T_i}^{-1} X_{T_i}\rangle_{{\bf Q},{\cal F}_t}=
 \nonumber\\
 &&K^{-1}B_t\left\langle P^{-1}(T_{i-1},T_i)B_{T_i}^{-1}(K-P(T_{i-1},T_i))^+
 \right\rangle_{{\bf Q},{\cal F}_t}=\nonumber\\
 &&K^{-1}B_t\left\langle B_{T_{i-1}}^{-1}(K-P(T_{i-1},T_i))^+
 \right\rangle_{{\bf Q},{\cal F}_t}~,
\end{eqnarray}
where in the last line we have used the tower law. Thus, the value of a caplet
is just the price of $(1+k\delta)$ put options on the $T_i$-bond with
the strike price $K$ and the exercise date $T_{i-1}$.

{}A {\em floor} contract works similarly, we receive a premium for agreeing to
never pay less than some fixed rate $k$. That is, we pay an extra amount
\begin{equation}
 \delta(k-L(T_{i-1}))^+
\end{equation}
at time $T_i$. There is a {\em floor-cap parity}. Thus, the worth of a
floorlet less the cost of a caplet with the same fixed rate $k$ is
\begin{equation}
 B_t\langle B_{T_i}^{-1}\delta(k-L(T_{i-1}))\rangle_{{\bf Q},{\cal F}_t}=
 (1+\delta k)P(t,T_i)-P(t,T_{i-1})~,
\end{equation}
where, once again, we have used the tower law. Note that
\begin{equation}
 \sum_{i=1}^n\left[(1+\delta k)P(t,T_i)-P(t,T_{i-1})\right]=P(t,T_n)-
 P(t,T_0)+k\delta\sum_{i=1}^n P(t,T_i)~,
\end{equation}
which for time $t\leq T_0$ is the value of the forward
swap with maturity $T\geq T_n$ and the fixed rate $k$.

\subsection{Swaptions}

{}A swaption is an option to enter into a swap on a future date at a given
rate $k$. The worth of this option at time $T_0$ is
\begin{equation}
 \left(P(T_0,T_n)+k\delta\sum_{i=1}^n P(T_0,T_i)-1\right)^+~,
\end{equation}
which is nothing but a call option struck at \$1 on a $T_n$-bond with the
coupon rate $k$. This can be understood from the fact that a swap is just a
coupon bond less a floating bond (the latter always has par value). If you
receive a fixed on a swap, you have a long position in the bond market, so
a swaption looks like a coupon bond option.

\section{The General Multi-factor Log-Normal Model}

{}Consider an HJM model with the factorizable volatility surfaces:
\begin{equation}
 \sigma^i(t,T)=x^i(t)y^i(T)~.
\end{equation}
Then we have
\begin{equation}
 \Sigma^i(t,T)=-x^i(t)\int_t^T y^i(u)du\equiv -x^i(t)Y^i(t,T)~,
\end{equation}
and
\begin{equation}
 d_t f(t,T)=\sum_{i=1}^n y^i(T)x^i(t)d{\widetilde W}^i_t+
 \sum_{i=1}^n y^i(T)[x^i(t)]^2 Y^i(t,T)dt~.
\end{equation}
The market completeness condition requires that the matrix
$A_t=(A_t^{ij})\equiv(Y^i(t,T^j))$ be non-singular for all $t<T_1$ for every
set of $n$ maturities $T_1,\dots,T_n$.

{}Note that
\begin{equation}
 f(t,T)=f(0,T)+\sum_{i=1}^n y^i(T)\int_0^t x^i(s)d{\widetilde W}_s+
 \sum_{i=1}^n y^i(T)\int_0^t[x^i(s)]^2 Y^i(s,T)ds~.
\end{equation}
This gives:
\begin{eqnarray}
 -\ln(P(t,T))=&&\int_t^T f(0,u)du +\sum_{i=1}^n Y^i(t,T)\int_0^t x^i(s)
 d{\widetilde W}_s +\nonumber\\
 &&{1\over 2}\sum_{i=1}^n \int_0^t[x^i(s)]^2 \left(
 [Y^i(s,T)]^2-[Y^i(s,t)]^2\right)ds~.
\end{eqnarray}
Note that
\begin{eqnarray}
 &&-\ln(F_t(\tau,T))=\int_\tau^T f(0,u)du +\sum_{i=1}^n Y^i(\tau,T)
 \int_0^t x^i(s) d{\widetilde W}_s +\nonumber\\
 &&\,\,\,\,\,\,\,{1\over 2}\sum_{i=1}^n \int_0^t[x^i(s)]^2 \left(
 [Y^i(s,T)]^2-[Y^i(s,\tau)]^2\right)ds~.
\end{eqnarray}
On the other hand,
\begin{eqnarray}
 -\ln(P(\tau,T))=&&\int_\tau^T f(0,u)du +\sum_{i=1}^n Y^i(\tau,T)
 \int_0^\tau x^i(s)
 d{\widetilde W}_s +\nonumber\\
 &&{1\over 2}\sum_{i=1}^n \int_0^\tau[x^i(s)]^2 \left(
 [Y^i(s,T)]^2-[Y^i(s,\tau)]^2\right)ds~.
\end{eqnarray}
This implies that
\begin{eqnarray}
 -\ln\left({P(\tau,T)\over F_t(\tau,T)}\right)=
 &&\sum_{i=1}^n Y^i(\tau,T)
 \int_t^\tau x^i(s)
 d{\widetilde W}_s +\nonumber\\
 &&{1\over 2}\sum_{i=1}^n \int_t^\tau[x^i(s)]^2 \left(
 [Y^i(s,T)]^2-[Y^i(s,\tau)]^2\right)ds~.
\end{eqnarray}
Let us now go to the $\tau$-forward measure. The corresponding Brownian
motions are
\begin{equation}
 d{\widehat W}^i_t=d{\widetilde W}^i_t-\Sigma^i(t,\tau)dt=d{\widetilde W}_t+
 x^i(t)Y^i(t,\tau)dt~.
\end{equation}
This then implies that
\begin{eqnarray}
 &&-\ln\left({P(\tau,T)\over F_t(\tau,T)}\right)=\nonumber\\
 &&\,\,\,\,\,\,\,\sum_{i=1}^n Y^i(\tau,T)
 \int_t^\tau x^i(s)
 d{\widehat W}_s +
 {1\over 2}\sum_{i=1}^n [Y^i(\tau,T)]^2 \int_t^\tau[x^i(s)]^2 ds~.
\end{eqnarray}
The price of a call option then is given by
\begin{eqnarray}
 &&V_t=P(t,\tau)\langle(P(\tau,T)-k)^+\rangle_{{\bf P}_\tau,{\cal F}_t}=
 \nonumber\\
 &&\,\,\,\,\,\,\,P(t,\tau)\left[F_t(\tau,T)\Phi\left({\ln\left({F_t(\tau,T)\over k}\right)
 \over
 \sqrt{v(t,\tau,T)}}+{\sqrt{v(t,\tau,T)}\over 2}\right)\right.\nonumber\\
 &&\,\,\,\,\,\,\,\left.-k\Phi\left({\ln\left({F_t(\tau,T)\over k}\right)
 \over
 \sqrt{v(t,\tau,T)}}-{\sqrt{v(t,\tau,T)}\over 2}\right)\right]~,
\end{eqnarray}
where
\begin{equation}
 v(t,\tau,t)\equiv\sum_{i=1}^n [Y^i(\tau,T)]^2 \int_t^\tau[x^i(s)]^2 ds~.
\end{equation}
Once again, we have a Black-Scholes-like formula as this model is log-normal.

\subsection{The Brace-Gatarek-Musiela (BGM) Model}

{}Let
\begin{equation}
 L(t,T)\equiv {1\over \delta}\left[{P(t,T)\over P(t,T+\delta)}-1\right]=
 {1\over \delta}\left[\exp\left(\int_T^{T+\delta}f(t,u)du\right)-1\right]~.
\end{equation}
Note that $L(t,T)$ is the $\delta$-period forward {\small LIBOR} rate for
borrowing at time $T$. Also,
\begin{equation}
 L(T)\equiv L(T,T) = {1\over \delta}\left[{1\over P(T,T+\delta)}-1\right]
\end{equation}
is the instantaneous {\small LIBOR} rate.

{}In the BGM model the volatility surfaces are restricted as follows:
\begin{equation}
 \int_T^{T+\delta} \sigma^i(t,u)du={\delta L(t,T)\over{1+\delta L(t,T)}}~
  \gamma^i (t,T)~,
\end{equation}
where $\gamma^i (t,T)$ are some deterministic functions. Under the martingale
measure ${\bf Q}$ we have:
\begin{eqnarray}
 &&d_t L(t,T)={1\over \delta}\exp\left(\int_T^{T+\delta}f(t,u)du\right)\times\nonumber\\
&&\,\,\,\,\,\,\,\left[\int_T^{T+\delta}d_tf(t,u)du+{1\over 2}
 \left(\int_T^{T+\delta}d_tf(t,u)du\right)^2
 \right]=
 \nonumber\\
 &&\,\,\,\,\,\,\,{1\over \delta}[1+\delta L(t,T)]\times\nonumber\\
 &&\,\,\,\,\,\,\,\sum_{i=1}^n
 \left[d{\widetilde W}^i_t\int_T^{T+\delta}\sigma^i(t,u)du -
 \Sigma^i(t,T+\delta)
 \left(\int_T^{T+\delta}\sigma^i(t,u)du\right) dt
 \right]=
 \nonumber\\
 &&\,\,\,\,\,\,\,L(t,T)\sum_{i=1}^n \gamma^i(t,T)\left[d{\widetilde W}^i_t -
 \Sigma^i(t,{T+\delta})dt\right]~.
\end{eqnarray}
This implies that under the $(T+\delta)$-forward measure $L(t,T)$ is a
martingale:
\begin{equation}
 d_tL(t,T)=L(t,T)\sum_{i=1}^n\gamma^i(t,T)d{\widehat W}^i_t~.
\end{equation}
Moreover, it is log-normally distributed under ${\bf P}_{T+\delta}$.
This enables us to price certain options.

{}Suppose a payment at time $T_{i+1}$ depends on the instantaneous {\small LIBOR} rate
at time $T_i$: $X_{T_{i+1}}=f(L(T_i))$. Then the value of the payment at time
$t$ is given by:
\begin{equation}
 V_t=P(t,T_{i+1})\langle f(L(T_i))\rangle_{{\bf P}_{T_{i+1}},{\cal F}_t}~.
\end{equation}
As an example consider a caplet payoff $f(L(T_{i-1}))=\delta(L(T_{i-1})-k)^+$.
Then we have the following Black-Scholes-like formula for its price:
\begin{eqnarray}
 &&V_t = \delta
 P(t,T_i)\left[L(t,T_{i-1})\Phi\left({\ln\left({L(t,T_{i-1})\over k}\right)
 \over
 \sqrt{\zeta(t,T_{i-1})}}+{\sqrt{\zeta(t,T_{i-1})}\over 2}\right)\right.\nonumber\\
 &&\,\,\,\,\,\,\,\left.-k\Phi\left({\ln\left({L(t,T_{i-1})\over k}\right)
 \over
 \sqrt{\zeta(t,T_{i-1})}}-{\sqrt{\zeta(t,T_{i-1})}\over 2}\right)\right]~,
\end{eqnarray}
where
\begin{equation}
 \zeta(t,T)\equiv\sum_{j=1}^n \int_t^T\left[\gamma^j(s,T)\right]^2 ds
\end{equation}
is the log-variance of $L(T,T)$ given ${\cal F}_t$.

\section{Foreign Currency Interest-rate Models}

{}Suppose we have a dollar zero-coupon bond $P(t,T)$ as well as the dollar
cash bond $B_t$, a sterling zero-coupon bond $Q(t,T)$ as well as the sterling
cash bond $D_t$, and the exchange rate $C_t$, which is the value in dollars
of one pound.

{}A multi-factor model for this market is given by:
\begin{eqnarray}
 &&d_t f(t,T)=\sum_{i=1}^n \sigma^i(t,T)dW^i_t+\alpha(t,T)dt~,\\
 &&d_t g(t,T)=\sum_{i=1}^n \tau^i(t,T)dW^i_t+\beta(t,T) dt~,\\
 &&dC_t=C_t\left[\sum_{i=1}^n\rho^i_t dW^i_t +\lambda_t dt\right]~,
\end{eqnarray}
where $f(t,T)$ and $g(t,T)$ are the dollar respectively sterling
forward rates. The dollar tradable securities in this market consist of
$B_t$, $P(t,T)$, $C_t Q(t,T)$, and $C_t D_t$. To price various derivative
securities, we must make the discounted versions of these processes into
martingales under a single measure ${\bf Q}$. This, as usual,
gives certain constraints on the drifts.

\section{Quantos}

{}Consider the following {\em quanto} model. The sterling stock price $S_t$ and the
value of one pound in dollars $C_t$ follow the processes:
\begin{eqnarray}
 &&S_t=S_0\exp\left(\sigma_1 W_1(t)+\mu t\right)~,\\
 &&C_t=C_0\exp\left(\rho\sigma_2 W_1(t)+\sqrt{1-\rho^2}\sigma_2W_2(t)+\nu t
 \right)~.
\end{eqnarray}
In addition we have a dollar cash bond $B_t=\exp(rt)$ and a sterling cash
bond $D_t=\exp(ut)$.

{}The dollar tradables are $B_t$, $C_tD_t$ and $C_tS_t$. The discounted
processes for the last two tradables are $Y_t=B_t^{-1}C_t D_t$ and
$Z_t=B_t^{-1}C_t S_t$:
\begin{eqnarray}
 &&Y_t=C_0\exp\left(\rho\sigma_2 W_1(t)+\sqrt{1-\rho^2}\sigma_2W_2(t)+
 [\nu+u-r]t\right),\\
 &&Z_t=C_0 S_0\exp(\left([\sigma_1+\rho\sigma_2]W_1(t)+
 \sqrt{1-\rho^2}\sigma_2 W_2(t)+[\mu+u-r]t\right).
\end{eqnarray}
Under the measure ${\bf Q}$ that makes both of
these into martingales we have:
\begin{eqnarray}
 &&Y_t=C_0\exp\left(\rho\sigma_2 {\widetilde W}_1(t)+
 \sqrt{1-\rho^2}\sigma_2{\widetilde W}_2(t)-
 {1\over 2}\sigma_2^2 t\right)~,\\
 &&Z_t=C_0 S_0\exp(\left([\sigma_1+\rho\sigma_2]{\widetilde W}_1(t)+
 \sqrt{1-\rho^2}\sigma_2 {\widetilde W}_2(t)-\right.\nonumber\\
 &&\,\,\,\,\,\,\,\left.{1\over 2}
 \left[\sigma_1^2+\sigma_2^2+2\sigma_1\sigma_2\rho\right]t\right)~.
\end{eqnarray}
We, therefore, have
\begin{eqnarray}
 &&C_t=C_0\exp\left(\rho\sigma_2 {\widetilde W}_1(t)+
 \sqrt{1-\rho^2}\sigma_2{\widetilde W}_2(t)+
 \left[r-u-{1\over 2}\sigma_2^2\right] t\right)~,\\
 &&S_t=S_0\exp\left(\sigma_1{\widetilde W}_1(t)+
 \left[u-{1\over 2}\sigma_1^2-\sigma_1\sigma_2\rho\right]t\right)~.
\end{eqnarray}
In quanto contracts the stock price is quoted in the ``wrong'' currency, in
this case in dollars. We can then price quanto contracts as follows.

\subsection{A Forward Quanto Contract}

{}Consider a forward quanto contract for buying the stock at time $T$ for
a pre-agreed dollar amount $k$. At time $t=0$ the price of this contract is
given by
\begin{equation}
 V_0=e^{-rT}\langle (S_T-k)\rangle_{\bf Q}=e^{-rT}(F_Q-k)~,
\end{equation}
where
\begin{equation}
 F_Q\equiv F\exp(-\sigma_1\sigma_2\rho T)~,
\end{equation}
and $F\equiv S_0\exp(uT)$ is the forward price in the local currency (that
is, in pounds). To give the forward quanto contract zero initial value, we
must set $k=F_Q$, which is the forward quanto price.

{}All other quanto contracts now have a familiar Black-Scholes form with the
forward price given by the forward quanto price $F_Q$.

\section{Optimal Hedge Ratio}

{}Suppose we are hedging an asset using a futures contract. Let $S$ be the
spot price of the asset, and let $F$ be the price of the futures contract. The
{\em basis} is defined as $b=S-F$.

{}Let $\Delta S$ and $\Delta F$ be the change in spot respectively futures
price during the life of the hedge. Also, let $\sigma_S$, $\sigma_F$ and
$\rho$ be the standard deviation of $S$, the standard deviation of $F$ and
the correlation between $S$ and $F$, respectively. Finally, let $h$ be the
hedge ratio (the size of the position taken in the futures contracts to the
size of the exposure).

{}The change in the value of the hedger's position during the life of the
hedge is
\begin{equation}
 \pm(\Delta S-h\Delta F)~,
\end{equation}
where plus stands for the position long in the asset and short in the futures,
while minus stands for the position short in the asset and long in the futures.
In either case the variance is
\begin{equation}
 v=\sigma_S^2+h^2\sigma_F^2-2h\sigma_S\sigma_F\rho~.
\end{equation}
Minimizing this expression gives
\begin{equation}
 h=\rho ~{\sigma_S\over\sigma_F}~,
\end{equation}
which gives the optimal hedge ratio.

{}Typically in the hedging strategy the futures contract has delivery date
close to but later than the expiration of the hedge. It has to be close so that
the basis risk is minimized. It is usually chosen later so that the erratic
nature of the futures prices during the delivery months does not affect the
hedge. If at a given time there is no liquid futures contract that matures
later than the expiration of the hedge, one can use the strategy of rolling the
hedge forward. This strategy works well if there is a close correlation between
changes in the futures prices and the changes in the spot prices.

\section*{Acknowledgments}\addcontentsline{toc}{section}{Acknowledgments}

{}I would like to thank everyone who took the course back in 2002 and made it a success. I am especially grateful to my then Ph.D. students at the C.N. Yang Institute for Theoretical Physics Olindo Corradini, Alberto Iglesias and Peter Langfelder for their enthusiastic participation in the course. I am indebted to Yan Vtorov, among so many other things, for introducing me to Baxter and Rennie's book, which inspired me to give this course.

\appendix

\section{Some Fun Questions}

{}{\bf Question 1.} Two ropes burn inhomogeneously (different lengths, thicknesses), each in 1 hour. You need to measure 45 minutes. How?

{}{\bf Answer.} Light both ends of rope A and one end of rope B. Rope A will burn out in exactly 30 minutes. At that time light the second end of rope B. When it burns out, that's the 45 minute mark.

\bigskip

{}{\bf Question 2.} You have two jars, 5 liters (jar A) and 3 liters (jar B). How do you pour 4 liters of water into jar A?

{}{\bf Answer.} The following sequence does the trick:\\
Jar A: 5 2 2 0 5 4\\
Jar B: 0 3 0 2 2 3

\section{Quiz 1}

{}{\bf Problem 1.} If a family has two children and there is a boy in the
family, what is the probability that there is a girl?

{}{\bf Answer.} 2/3.

{}{\bf Solution.} Let ${\bf B}$ stand for a boy, while ${\bf G}$ stand for a
girl. Then in a family with two children {\em a priori} we have the following
{\em four} possibilities:
\begin{eqnarray}
 &&{\bf B}~~~{\bf B}~,\\
 &&{\bf B}~~~{\bf G}~,\\
 &&{\bf G}~~~{\bf B}~,\\
 &&{\bf G}~~~{\bf G}~.
\end{eqnarray}
Since we know that in the aforementioned family there is a boy, the last of
the above four possibilities cannot be the case. This leaves us with the
first 3 possibilities, among which we have 2 possibilities that there is a girl
in this family. Thus, the probability that there is a girl in the family is
\begin{equation}
 P_{\rm{\scriptstyle{cond}}}={2\over 3}~.
\end{equation}
This is an example of a conditional probability, which differs from the naive
probability $P=1/2$.

\bigskip

{}{\bf Problem 2.} If you have two stocks and they both have the same expected
return, but one has volatility 20\% and the other has volatility of 30\%, and
they have a 50\% correlation, how should I allocate a fixed sum of money
between the two stocks so as to minimize my risk?

{}{\bf Answer.} 6/7 in the first stock, 1/7 in the second stock.

{}{\bf Solution.} Since the expected returns for the two stocks are the same,
we assume that they have the same drift. Then the risk for
a portfolio containing these stocks in some proportion is minimized by
minimizing the volatility of the portfolio. Thus, let the portfolio
contain $X$ amount of stock 1 and $1-X$ amount of stock 2, where
$0\leq X\leq 1$ is the fraction of stock 1 in the portfolio. Then the
volatility of the portfolio is
\begin{equation}
 \sigma^2=X^2\sigma_1^2+(1-X)^2\sigma_2^2+2X(1-X)\sigma_1\sigma_2\rho~,
\end{equation}
where $\sigma_1=20\%$ is the volatility of stock 1, $\sigma_2=30\%$ is the
volatility of stock 2, and $\rho=50\%$ is the correlation. Minimizing
$\sigma^2$ we obtain:
\begin{equation}
 X={{\sigma_2^2-\rho\sigma_1\sigma_2}\over {\sigma_1^2+\sigma_2^2-2\rho
 \sigma_1\sigma_2}}~.
\end{equation}
This gives $X=6/7$. We should therefore invest $6/7$ of our money into stock 1,
and $1/7$ into stock 2.

\bigskip

{}{\bf Problem 3.} Suppose there is an infinite straight beach and there is a
lighthouse 1 mile offshore. The light rotates at 1 revolution per minute. How
fast is the image of the beam on the beach, {\em i.e.} the ``white dot'',
moving along the beach when that white dot is exactly 3 miles from the
lighthouse?

{} {\bf Answer.} Approximately 56.5 miles/min.

{}{\bf Solution.} To solve this problem, it is useful to visualize it via
Fig.1. The horizontal line is the beach, the center of the circle is the
lighthouse, the vertical distance $Y=1~{\rm mile}$, while the radius of the
circle is $R=3~{\rm miles}$. The position $X$ of the white dot
along the beach is given by (the origin of the $X$-axis, which is directed from
left to right, is chosen so that
when the beam is perpendicular to the beach $X=0$)
\begin{equation}
 X=Y \tan(\theta)~.
\end{equation}

\begin{figure}
\centerline{\epsfxsize 11.truecm \epsfysize 11.truecm\epsfbox{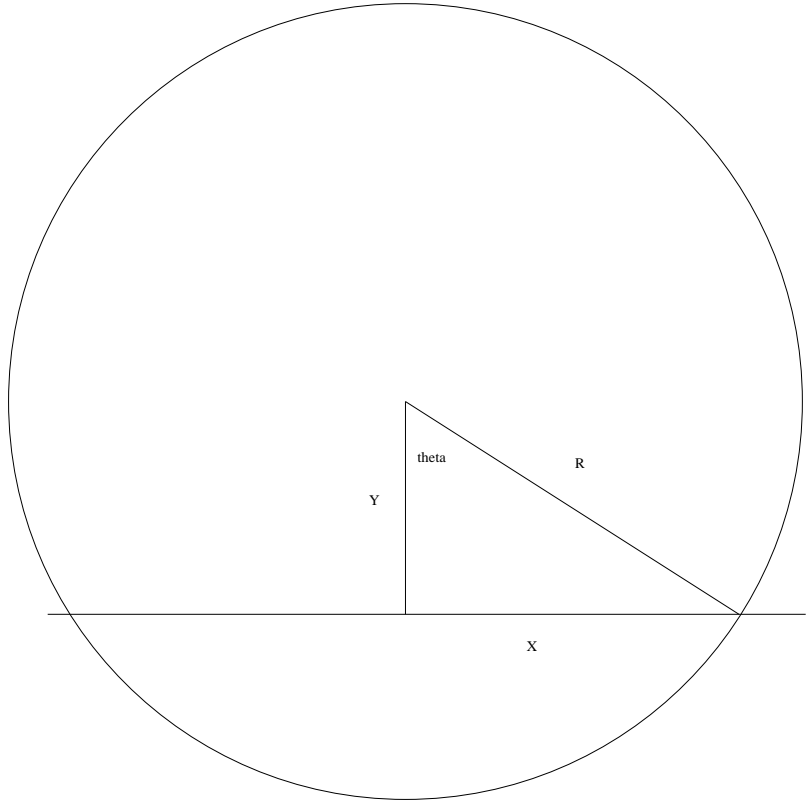}}
\caption[Figure for Problem 3 in Quiz 1]{Figure for Problem 3 in Quiz 1. The light emanating from the lighthouse is
for definiteness assumed to be rotating counterclockwise.}
\end{figure}

The angle theta is given by ($t=0$ corresponds to $X=0$, and we are working
within the first quarter of the period $T=1~{\rm min.}$, that is, within
the first 15 seconds)
\begin{equation}
 \theta=\omega t~,
\end{equation}
where $\omega=2\pi f=2\pi~{\rm rev./min}$ is the angular velocity,
and $f=1/T=1~{\rm rev./min}$ is the frequency
of the circular motion of the light beam
emanating from the lighthouse. Now, the velocity of the white dot along the
beach in the $X$-direction is given by
\begin{equation}
 V_X={dX\over dt}={Y\omega\over\cos^2(\theta)}=
 {R^2\omega\over Y}=2\pi f {R^2\over Y}\approx 56.5~{\rm miles/min.}
\end{equation}
In the last line we have used the fact that $\cos(\theta)=Y/R$.

\bigskip

{}{\bf Problem 4.} Suppose $X$ is a normal random variable with mean 0 and
variance $v$, what is the expected value of $e^X$?

{}{\bf Answer.} $\exp\left({v\over 2}\right)$.

{}{\bf Solution.} The probability distribution for the variable $X$ is given
by:
\begin{equation}
 P(X)={1\over \sqrt{2\pi v} }\exp\left(-{X^2\over 2v}\right)~.
\end{equation}
Note that
\begin{eqnarray}
 &&\int_{-\infty}^{\infty} dX~P(X)=1~,\\
 &&\langle X^2\rangle\equiv\int_{-\infty}^{\infty} dX~X^2~P(X)=v~.
\end{eqnarray}
Now, the expected value of $e^X$ is
\begin{eqnarray}
 \langle e^X\rangle &\equiv&\int_{-\infty}^{\infty} dX~e^X~P(X)=\nonumber\\
 &=&{1\over \sqrt{2\pi v}}\int_{-\infty}^{\infty} dX~\exp\left(
 X-{X^2\over 2v}\right)=\nonumber\\
 &=&\exp\left({v\over 2}\right)
 {1\over \sqrt{2\pi v} }\int_{-\infty}^{\infty} dX~\exp\left(
 -\left[{X\over \sqrt{2v}}-\sqrt{v\over 2}\right]^2\right)=\nonumber\\
 &=&\exp\left({v\over 2}\right)
 {1\over \sqrt{2\pi v} }\int_{-\infty}^{\infty} dY~\exp\left(
 -{Y^2\over 2v}\right)=\nonumber\\
 &=&\exp\left({v\over 2}\right)~.
\end{eqnarray}
In the last line we have used the following change of variables:
$Y\equiv X-v$.

\bigskip

{}{\bf Problem 5.} What is the integral of $\sec(x)$ from $x=0$ to $x=\pi/6$?

{}{\bf Answer.} ${1\over 2}\ln(3)\approx .55$.

{\bf Solution.} This integral is computed in the following standard way:
\begin{eqnarray}
 &&\int_0^{\pi/6} \sec(x) dx=\int_0^{\pi/6} {dx\over \cos(x)}=
 \int_0^{\pi/6} {\cos(x)dx\over \cos^2(x)}=
 \int_0^{\pi/6} {d\sin(x)\over {1-\sin^2(x)}}=\nonumber\\
 &&\,\,\,\,\,\,\,{1\over 2} \int_0^{\pi/6}d\sin(x)\left[{1\over{1-\sin(x)}}+
 {1\over{1+\sin(x)}}\right]=\nonumber\\
 &&\,\,\,\,\,\,\,{1\over 2}\left.\left[-\ln(1-\sin(x))+\ln(1+\sin(x))\right]
 \right|_0^{\pi/6}={1\over 2}\left.\ln{{1+\sin(x)}\over{1-\sin(x)}}
 \right|_0^{\pi/6}=\nonumber\\
 &&\,\,\,\,\,\,\,{1\over 2}\ln(3)\approx .55~.
\end{eqnarray}

\bigskip

{}{\bf Problem 6.} If you are solving a parabolic partial differential equation
by using the explicit finite difference method, is it worse to have too many
time steps or too fine a grid in the space dimension?

{}{\bf Answer.} For stability of the algorithm it is worse to have too fine
a grid in the space dimension.

{}{\bf Solution.} For definiteness let us consider the simplest example of
a parabolic PDE, the diffusion equation in one space dimension with a constant
diffusion coefficient $D>0$:
\begin{equation}
 {\partial u\over \partial t}=D{\partial^2 u\over \partial x^2}~.
\end{equation}
Let us consider the
FTCS (Forward Time Centered Space) representation, which is an
explicit finite difference scheme:
\begin{equation}\label{diff}
 {{u_j^{n+1}-u_j^n}\over \Delta t}=D\left[{{u^n_{j+1}-2u^n_j+u^n_{j-1}}\over
 (\Delta x)^2}\right]~.
\end{equation}
Here the subscript $j$ corresponds to the discretized $x$ coordinate, while
the superscript $n$ corresponds to the discretized time $t$.

{}An important point in solving such equations numerically, as is generally
the case when solving initial value (Cauchy) problems, is stability of the
algorithm. Here the von Neumann stability analysis is particularly convenient.
Thus, we look for eigenmodes (of the difference equation) of the form:
\begin{equation}\label{vonN}
 u^n_j=(\xi)^n e^{ikj\Delta x}~,
\end{equation}
where $k$ is the wave number, and $\xi=\xi(k)$, which is called the
amplification factor, is a complex number. The
difference equation is unstable, in particular, it has exponentially growing
modes, if $|\xi(k)|>1$ for some $k$.

{}In the case of the diffusion equation (\ref{diff}) we have the following
solution for the amplification factor in (\ref{vonN}):
\begin{equation}
 \xi=1-{4D\Delta t\over (\Delta x)^2}\sin^2\left({k\Delta x\over 2}\right)~.
\end{equation}
The stability requirement $|\xi|\leq 1$ then implies the following condition:
\begin{equation}\label{cond}
 {2D\Delta t\over (\Delta x)^2}\leq 1~.
\end{equation}
An intuitive interpretation of this restriction is clear: the maximum allowed
time step $\Delta t$ (up to a numerical factor of order 1) is the diffusion
time across a cell of width $\Delta x$.

{}Thus, as we see, for a given size of the spatial grid there is
a minimum allowed number of time steps, and if we, say, decrease the former
by a factor of 10, then the latter must be increased by a factor of 100.
So, for the stability of the algorithm it appears to be worse to have
too fine a grid in the spatial dimension.

{}However, in practice the conclusions one might draw from the above discussion
in general are not particularly useful. The point is that usually we are
interested in modeling accurately the evolution of features with spatial scales
$L\gg \Delta x$. The diffusion time across a spatial scale of size $L$ is
of order
\begin{equation}
 T\sim {L^2\over D}~.
\end{equation}
If we are limited to time steps satisfying (\ref{cond}), we will need to evolve
through of order $L^2/(\Delta x)^2$ steps
before interesting things start to happen
on the scale $L$. This number of steps, however, is usually too large
(prohibitive). This is why in practice one usually appeals to either fully
({\em e.g.}, backward time) or partially implicit ({\em e.g.}, Crank-Nicholson)
schemes that do not suffer from severe stability restrictions such as
(\ref{cond}).

\bigskip

{}{\bf Problem 7.} Suppose 2 teams play a series of up to 7 games in which the
first team to win 4 games wins the series and then no other games are played.
Suppose that you want to bet on each individual game in such a way that when
the series ends you will be ahead \$100 if your team wins the series, or behind
by exactly \$100 if your team loses the series, no matter how many games it
takes. How much would you bet on the first game?

{}{\bf Answer.} $\${125\over 4}= 31$ dollars and 25 cents.

{}{\bf Solution.} To solve this problem we can draw a binary tree and work
backwards. There are two observations that simplify the analysis. Thus, let us
put ``$+$'' if our team wins, and put ``$-$'' if our team loses. Then
if we have a slot with 3 $+$'s and 3 $-$'s, no matter in what order, there is
one more game to be played, which is deciding for the series. It is then clear
that before that game, that is, after the sixth game, we must break even, and
on the seventh game we must bet \$100. Also, we can restrict our attention
to only a half of the binary tree, say, the half that corresponds to our team
winning the first game -- indeed, the other half is the same as this half up
to exchanging $+$'s and $-$'s.

\section{Quiz 2}

{}{\bf Problem 1.} What is the expected minimum number of coin tosses you
would need to make in order to get 3 heads in a row?

{}{\bf Answer.} 14.

{}{\bf Solution.} For any finite number of coin tosses there is a finite
probability that we do {\em not} get 3 heads in a row. Therefore, there is
no finite minimum number of coin tosses that would guarantee 3 heads in a row.

{}In fact, the probability $P(N)$ that we do get 3 heads in a row grows
with the number of coin tosses $N$. In particular, $P(N)\rightarrow 1$ as
$N\rightarrow\infty$, so that the set $\{P(N)\}$ is {\em not} a measure
for defining an average number of coin tosses we need to make to get 3 heads
in a row. However, as we will see below, we can define a {\em conditional}
probability ${\widetilde P}(N)$ such that $\{{\widetilde P}(N)\}$ is an
appropriate measure. To do this, let us first study some properties of
the probabilities $P(N)$.

{}The probability $P(N)$ can be determined as follows. Let $Q(N)\equiv 1-P(N)$
(this is the probability that we do not get 3 heads in a row). Then
we have $P(0)=P(1)=P(2)=0$, $Q(0)=Q(1)=Q(2)=1$.
For $N\geq 3$ we have non-zero $P(N)$.
Thus, for instance, $P(3)=1/8$, and $Q(3)=7/8$. It is then not difficult to see
that
\begin{equation}
 P(N)={1\over 8}\left[1+{1\over 2}\sum_{n=0}^{N-4} Q(n)\right]~,~~~N\geq 4~.
\end{equation}
Equivalently, we have
\begin{equation}\label{Q(N)}
 Q(N)={1\over 16}\left[14-\sum_{n=0}^{N-4} Q(n)\right]~,~~~N\geq 4~.
\end{equation}
Note that
\begin{equation}\label{sumQ}
 \sum_{n=0}^\infty Q(n)=14~,
\end{equation}
so that $Q(N)\rightarrow 0$ and $P(N)\rightarrow 1$ as $N\rightarrow \infty$.

{}Next, we define the {\em conditional} probability ${\widetilde P}(N)$ as
the probability of getting 3 heads in a row with $N$ coin tosses such that
we do {\em not} get 3 heads in a row until the last (that is, $N$th)
coin toss. It is not difficult to see that
\begin{eqnarray}
 &&{\widetilde P}(0)={\widetilde P}(1)={\widetilde P}(2)=0~,\label{P012}\\
 &&{\widetilde P}(3)={1\over 8}~,\\
 &&{\widetilde P}(N)={1\over 16}Q(N-4)~,~~~N\geq 4.
\end{eqnarray}
Note that ${\widetilde P}(N)\rightarrow 0$ as $N\rightarrow\infty$. In fact,
using (\ref{sumQ}) we have
\begin{equation}
 \sum_{N=0}^\infty {\widetilde P}(N)=\sum_{N=3}^\infty {\widetilde P}(N)=
 {1\over 8}+{1\over 16}\sum_{n=0}^\infty Q(n)=1~.
\end{equation}
Thus, the set ${\widetilde {\bf P}}\equiv\{{\widetilde P}(N)\}$ is an
appropriate measure for computing an average number of coin tosses.

{}This average number is defined as
\begin{equation}\label{N*0}
 N_*\equiv\langle N\rangle_{\widetilde{\bf P}}\equiv
 \sum_{N=0}^\infty N{\widetilde P}(N)=\sum_{N=3}^\infty N{\widetilde P}(N)~,
\end{equation}
where we took into account (\ref{P012}).

{}We can rewrite (\ref{N*0}) as follows:
\begin{eqnarray}
 N_*&=&{3\over 8}+\sum_{N=4}^\infty N{\widetilde P}(N)=\nonumber\\
    &=&{3\over 8}+{1\over 16} \sum_{N=4}^\infty NQ(N-4)=\nonumber\\
    &=&{3\over 8}+{1\over 16}\left[\sum_{N=4}^\infty (N-4)Q(N-4)+
        4\sum_{N=4}^\infty Q(N-4)\right]=\nonumber\\
    &=&{3\over 8}+{1\over 16}\left[\sum_{n=0}^\infty nQ(n)+
     4\sum_{n=0}^\infty Q(n)\right]=\nonumber\\
    \label{N*}
    &=&{31\over 8}+{1\over 16}\sum_{n=0}^\infty nQ(n)~.
\end{eqnarray}
Next, we can compute the last term in the last line above as follows. From
(\ref{Q(N)}) it follows that
\begin{equation}
 Q(n)=Q(n-1)-{1\over 16}Q(n-4)~,~~~n\geq 4~.
\end{equation}
Using this formula, we obtain:
\begin{eqnarray}
 &&\sum_{n=0}^\infty nQ(n)=Q(1)+2Q(2)+3Q(3)+\sum_{n=4}^\infty nQ(n)=\nonumber\\
  &&\,\,\,\,\,\,\,=Q(1)+2Q(2)+3Q(3)+\sum_{n=4}^\infty nQ(n-1)-{1\over 16}\sum_{n=4}^\infty
  nQ(n-4)=\nonumber\\
  &&\,\,\,\,\,\,\,=Q(1)+2Q(2)+3Q(3)+\sum_{n=4}^\infty (n-1)Q(n-1)+
  \sum_{n=4}^\infty Q(n-1)-\nonumber\\
  &&\,\,\,\,\,\,\,-{1\over 16}\sum_{n=4}^\infty (n-4)Q(n-4)-{1\over 4}\sum_{n=4}^\infty
  Q(n-4)=\nonumber\\
  &&\,\,\,\,\,\,\,=3Q(3)-Q(0)-Q(1)-Q(2)+{15\over 16}\sum_{n=0}^\infty nQ(n) +{3\over 4}
  \sum_{n=0}^\infty Q(n)~.
\end{eqnarray}
This implies that
\begin{equation}
 {1\over 16}\sum_{n=0}^\infty nQ(n)=3Q(3)-Q(0)-Q(1)-Q(2)+{3\over 4}
 \sum_{n=0}^\infty Q(n)={81\over 8}~.
\end{equation}
Plugging this into (\ref{N*}), we finally obtain:
\begin{equation}
 N_*=14~.
\end{equation}

{}{\bf Note.} This average number $N_*$ is the same as the sum
\begin{equation}
 \sum_{n=0}^\infty Q(n)~,
\end{equation}
which is not a coincidence.

\bigskip

{}{\bf Problem 2.} Suppose that $x$ is a Brownian motion with no drift and
unit variance, {\em i.e.} $dx=dz$. If $x$ starts at 0, what is the probability
that $x$ hits 3 before hitting $-5$?

{}{\bf Answer.} $5/8$.

{}{\bf Solution.} Let $P(x_0;x_1;x_2)$ denote the probability that starting
at $x_0$ the Brownian motion $x$ hits $x_1$ before it hits $x_2$, where
$x_1\not=x_2$. By definition, $P(x_0;x_0;x_2)=1$, and $P(x_0;x_1;x_0)=0$.
Clearly, we have
\begin{equation}\label{P}
 P(x_0;x_1;x_2)+P(x_0;x_2;x_1)=1~.
\end{equation}
We need to determine $P(0;3;-5)$. According to (\ref{P}), we have
\begin{equation}
 P(0;3;-5)=1-P(0;-5;3)~.
\end{equation}
Here $P(0;-5;3)$ is the probability that starting at $0$ $x$ hits $-5$ before
it hits 3. Since Brownian motion is continuous, to hit $-5$
$x$ must first hit $-3$, so we have
\begin{equation}
 P(0;-5;3)=P(0;-3;3)P(-3;-5;3)~.
\end{equation}
Note that, due to the symmetry under $x\rightarrow -x$, we have
\begin{equation}
 P(0;-3;3)=P(0;3;-3)={1\over 2}~.
\end{equation}
On the other hand, since Brownian motion is independent of the previous
history, we have
\begin{equation}
 P(-3;-5;3)=P(0;-2;6)~,
\end{equation}
so that
\begin{equation}
 P(0;-5;3)={1\over 2}P(0;-2;6)~.
\end{equation}
Now we can use the above trick repeatedly until we obtain a desired result.
Thus, we have:
\begin{eqnarray}
 &&P(0;-2;6)=1-P(0;6;-2)~,\\
 &&P(0;6;-2)=P(0;2;-2)P(2;6;-2)={1\over 2}P(2;6;-2)~,\\
 &&P(2;6;-2)=P(0;4;-4)={1\over 2}~,\\
 &&P(0;6;-2)={1\over 4}~,\\
 &&P(0;-2;6)={3\over 4}~.
\end{eqnarray}
Thus, we have
\begin{equation}
 P(0;-5;3)={3\over 8}~,
\end{equation}
and
\begin{equation}
 P(0;3;-5)={5\over 8}~.
\end{equation}
So the probability that starting at 0 $x$ hits 3 before it hits $-5$ is
$5/8$.

{}{\bf Note.} This result is independent of the variance $v$ of $x$,
which is not
surprising as the actual variance of the corresponding probability
distribution at time $t$ is $vt$, and the answer to the
question stated in this problem cannot possibly involve any time interval.
Another way of stating this is that the variance $v$ is a dimensionful
quantity (it has dimension of inverse time assuming that $x$ is dimensionless),
so it cannot enter into a dimensionless quantity such as probability since
there are no other dimensionful quantities in this problem.

\bigskip

{}{\bf Problem 2a.} In Problem 2, what if the drift is $m$, {\em i.e.} $dx=
m~dt+dz$?

{}{\bf Answer.} The probability that starting at 0 $x$ hits 3 before it hits
$-5$ in this case equals
\begin{equation}
 {{\exp(9m)+\exp(5m)+\exp(m)+2\cosh(3m)}\over 8\cosh(2m)\cosh(3m)
 \cosh(4m)}~.
\end{equation}

{}{\bf Solution.} Note that now we have two dimensionful quantities, namely,
the variance $v$ and the drift $m$. Out of these we can form the following
dimensionless combination (assuming that $x$ is dimensionless):
$m/v$. This can now enter non-trivially into
various probabilities.

{}In the presence of the drift $m$ our discussion in Problem 2 is modified
as follows. Note that in Problem 2 we used the fact that a path $x(t)$ with
$x(0)=0$ and $x(T)=x_T$ was as probable as the path $-x(t)$. This, in
particular, implied that $P(0;x_1;-x_1)=P(0;-x_1;x_1)=1/2$. To avoid
confusion, in the presence of the drift $m$ we will denote all probabilities
via $Q$ instead of $P$. Then we have (here we are taking into account that
the variance of $x$ $v=1$ in the appropriate units of time)
\begin{equation}\label{Qr}
 Q(0;x_1;-x_1)=\exp(2mx_1)Q(0;-x_1;x_1)~.
\end{equation}
This can be seen by using the continuous version of the Radon-Nikodym
derivative and the Cameron-Martin-Girsanov theorem. Thus, we have
\begin{equation}\label{Qrr}
 Q(0;x_1;-x_1)={1\over{1+\exp(-2mx_1)}}={\exp(mx_1)\over2\cosh(mx_1)}~.
\end{equation}
All the other probabilities will reduce to probabilities of this type.

{}Before we obtain $Q(0;3;-5)$, we would like to give a simple derivation
of (\ref{Qr}) and (\ref{Qrr}). Thus, let us assume that $x$ has dimension of
length. Then the variance $v$ of $x$ has dimension
${\rm length}^2/{\rm time}$, while
the drift $m$ has dimension of ${\rm length}/{\rm time}$, so that the ratio
$m/v$ has dimension of $1/{\rm length}$. Just on dimensional grounds it is then
clear that the ratio
\begin{equation}\label{f}
 {Q(0;x_1;-x_1)\over Q(0;-x_1;x_1)}=f\left({mx_1\over v}\right)~,
\end{equation}
where $f(y)$ is a dimensionless function of a dimensionless variable $y$.
In the following we will set $v=1$ (in the appropriate units).

{}Now, from (\ref{f}) it follows that $f(-mx_1)=1/f(mx_1)$, that is,
$f(-y)=1/f(y)$. This implies that
\begin{equation}
 f(y)=\exp\left[g(y)\right]~,
\end{equation}
where $g(y)$ is an odd function of $y$: $g(-y)=-g(y)$.

{}To further constrain $g(y)$, consider the following trick. Thus, we have:
\begin{eqnarray}
 &&Q(0;2x_1;-2x_1)=Q(0;x_1;-2x_1)Q(x_1;2x_1;-2x_1)=\nonumber\\
 &&\,\,\,\,\,\,\,Q(0;x_1;-2x_1)
 Q(0;x_1;-3x_1)~,\\
 &&Q(0;x_1;-3x_1)=1-Q(0;-3x_1;x_1)~,\\
 &&Q(0;-3x_1;x_1)=Q(0;-x_1;x_1)Q(-x_1;-3x_1;x_1)=\nonumber\\
 &&\,\,\,\,\,\,\,Q(0;-x_1;x_1)
 Q(0;-2x_1;2x_1)~,\\
 &&Q(0;-2x_1;2x_1)=1-Q(0;2x_1;-2x_1)~.
\end{eqnarray}
Putting all of this together, we obtain:
\begin{equation}
 Q(0;2x_1;-2x_1)=Q(0;x_1;-2x_1)\left\{1-Q(0;-x_1;x_1)\left[1-Q(0;2x_1;-2x_1
 \right]\right\}~,
\end{equation}
that is,
\begin{eqnarray}
 Q(0;2x_1;-2x_1)&=&{Q(0;x_1;-2x_1)\left[1-Q(0;-x_1;x_1)\right]\over
 {1-Q(0;x_1;-2x_1)Q(0;-x_1;x_1)}}=\nonumber\\
 \label{Q22}
 &=&{Q(0;x_1;-2x_1)Q(0;x_1;-x_1)\over
 {1-Q(0;x_1;-2x_1)Q(0;-x_1;x_1)}}~.
\end{eqnarray}
This expression can be further reduced using the following trick:
\begin{eqnarray}
 &&Q(0;x_1;-2x_1)=1-Q(0;-2x_1;x_1)~,\\
 &&Q(0;-2x_1;x_1)=Q(0;-x_1;x_1)Q(-x_1;-2x_1;x_1)= \nonumber\\
 &&\,\,\,\,\,\,\,Q(0;-x_1;x_1)Q(0;-x_1;2x_1)~,\\
 &&Q(0;-x_1;2x_1)=1-Q(0;2x_1;-x_1)~,\\
 &&Q(0;2x_1;-x_1)=Q(0;x_1;-x_1)Q(x_1;2x_1;-x_1)=\nonumber\\
 &&\,\,\,\,\,\,\,Q(0;x_1;-x_1)Q(0;x_1;-2x_1)~.
\end{eqnarray}
Putting all of this together, we obtain:
\begin{equation}
 Q(0;x_1;-2x_1)=1-Q(0;-x_1;x_1)\left[1-Q(0;x_1;-x_1)Q(0;x_1;-2x_1)\right]~,
\end{equation}
that is,
\begin{eqnarray}
 Q(0;x_1;-2x_1)&=&{{1-Q(0;-x_1;x_1)}\over {1-Q(0;-x_1;x_1)Q(0;x_1;-x_1)}}=
 \nonumber\\
 &=&{Q(0;x_1;-x_1)\over {1-Q(0;x_1;-x_1)Q(0;-x_1;x_1)}}~.
\end{eqnarray}
Plugging this into (\ref{Q22}), we obtain:
\begin{equation}
 Q(0;2x_1;-2x_1)={\left[Q(0;x_1;-x_1)\right]^2\over {1-2Q(0;x_1;-x_1)
 Q(0;-x_1;x_1)}}~.
\end{equation}
{}From this expression it immediately follows that
\begin{equation}
 {Q(0;2x_1;-2x_1)\over Q(0;-2x_1;2x_1)}=
 \left[{Q(0;x_1;-x_1)\over Q(0;-x_1;x_1)}\right]^2~.
\end{equation}
This then implies that the function $f(y)$ has the following property:
\begin{equation}
 f(2y)=\left[f(y)\right]^2~,
\end{equation}
that is,
\begin{equation}
 g(2y)=2g(y)~.
\end{equation}
In fact, the function $g(y)$ has the property that for an arbitrary real
number $\lambda$
\begin{equation}
 g(\lambda y)=\lambda g(y)~,
\end{equation}
that is, $g(y)$ is a homogeneous linear function of $y$:
\begin{eqnarray}
 &&g(y)=\kappa y~,\\
 &&f(y)=\exp(\kappa y)~,
\end{eqnarray}
where $\kappa$ is a coefficient which still needs to be fixed.

{}Finally, let us fix $\kappa$. We have:
\begin{eqnarray}
 &&Q(0;x_1;-x_1)=\exp(\kappa m x_1)Q(0;-x_1;x_1)~,\\
 \label{Q55}
 &&Q(0;x_1;-x_1)={\exp\left({\kappa\over 2}~m x_1\right)\over 2\cosh
 \left({\kappa\over 2}~m x_1\right)}~.
\end{eqnarray}
It is convenient to consider the case of small $x_1$.
Then we can consider a discrete
version of the above Brownian motion:
\begin{equation}
 \Delta x =\Delta z+m\Delta t~,
\end{equation}
where
\begin{equation}
 \Delta z=\epsilon \sqrt{\Delta t}~.
\end{equation}
Here $\epsilon$ is a normally distributed random variable with a mean of
zero and unit variance. Then the corresponding binomial model will, at any
given value of discrete time, have a step up $U$ and a step down $D$ with
the probabilities $p$ and $1-p$, respectively. The mean and the variance
of $\Delta x$ are given by
\begin{eqnarray}
 &&\langle \Delta x\rangle = pU+(1-p)D~,\\
 &&\langle (\Delta x)^2\rangle =pU^2+(1-p)D^2~.
\end{eqnarray}
On the other hand, we know that (in the second equation below we are neglecting
a term of order $(\Delta t)^2$)
\begin{eqnarray}
 &&\langle \Delta x\rangle = m\Delta t~,\\
 &&\langle (\Delta x)^2\rangle =\Delta t~.
\end{eqnarray}
This gives two equations for three unknowns $U,D,p$. We, therefore, have some
freedom in choosing our binary model. As will become clear in a moment,
for our purposes here it is convenient to choose $D=-U$. Then we have:
\begin{eqnarray}
 &&U=-D=\sqrt{\Delta t}~,\\
 \label{p}
 &&p={1\over 2}\left[1+m\sqrt{\Delta t}\right]~.
\end{eqnarray}
Now consider $x_1=U=\sqrt{\Delta t}$ in (\ref{Q55}). In the context of the
above binomial model it is clear that $Q(0;x_1;-x_1)$ in this case
is nothing but the
probability that $x$ will make a step up (while $Q(0;-x_1;x_1)$ is the
probability that $x$ will make a step down). That is,
\begin{equation}
 Q(0;x_1;-x_1)=Q(0;U;-U)=p~.
\end{equation}
On the other hand, from (\ref{Q55}) we have (here we are neglecting
the ${\cal O}(U^2)$ terms)
\begin{equation}
 Q(0;U;-U)={1\over 2}\left[1+{\kappa\over 2}~mU\right]=
 {1\over 2}\left[1+{\kappa\over 2}~m\sqrt{\Delta t}\right]~.
\end{equation}
Comparing this with (\ref{p}) we obtain $\kappa=2$.

{}Finally, let us compute $Q(0;3;-5)$.
As in Problem 2 we proceed as follows.
We have
\begin{eqnarray}
 Q(0;3;-5)&=&1-Q(0;-5;3)~,\\
 Q(0;-5;3)&=&Q(0;-3;3)Q(-3;-5;3)=Q(0;-3;3)Q(0;-2;6)=\nonumber\\
 &=&{\exp(-3m)\over 2\cosh(3m)}
 ~Q(0;-2;6)~,\\
 Q(0;-2;6)&=&1-Q(0;6;-2)~,\\
 Q(0;6;-2)&=&Q(0;2;-2)Q(2;6;-2)=Q(0;2;-2)Q(0;4;-4)=\nonumber\\
 &=&{\exp(6m)\over 4\cosh(2m)\cosh(4m)}~.
\end{eqnarray}
Putting all of this together, we obtain:
\begin{eqnarray}
 Q(0;3;-5)&=&1-{\exp(-3m)\over 2\cosh(3m)}\left[1-
 {\exp(6m)\over 4\cosh(2m)\cosh(4m)}\right]~,\nonumber\\
 &=&{{\exp(9m)+\exp(5m)+\exp(m)+2\cosh(3m)}\over 8\cosh(2m)\cosh(3m)
 \cosh(4m)}~.
\end{eqnarray}

{}{\bf Note.} The binary tree approach gives us an immediate answer to Problem 2 above, where we have no drift. Thus, consider the probability $P(0; a; -b)$ (in the notations of Problem 2) with $a,b > 0$. Since there is no drift, and since the Brownian motion has no scale, it is clear that
\begin{equation}
 P(0;a;-b) = f(\lambda)~,
\end{equation}
where $f$ is some function and $\lambda\equiv a/b$. Furthermore, we have
\begin{eqnarray}
 &&P(0;a;-b) + P(0;-b;a)=1~,\\
 &&P(0;b;-a) = P(0;-b;a)~,
\end{eqnarray}
so
\begin{equation}
 f(\lambda) + f(1/\lambda) = 1~.
\end{equation}
As above let us now consider a binary tree with a step up $U$ (probability $p$), and a step down $D$ (probability $1-p$). The driftlessness implies that
\begin{equation}
 pU + (1-p)D = 0~,
\end{equation}
so $p = -D/(U-D)$, $U$ is positive, and $D$ is negative. Furthermore, $P(0;U;D) = p$. However, above we established that $P(0;U;D) = f(\lambda)$, where $\lambda = -U/D$. This then implies that
\begin{equation}
 f(\lambda) = {1\over{1+\lambda}}
\end{equation}
and
\begin{equation}
 P(0;a,-b) = {b\over{a + b}}~.
\end{equation}
So, $P(0;3;-5) = 5/8$.

\bigskip

{}{\bf Problem 3.} If $X$, $Y$ and $Z$ are 3 random variables such that $X$ and
$Y$ are 90\% correlated and $Y$ and $Z$ are 80\% correlated, what is the
minimum correlation that $X$ and $Z$ can have?

{}{\bf Answer.} The minimum possible correlation between $X$ and $Z$ is
approximately $45.8\%$. (The maximum possible correlation between
$X$ and $Z$ is approximately $98.2\%$.)

{}{\bf Solution.} For notational convenience let us introduce the
following notation: $X_1\equiv X$, $X_2\equiv Y$, $X_3\equiv Z$. We can write
$X_i$, $i=1,2,3$, as linear combinations of some independent random variables
$P_i$ with unit variances and zero correlations:
\begin{equation}
 \langle P_i P_j\rangle=\delta_{ij}~,
\end{equation}
where $\langle A\rangle$ denotes the expectation value of $A$. Thus,
\begin{equation}
 X_i=\Lambda_{ij} P_j~,
\end{equation}
where $\Lambda_{ij}$ are real coefficients, and
summation over repeated indices is implicit.
Then we have:
\begin{equation}
 M_{ij}\equiv \langle X_iX_j\rangle=\Lambda_{ik}\Lambda_{jk}~,
\end{equation}
or in the matrix form
\begin{equation}\label{M}
 M=\Lambda\Lambda^{T}~,
\end{equation}
where superscript $T$ denotes transposition.

{}From (\ref{M}) we have the following condition:
\begin{equation}
 \det(M)=\left[\det(\Lambda)\right]^2\geq 0~.
\end{equation}
On the other hand, we have
\begin{equation}
 M=\left(
 \begin{array}{ccc}
 \sigma_1^2 & \rho_{12}\sigma_1\sigma_2 & \rho_{13}\sigma_1\sigma_3 \cr\\
 \rho_{12}\sigma_1\sigma_2 & \sigma_2^2 & \rho_{23}\sigma_2\sigma_3 \cr\\
 \rho_{13}\sigma_1\sigma_3 & \rho_{23}\sigma_2\sigma_3 & \sigma_3^2 \cr
 \end{array}\right)~,
\end{equation}
where $\sigma_i^2$ is the variance of the random variable $X_i$, and
$\rho_{ij}$, $i\not= j$ is the correlation between the variables $X_i$ and
$X_j$. In terms of $\sigma_i$ and $\rho_{ij}$ we have:
\begin{equation}
 \det(M)=\sigma_1^2\sigma_2^2\sigma_3^2\left[1+2\rho_{12}\rho_{23}\rho_{13}-
 \rho_{12}^2-\rho_{23}^2-\rho_{13}^2\right]~.
\end{equation}
Since $M$ must be positive semi-definite, we have the following condition:
\begin{equation}\label{condrho}
 1+2\rho_{12}\rho_{23}\rho_{13}-
 \rho_{12}^2-\rho_{23}^2-\rho_{13}^2\geq 0~.
\end{equation}
The roots of the corresponding quadratic equation for $\rho_{13}$
\begin{equation}
 \rho_{13}^2-2\left(\rho_{12}\rho_{23}\right)\rho_{13}+\left(\rho_{12}^2+
 \rho_{23}^2-1\right)=0
\end{equation}
are given by
\begin{equation}
 \rho_{13}^{\pm}=\rho_{12}\rho_{23}\pm\sqrt{1+\rho_{12}^2\rho_{23}^2-
 \rho_{12}^2-\rho_{23}^2} = \rho_{12}\rho_{23}\pm {\overline\rho}_{12}{\overline\rho}_{23}~,
\end{equation}
where
\begin{eqnarray}
 &&{\overline\rho}_{12}\equiv\sqrt{1-\rho_{12}^2}~,\\
 &&{\overline\rho}_{23}\equiv\sqrt{1-\rho_{23}^2}~.
\end{eqnarray}
It is not difficult to see that for any values of $\rho_{12}$ and $\rho_{23}$
between $-1$ and 1, we have $\rho_{13}^-\geq -1$, and $\rho_{13}^+\leq 1$.

{}Next, to satisfy the condition (\ref{condrho}), we must have
\begin{equation}
 \rho_{13}^-\leq\rho_{13}\leq\rho_{13}^+~,
\end{equation}
so that the minimum possible correlation $\rho_{13}$ is
\begin{equation}
 \left(\rho_{13}\right)_{\rm{\scriptstyle{min}}}=\rho_{13}^-~.
\end{equation}
In our case $\rho_{12}=.9$, and $\rho_{23}=.8$, so $\rho_{13}^-\approx .458$,
so that the minimum possible correlation between $X$ and $Z$ is approximately
$45.8\%$. (Similarly, $\rho_{13}^+\approx .982$, so that the maximum
possible correlation between $X$ and $Z$ is approximately $98.2\%$.)

\bigskip

{}{\bf Problem 4.} Suppose two cylinders each with radius 1 intersect at right
angles and their centers also intersect. What is the volume of the
intersection?

{}{\bf Answer.} 16/3.

{}{\bf Solution.} This problem can be solved in the following standard way.
Let one of the cylinders have its axis along the $z$-axis, while the other
one along the $y$-axis. Then the boundary ${\cal B}=\partial{\cal M}$
of the intersection ${\cal M}$ is described by
the following set of equations:
\begin{eqnarray}
 &&x^2+y^2=1~,\\
 &&x^2+z^2=1~.
\end{eqnarray}
To find the volume of the intersection we must compute the integral
\begin{equation}
 V_{\cal M}=\int_{\cal M} dxdydz~.
\end{equation}
To compute this integral, it is convenient to divide the intersection into
8 octants, compute the volume of any one octant, and multiply the
answer by 8. This is because in computing the above integral we will encounter
a square root, for which we will have to choose an appropriate branch (which
corresponds to choosing an octant, or, more precisely, a set of octants). To
avoid this, we can use the symmetry of the problem, and compute the volume of
an individual octant.

{}Thus, let us compute the volume of the octant for which $0\leq x,y,z\leq 1$.
The corresponding integral is given by
\begin{equation}
 V_1=\int_{{\cal D}_1} dxdy \int_0^{\sqrt{1-x^2}} dz=
 \int_{{\cal D}_1} dxdy \sqrt{1-x^2}~,
\end{equation}
where ${\cal D}_1$ is the first quarter of the disk of unit radius in the $xy$
plane: $x^2+y^2\leq 1$, $0\leq x,y\leq 1$. To compute $V_1$ let us change the
$x,y$ integration to that over the corresponding polar coordinates:
\begin{equation}
 x=\rho\cos(\phi)~,~~~y=\rho\sin(\phi)~.
\end{equation}
In the polar coordinates ${\cal D}_1$ is given by $0\leq \rho\leq 1$,
$0\leq \phi\leq \pi/2$. Thus, our integral becomes:
\begin{eqnarray}
 V_1&=&\int_0^{\pi/2}d\phi\int_0^1 d\rho~\rho\sqrt{1-\rho^2\cos^2(\phi)}=
 \nonumber\\
 &=&\int_0^{\pi/2}d\phi\left(-{1\over 3\cos^2(\phi)}\right)\left.
 \left(1-\rho^2\cos^2(\phi)\right)^{3/2}\right|_0^1=\nonumber\\
 &=&{1\over 3}\int_0^{\pi/2}d\phi~{{1-\sin^3(\phi)}\over \cos^2(\phi)}=
 {1\over 3}\int_0^{\pi/2}d\phi\left[{1\over\cos^2(\phi)}-
 {\sin(\phi)\over\cos^2(\phi)}+\sin(\phi)\right]=
 \nonumber\\
 &=&{1\over 3}\left.\left[\tan(\phi)-{1\over\cos(\phi)}-\cos(\phi)\right]
 \right|_0^{{\pi\over 2}-\epsilon}~.
\end{eqnarray}
In the last line we have introduced an infinitesimal shift in the
upper integration limit ($\epsilon>0$)
to carefully treat the fact that $\tan(\phi)$ as well
as $1/\cos(\phi)$ blow up as $\phi\rightarrow\pi/2$. At the end of the day we
will take $\epsilon\rightarrow 0$.

{}Thus, we have
\begin{equation}
 V_1={2\over 3}~,
\end{equation}
and the volume of the intersection is
\begin{equation}
 V_{\cal M}=8V_1={16\over 3}~.
\end{equation}
Note that this volume is somewhat larger than the volume $V_{\rm{\scriptstyle{
ball}}}={4\pi\over 3}$ of a unit ball, which is consistent with the fact that
the intersection ${\cal M}$ contains a unit ball centered at the center of the
intersection.

\bigskip

{}{\bf Problem 5.} Consider the following C program for producing Fibonacci
numbers:

\medskip
\noindent
{\tt int Fibonacci(int n)}\\
{\{}\\
\indent
{\tt
{}if (n<=0 || n==1)\\
\indent
{}~~~return 1;\\
\indent
else\\
\indent
{}~~~return Fibonacci(n-1)+Fibonacci(n-2);\\
}
\}

\medskip
\noindent
If for some large {\tt n}, it takes 100 seconds to compute {\tt Fibonacci(n)},
how long will it take to compute {\tt Fibonacci(n+1)}, to the nearest second?

{}{\bf Answer.} Approximately ${{1+\sqrt{5}}\over 2}\times 100~{\rm seconds}
\approx 162~{\rm seconds}$.

{}{\bf Solution.} The above program does the following. Let $F(n)$ be
{\tt Fibonacci(n)}. Then the first step sets $F(n\leq 1)=1$. For $n>1$ the
second step computes $F(n)$ via
\begin{equation}
 F(n)=F(n-1)+F(n-2)~.
\end{equation}
So the entire process can be viewed as a binary tree where the top of this
tree is $F(n)$, which is computed by adding two numbers $F(n-1)$ and $F(n-2)$,
$F(n-1)$ is computed by adding $F(n-2)$ and $F(n-3)$, while $F(n-2)$ is
computed by adding $F(n-3)$ and $F(n-4)$, and so on. A particular branch ends
if we hit $F(1)$ or $F(0)$. From this binary tree we see that the time $T(n)$
that it takes to compute $F(n)$ is given by
\begin{equation}\label{T}
 T(n)=T(n-1)+T(n-2)+\Delta(n)~,
\end{equation}
where $\Delta(n)$ is the time required to call $F(n-1)$ and $F(n-2)$, and then
add them. We do not have enough information to determine $\Delta(n)$. However,
as we will see in a moment, we actually do {\em not} need it. All we need is
that for large $n$ the ratio $\Delta(n)/T(n)$ goes to zero, which is a
reasonable assumption.

{}From (\ref{T}) we have
\begin{equation}
 {T(n+1)\over T(n)}=1+{T(n-1)\over T(n)}+{\Delta(n+1)\over T(n)}~,
\end{equation}
or, equivalently,
\begin{equation}\label{Q}
 Q(n)=1+{1\over Q(n-1)}+{\Delta(n+1)\over T(n)}~,
\end{equation}
where
\begin{equation}
 Q(n)\equiv {T(n+1)\over T(n)}~.
\end{equation}
The last term in (\ref{Q}) goes to zero for large $n$. It is then clear that
$Q(n)$ has a finite non-zero limit as $n\rightarrow\infty$, call it $Q_*$.
From (\ref{Q}) we see that
\begin{equation}
 Q_*=1+{1\over Q_*}~.
\end{equation}
Solving this equation (and keeping the positive root), we obtain
\begin{equation}
 Q_*={{1+\sqrt{5}}\over 2}\approx 1.62~.
\end{equation}
Thus, if for some large $n$ it takes $T(n)=100~{\rm seconds}$ to compute
$F(n)$, then it takes $T(n+1)\approx Q_* T(n)\approx 162~{\rm seconds}$
to compute $F(n+1)$.

\bigskip

{}{\bf Problem 6.} Show that $p^2-1$ is divisible by 24 if $p$ is a prime number, $p > 3$.

{}{\bf Solution.} $p = 2n+1~(n > 1) \Rightarrow p^2-1 = 4n(n+1) \Rightarrow p^2-1$ is divisible by 8. $p=3m\pm 1~(m \geq 2)\Rightarrow p^2-1 =3m(3m\pm 2) \Rightarrow p^2-1$ is divisible by 3.

\bigskip

{}{\bf Problem 7.} You have $N$ random variables taking values between 0 and 1. What is the expected value of the smallest one.

{}{\bf Answer.} $1/(N+1)$.

{}{\bf Solution.} Let
\begin{equation}
 I_N(a)\equiv \int_0^1 dx_1\dots dx_N \mbox{min}\left(x_1, \dots, x_N, a\right)~.
\end{equation}
What we need to compute is $I_N(1)$. We have
\begin{eqnarray}
 &&I_N(a) = \nonumber\\
 && = \int_0^1 dx_1\dots dx_N ~\mbox{min}\left(x_1, \dots, x_N, a\right) = \nonumber\\
 && = \int_0^1 dx_1\dots dx_N \left\{\mbox{min}\left(x_1, \dots, x_N\right) - \left(\mbox{min}\left(x_1, \dots, x_N\right) - a\right)^+\right\} = \nonumber\\
 && = I_N(1) - \int_a^1 dx_1\dots dx_N \left(\mbox{min}\left(x_1, \dots, x_N\right) - a\right) = \nonumber\\
 && = I_N(1) - \int_0^{1-a} d{\widetilde x}_1\dots d{\widetilde x}_N ~\mbox{min}\left({\widetilde x}_1, \dots, {\widetilde x}_N\right) =\nonumber\\
 && = I_N(1) - (1-a)^{N+1}\int_0^1 dy_1\dots dy_N ~\mbox{min}\left(y_1, \dots, y_N\right) =\nonumber\\
 && = I_N(1)\left[1 - (1-a)^{N+1}\right]~.
\end{eqnarray}
Furthermore,
\begin{eqnarray}
 I_{N+1}(1) = \int_0^1 da~I_N(a) = I_N(1)\left[1 - \int_0^1 da~(1-a)^{N+1}\right] = I_N(1)~{{N+1}\over{N+2}}~,
\end{eqnarray}
which recursion relation together with $I_1(1) = 1/2$ then implies that $I_N = 1/(N+1)$.

\bigskip

{}{\bf Problem 7a.} You have $N$ cars entering a one-lane highway at random speeds. What is the expected number of clusters?

{}{\bf Answer.} The expected number of clusters $E_N$ is given by the $N$-th harmonic number $H_N = \sum_{k=1}^N 1/k = \ln(N) + \gamma + {\cal O}(1/N)$, where $\gamma$ is the Euler constant.

{}{\bf Solution.} If we have $N$ cars on the highway and an $(N+1)$-th car enters, if its speed is lower than the expected minimum speed of the $N$ cars then it will form another cluster, otherwise it'll join the last existing cluster. Assuming all cars travel at random speeds between 0 and 1, the expected minimum speed of the $N$ cars is $1/(N+1)$ (see Problem 7), so the probability that the speed of the $(N+1)$-th car is lower is $P_{N+1} = 1/(N+1)$, and we have $E_{N+1} = E_N + P_{N+1}$, from which recursion relation together with $E_1=1$ it follows that $E_N=H_N$.

\section*{Bibliography}\addcontentsline{toc}{section}{Bibliography}

$\bullet$ M. Baxter and A. Rennie, ``Financial Calculus: An Introduction to Derivative Pricing",
Cambridge University Press (1996), 233 pp; also see references therein.\\
\noindent
$\bullet$ Z. Kakushadze, ``Path Integral and Asset Pricing", SSRN Working Papers Series, http://ssrn.com/abstract=2506430; also see references therein.

\end{document}